\documentclass[english,11pt,a4paper,{oneside}]{book}
\usepackage[utf8]{inputenc}
\usepackage[english]{babel}
\usepackage{setspace}
\usepackage{graphicx}
\usepackage{caption}
\usepackage{amsmath}
\usepackage{amsthm}
\usepackage{amssymb}
\usepackage{mathtools}
\usepackage{bigints}
\usepackage{makeidx}\usepackage[T1]{fontenc}
\usepackage{mathrsfs}
\usepackage{csquotes}
\usepackage{upgreek}
\usepackage{multicol}
\usepackage{float}

\usepackage{titlesec}
\newcommand{\hsp}{\hspace{20pt}}
\titleformat{\chapter}[hang]{\Huge\bfseries}{\thechapter\hsp{|}\hsp}{0pt}{\Huge\bfseries}

\usepackage{mathrsfs}
\theoremstyle{plain}

\usepackage{emptypage}

\usepackage{sectsty}

\usepackage{titlesec}

\titleformat{\chapter}[display]
  {\bfseries\Large}
  {\filright\MakeUppercase{\chaptertitlename} \Huge\thechapter}
  {1ex}
  {\titlerule\vspace{1ex}\filleft}
  [\vspace{1ex}\titlerule]

\newcommand{\eq}[1]{\begin{eqnarray} #1 \end{eqnarray}}

\newcommand{\al}[1]{\begin{align} #1 \end{align}}

\newcommand{\q}[1]{\mathbf{q}}
\newcommand{\p}[1]{\mathbf{p}}
\newcommand{\D}[2]{\frac{\textrm{d}#1}{\textrm{d}#2}}
\newcommand{\DP}[2]{\frac{\partial#1}{\partial#2}}

\usepackage{fancyhdr}
\fancypagestyle{plain}{ \fancyhead{}
\fancyfoot[C]{\thepage}

}
\makeindex
\date{}
\author{Livio De Fabrizio}
\title{\textsc{Linear and Nonlinear} \\ \textsc{Kinetic Alfvén Wave Physics} \\ \textsc{in Cylindrical Plasmas}}

\begin{document}
\begin{figure}
\vspace*{-3.8cm}
\hspace{-3.5cm} \includegraphics{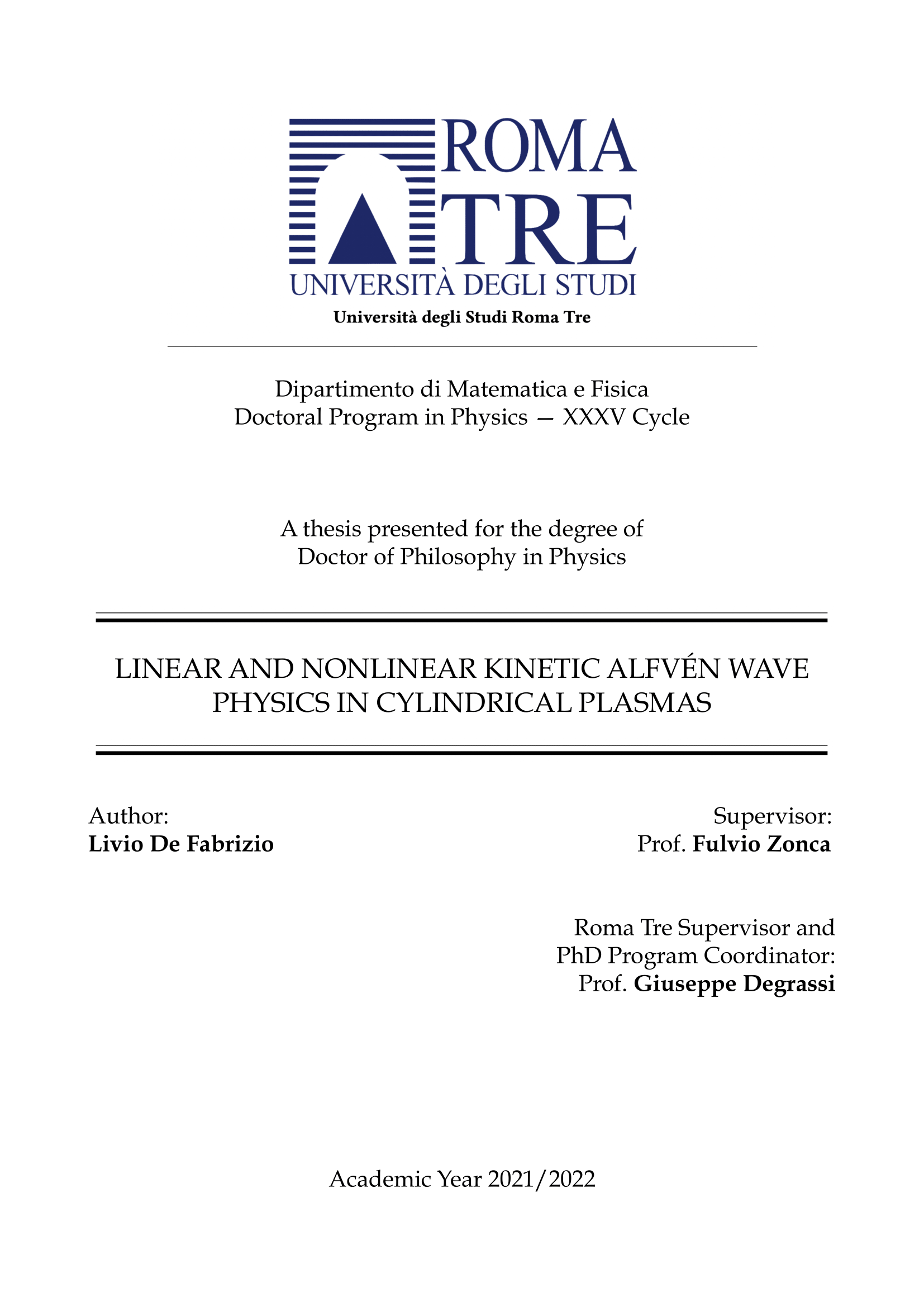}
\end{figure}
\thispagestyle{empty}
\frontmatter

\maketitle
\setlength{\baselineskip}{15pt}
\makeatletter

\cleardoublepage
\vspace*{\stretch{1}}
\markboth{}{}
\thispagestyle{empty}
\begin{flushright}
\itshape to my father
\end{flushright}
\vspace*{\stretch{3}}

\renewcommand\tableofcontents{\begingroup
    \if@twocolumn
      \@restonecoltrue\onecolumn
    \else
      \@restonecolfalse
    \fi
    \chapter*{\contentsname
        \@mkboth{%
           \MakeUppercase\contentsname}{\MakeUppercase\contentsname}}%
    \thispagestyle{empty}\pagestyle{empty}\@starttoc{toc}%
    \if@restonecol\twocolumn\fi
    \clearpage\endgroup
    }
\makeatother
\tableofcontents

\mainmatter

\chapter*{\textsc{introduction}}
\markboth{Introduction}{} 
\addcontentsline{toc}{chapter}{\textsc{introduction}}
\vspace{3cm}

The study of plasmas applies to a variety of fields, from plasma astrophysics to the physics of magnetospheres, including practical applications such as controlled thermonuclear fusion, plasma acceleration and industrial plasmas. Plasmas made of light element nuclei (e.g., deuterons) and electrons represent the natural choice to produce exothermic fusion reactions \cite{Stacey2012} which, being an analogue of those providing the life-sustaining energy of the sun and the other stars, are considered to be the most fundamental energy source of the universe. Controlled nuclear fusion would represent a clean and almost inexhaustible energy for mankind, potentially solving world's energy problem \cite{Chenhow}. The demonstration power plant (DEMO\footnote{https://www.euro-fusion.org/programme/demo/}) is the final aim of the EUROfusion research plan in magnetic confinement fusion, and targets achieving 300-500 MWe by 2050. Similar efforts are ongoing worldwide with broadly consistent objectives\footnote{https://www.iaea.org/bulletin/demonstration-fusion-plants}, while an international collaboration is working on the final construction phase of the International Thermonuclear Experimental Reactor (ITER\footnote{https://www.iter.org}).

Magnetized and non-collisional plasmas of space and fusion interest require non-linear kinetic theory. In particular, "fully kinetic" (FK) simulation schemes describe both ions and electrons by a Vlasov equation, which is often solved by particle-in-cell (PIC) techniques \cite{Birdsall2004,Dawson1983}, taking into account the dynamics of individual particles with their characteristic frequencies up to the electron cyclotron frequency $\Omega_e$, generally lower than or comparable to the electron plasma frequency. However, these schemes are expensive from the computational point of view and involve an extreme variety of spatiotemporal scales, explaining why reduced models have been introduced: the gyrokinetic (GK) model \cite{Frieman1982,Lee83,Lee87} uses an orbit-averaged description of the cyclotron motions, able to resolve physical processes by means of a discrete time interval larger than the ion cyclotron period. Also, when the frequencies $\omega$ are lower than or comparable to the ion cyclotron frequency $\Omega_i$, namely $\omega\ll\Omega_i\ll\Omega_e$, one can use the gyrokinetic approach for ions and simply neglect electrons gyromotion (drift-kinetic electrons), provided the spatial scale of interest is larger than the electron Larmor radius.

For the description of a variety of phenomena, a further, hybrid model has been recently developed, based on a GK description for the electrons \cite{Frieman1982,Hahm1988,Brizard1989} and on a FK description for the ions, called GeFi \cite{Lin2005}. This model is then suitable for frequencies which are, at most, intermediate between ion and electron cyclotron frequencies, for wavelengths perpendicular to the magnetic field which are small with respect to the parallel wavelength, $|k_\perp/k_\parallel|\gg1$, and for $|k_\perp\rho_e|\approx1$, where the electron Larmor radius $\rho_e$ is much smaller than the size of the system, $\rho_e\ll L$.

The GeFi model has only been validated in simple magnetic field geometries \cite{Zoncarev}. Meanwhile, GK codes constitute the leading numerical tools for advanced simulation studies in the complex magnetic field geometries characterizing strongly magnetized fusion plasmas \cite{Garbet2010,Lauber2013}. The main goal of the present Ph.D. thesis research is to provide a basis for bridging over between GK codes and the innovative approach proposed by GeFi by gradually extending the simple magnetic field geometries and nonuniform plasma equilibria that are typically investigated with this code. To this aim, we have identified the linear and nonlinear behavior of Alfvénic oscillations in a cylindrical plasma as the ideal test-bed. In fact, transverse shear Alfvén waves are fundamental electromagnetic oscillations in magnetized plasmas \cite{Alfven1942,Cramer2011}, which can importantly contribute in heating and transport processes of charged particles. The focus of this thesis work will be on establishing a test case to compare simulation results of the TRIMEG \cite{Lu2019,Lu2021} gyrokinetic code and of the STRUPHY \cite{Holderied2021} hybrid MHD-kinetic code with the linear and nonlinear theoretical predictions obtained by both analytical as well as numerical approaches on the generation of convective cells in a cylindrical magnetized plasma. In particular, the present thesis research will demonstrate that this problem includes in a nutshell important fundamental physics processes that are of relevance in both space and laboratory plasmas. Thus, the novel results presented here provide a first step to construct a clear and well-understood test case, where a first cylindrical version of GeFi could be tested against TRIMEG and STRUPHY codes. The possible future extension of this work will lead to the implementation of GeFi in realistic toroidal fusion plasmas equilibria as well as to the verification of the validity limits of the GK reduced description, which is the foundation of TRIMEG and of other GK codes.

Under Prof. Fulvio Zonca's supervision, this research work has been done at the Roma Tre University and at the ENEA center in Frascati (RM), has been supported by an international collaboration within the framework of CNPS\footnote{Center for Nonlinear Plasma Science, https://www.afs.enea.it/zonca/CNPS/}, and is connected with DTT\footnote{Divertor Tokamak Test facility,
https://www.dtt-project.it}.

The plan of the manuscript is as follows:
\begin{itemize}
\item Chapter 1 gives a general introduction to the plasma state, to the gyrokinetic model and to waves and oscillations in plasmas.
\item Chapter 2 focuses on Alfvén waves, which are the fundamental topic of this thesis work. To gradually introduce the underlying physics concepts, hydrodynamic waves in uniform plasmas are presented first and, then, the effect of plasma nonuniformity is discussed. In particular, we introduce the concept of spatial phase mixing and of resonant absorption, which naturally occur in nonuniform plasmas where shear Alfvén waves are characterized by a continuous spectrum and radial wavelength that decreases in time as $t^{-1}$. The spontaneous generation of increasingly shorter wavelengths leads us to introducing and discussing the gyrokinetic description that is necessary for analyzing kinetic Alfvén waves.
\item Chapter 3 introduces the cylindrical nonuniform magnetized plasma equilibrium that will be used throughout the rest of this work. In particular, we follow well-known existing literature to contextualize the behavior of shear Alfvén waves in the considered model plasma equilibrium. The properties of shear Alfvén waves launched by an external antenna are presented and the power transfer from the launched wave to the thermal plasma via resonant absorption is discussed in detail.
\item Chapter 4 revisits the ideal magneto-hydro-dynamic (MHD) analysis of chapter 3 using gyrokinetic theory. This illuminates the microscale phenomena that underlie the resonant absorption of the shear Alfvén wave launched by the external antenna at the radial location where the imposed antenna frequency matches the continuously varying frequency spectrum of the shear Alfvén wave in the considered nonuniform cylindrical plasma equilibrium. Consistent with earlier analyses \cite{Hasegawa1975,Hasegawa1976}, resonant absorption consists in the mode conversion to the short wavelength kinetic Alfvén wave, which is typically absorbed within a short distance from the resonant layer and within a few spatial oscillation period of the mode converted wave. This case of strong absorption was and has been the focus of existing studies so far due to its direct relevance for plasma heating \cite{Hasegawa1975,Hasegawa1976,Itoh1982,Itoh1983}. In this work, we also consider the opposite weak absorption case, where, even with a modest amplitude shear Alfvén wave launched by the antenna, large kinetic Alfvén waves can be excited inside the resonant (mode conversion) layer and yield a number of interesting nonlinear behaviors. As new novel results of the present work, it is demonstrated that the plasma region inside the mode conversion layer behaves as a "resonant cavity"; that is, when the antenna frequency matches a given discrete spectrum, the plasma response is stronger, exhibiting the clear behavior of a resonantly driven and weakly damped oscillator. This behavior is of particular interest for our scope to investigate nonlinear behaviors induced by large amplitude kinetic Alfvén waves, which can be fine-tuned in phase and amplitude by the external antenna.
\item Chapter 5 contains the theoretical analysis of nonlinear behaviors of kinetic Alfvén waves in the considered model plasma equilibrium and corresponding numerical solutions. The focus is on the spontaneous generation of convective cells by a large kinetic Alfvén wave. However, since the underlying physics is connected with the so-called parametric decay instability, this process is introduced first, following a recent review work \cite{Zoncarev} that allows us to qualitatively describe its essential elements. The importance of convective cells is then presented, along with a systematic derivation of the equations governing their nonlinear interaction with kinetic Alfvén waves. Maximization of nonlinear coupling demonstrates that optimal conditions for convective cell generation in cylindrical plasmas are those of a "theta pinch" (or "$\theta$-pinch") equilibrium, where the nonuniform magnetic field is directed along the symmetry axis only. This allows a short azimuthal wavelength convective cell to be excited most easily. Meanwhile, electron Landau damping can also be minimized by the possibility to achieve relatively high ratio of electron plasma kinetic energy density normalized to the magnetic energy density (in the order of 10\% or more), due to the intrinsic stability of this configuration. Earlier results of convective cell excitation by kinetic Alfvén waves in uniform plasmas \cite{Zonca2015} are recovered in the proper limiting case. The effect of plasma nonuniformity is shown to importantly affect the convective cell stability both qualitatively and quantitatively via the diamagnetic response \cite{Chen2022,Kaw1983}. As new novel result of this thesis work, it is shown that the convective cell rotates in the electron diamagnetic direction near the critical excitation threshold. Meanwhile, the convective cell growth is significantly stronger (typically up to an order of magnitude) than in uniform plasmas; and the unstable parameter region is significantly broader. Furthermore, the subtle interplay between nonlinearity and plasma nonuniformity shows that the plasma self-organization can be controlled by fine-tuning the amplitude of the antenna driven mode converted kinetic Alfvén wave. The convective cell (generally mixed) polarization and radial structure of the generated inductive parallel electric field are also computed self-consistently.
\item Chapter 6, finally, presents concluding remarks as well as a discussion of possible applications and outlook of the present Ph.D. thesis work.
\end{itemize}

\subsection*{Acknowledgments}

I wish to thank my supervisor, Prof. Fulvio Zonca, for his human and professional example. His knowledge, support, and patience have been basic ingredients for this thesis work and will always constitute a model for my life. I also express my appreciation to my Roma Tre advisor Prof. Giuseppe Degrassi and to my colleague Dr. Matteo Valerio Falessi for their fundamental presence and their experience, and to all the members of CNPS. Likewise, useful discussions with Dr. Zhixin Lu and Dr. Xin Wang are kindly acknowledged.

Then, I am extremely grateful to my parents, thanks to whom I am. My entire school and university careers would not have been possible without them, who raised, educated, encouraged, and financed me.

This thesis is dedicated to the memory of my father (1943-2016).

\chapter{\textsc{generalities on plasma physics}}

\vspace{100pt}

Since ancient times, mankind has been searching for the unity underlying the apparent multiplicity of things; in particular, the observation of nature suggested that the great variety of substances and phenomena all around can be considered as a variety of manifestations and combinations of few elementary entities, or elements. According to the Greek cultural tradition, the basic elements are four: earth, water, air, and fire. Actually, having developed neither an advanced chemistry nor a microscopic theory of matter, Greeks couldn't identify elementary substances as we mean today, i.e. chemical elements or, even further, elementary particles. They were rather able to identify the fundamental states of matter: solid, liquid, gas, and plasma (subsequently, Aristotle introduced, motivated by astronomical reasons, a fifth element, the aether or quintessence, which can be thought as the "state of absolute space", namely the spatial texture or, possibly, the cosmic vacuum). Any macroscopic object we may find will be constituted by parts that are in one ore more of these four states, so the reduction made by Greeks was a macroscopic one, pointing to constituent phases rather than to constituent particles.

Fire, and lightnings, are made of plasma, in the sense that they are spectacular manifestations of matter when in the plasma state. However, plasmas are not that common on Earth: on the contrary, they are almost all we can see when looking at the sky with eyes or telescopes; in fact, stars are essentially plasma objects, and plasma is a main constituent of interstellar medium, too. Therefore, it is no coincidence that plasma physics only began to develop in the twentieth century, though, as mentioned above, the existence of plasma phase had somehow been stated well before.

\section{The plasma state}

\begin{figure}[h]
	\centering
	\includegraphics[scale=0.20]{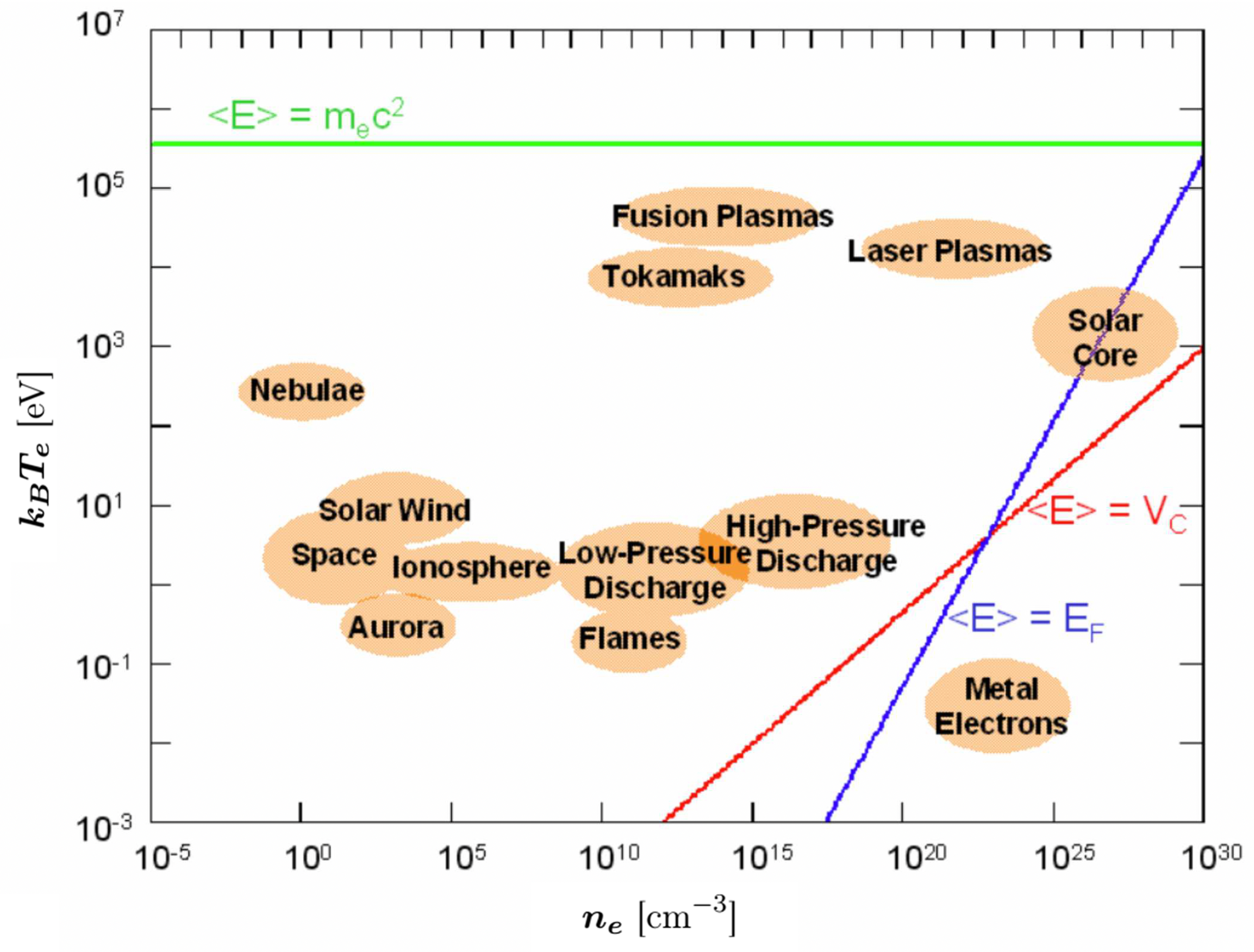}
	\caption{\emph{Fig.\ref{Phasediag}} A plasma phase diagram, showed as a function of (electron) density $n_e$ and (electron) temperature $k_BT_e$. Those to the right of the red line are the strongly-interacting plasmas, whose mean kinetic energy is lower or at most comparable to the interaction energy; to the right of the blue line, electrons occupy the energetic levels in the "Fermi sea" (metal electrons, white dwarfs).}\label{Phasediag}(Source: UniTus Master Lectures by F. Zonca\footnotemark)
\end{figure}

The reason for the plasma state to be considered the fourth state of matter is the following. When heated up to its melting point, a solid passes to the liquid state and, if temperature further increases up to the boiling point, a gas is obtained through evaporation. However, that's not the end of the story. When one insists in heating up towards very high temperatures, the gas molecules (not necessarily all) ionize: the result is a system which is almost neutral both as a whole and in all its wide enough parts, but now composed of charged particles whose collisions, i.e. gas-like, short-range interactions, though still more or less probable, are not as important as their medium- or long-range electromagnetic (essentially, Coulombian) interactions (and whose mean kinetic energy is greater than the mean interaction energy). In short, the result is a quasineutral gas-like system with collective behavior, or, a plasma. In the next subsections, this definition is discussed in some more detail; as we shall see, the fundamental quantities that characterize a plasma are the density of particles, $n$, and temperature, $T$ (generally, electron temperature, $T_e$). Fig.\ref{Phasediag} considers important examples of plasmas.

\vspace{0.5cm}
\footnotetext{https://www.afs.enea.it/zonca/Teaching/UniTus/lecture1.pdf}

\subsection{Quasineutrality: $L\gg\lambda_D$}

It has been stated that a plasma is almost neutral not only as a whole, as we expect a ionized gas to be, but also in its little constituent parts. The fact is, though the generic single particle is charged, in a large enough volume of plasma we expect to find a sufficient amount of both positively and negatively charged particles such that the resulting net charge $Q$ is almost vanishing. Namely, if there are two species of particles, one with charge $q_+$ and density (particle density or probability density) function $f_+$, and the other with charge $q_-$ and density $f_-$, then, for a large enough volume $V$, much smaller than the total volume of the system, \al{
Q=\int_V\left(f_+q_++f_-q_-\right)\,\textrm{d}V\approx0,\label{quasin}
} which generically expresses the condition of quasineutrality. The particles in this volume are so many that the charge of each particle is well screened, in any direction, by surrounding particles carrying opposite charge. The parameter able to describe this situation and to study quasineutrality and collective motion (subject of the next subsection) from a quantitative point of view is the Debye length, which is a common feature of any ionized system in thermal equilibrium.  Keeping in mind the Gauss-Maxwell equation $\nabla\cdot E=4\pi\sigma$, with $\sigma$ charge density, the equation for the potential $\phi$ due to a 
charge $q_0$ located at $\mathbf{x}_0$ is Poisson's equation \al{
\nabla^2\phi=-4\pi q_0\,\updelta(\mathbf{x}-\mathbf{x}_0),\label{Poisson}
} whose solution is $\phi=q_0/r$, being $r:=|\mathbf{x}-\mathbf{x}_0|$. When the potential is applied to a plasma, the particles contribute to the right-hand side with their positive and negative charge densities, and \eqref{Poisson} is extended as \al{
\nabla^2\phi=-4\pi q_0\,\updelta(\mathbf{x}-\mathbf{x}_0)-4\pi\left(q_+n_+-q_-n_-\right),
} or simply, by considering from now on the case of a plasma composed by ions of charge $q_i=Ze$, $Z\in\mathbb{N}$, and by electrons (of charge $q_e=-e$), \al{
\nabla^2\phi=-4\pi q_0\,\updelta(\mathbf{x}-\mathbf{x}_0)-4\pi e \left(Zn_i-n_e\right), \label{Poisson2}
} where $n_i$ and $n_e$ are the ionic and electronic densities, computed as the integral over velocity $\mathbf{v}$ of their Maxwell-Boltzmann distribution functions, \al{
n_{i,e}=\int A_{i,e}e^{-\frac{1}{k_BT_{i,e}}\left(\frac{1}{2}m_{i,e}v^2+q_{i,e}\phi\right)}\,\textrm{d}\mathbf{v}=n_{i,e}^\infty e^{-\frac{q_{i,e}\phi}{k_BT_{i,e}}},\label{MBdist}
} where $n_{i,e}^\infty=n_{i,e}(r\rightarrow\infty)=n_{i,e}(\phi\rightarrow0)$ are the densities far away from the source, namely the densities of the plasma particles when they are free. Noting that the mean kinetic energy of particles is larger than the mean interaction energy among particles themselves, such that $|q_{i,e}\phi|\ll k_BT_{i,e}$, the previous expression can be approximated as \al{
n_{i,e}\approx n_{i,e}^\infty\left(1-\frac{q_{i,e}\phi}{k_BT_{i,e}}\right)\label{densityapprox}
} and, introducing the \emph{Debye length} \al{
\lambda_D:=\sqrt{\frac{k_B}{4\pi e^2\left(\frac{Zn_i^\infty}{T_i}+\frac{n_e^\infty}{T_e}\right)}},
} equation \eqref{Poisson2} becomes \al{
\nabla^2\phi=-4\pi q_0\updelta(\mathbf{x}-\mathbf{x}_0)+\frac{4\pi e^2}{k_B}\left(\frac{Zn_i^\infty}{T_i}+\frac{n_e^\infty}{T_e}\right)\phi=-4\pi q_0\updelta(\mathbf{x}-\mathbf{x}_0)+\frac{\phi}{\lambda_D^2},
} that is \al{
\left(\frac{1}{\lambda_D^2}-\nabla^2\right)\phi=4\pi q_0\updelta(\mathbf{x}-\mathbf{x}_0),
} whose solution is \al{
\phi=\frac{q_0}{r}e^{-\frac{r}{\lambda_D}}.\label{PoissonSolution}
} So, the \textit{bare} potential, that \textsl{per se} decreases as the inverse of distance, is further screened by the presence of the other charges, thus decaying at a rate regulated by the Debye length, which represents the length over which the potential can be considered essentially screened: this feature is known as \emph{Debye shielding}. The Debye length is, in particular, a distance at and beyond which $|q_{i,e}\phi|\ll k_BT_{i,e}$ holds: indeed, this was exactly the basic assumption for obtaining \eqref{densityapprox} and, then, \eqref{PoissonSolution}. At a shorter distance $r^*$, such that \al{
e\phi(r^*)\approx k_BT_e,
} the electrons around the source charge are not enough to have complete screening efficiency, while their kinetic energy is less important than or comparable with the particle interaction energy. It is no surprise, then, that the Debye length increases with temperature and decreases with density. When, in any direction, the length $L$ of the plasma is very large compared to the Debye length, $L\gg\lambda_D$ \cite{Chenint}, and so the volume of the system is much larger than the Debye volume $V_D\sim\lambda_D^3$, the Debye scales may be considered infinitesimal and could be ignored when studying physical processes occurring at longer length scales. As a consequence, the charge density screening may be considered effective at any arbitrary point and allows for the approximation $n_i\approx n_e=n$, where, for the simple case of singly charged ions, $n$ is a common density value dubbed the \emph{plasma density}. Reconsidering the quasi-neutrality condition expressed in a generic way by \eqref{quasin}, where now $V_D$ is approaching zero, $V$ can be taken as any small but finite volume, and the quasi-neutrality condition becomes one of the fundamental governing equations for plasma response at $k^2\lambda_D^2\ll1$, with $k$ the relevant characteristic wave number.

Since $m_i\gg m_e$, the motion of ions can often be neglected with respect to that of the electrons, so one treats ions as stationary neutralizing background: $n_i$, according to \eqref{MBdist}, can be neglected and the only density entering in \eqref{Poisson2} is $n_e$; still, \eqref{PoissonSolution} holds, but now with the Debye length \al{
\lambda_D:=\sqrt{\frac{k_BT_e}{4\pi e^2n_e^\infty}}.
}

\subsection{Collective motion: $N_D\gg1$ and $\omega_p\tau\gg1$}\label{collective}
As kinetic, statistical property, the screening effect holds provided \cite{Chenint} a high number $N_D\gg1$ of particles is contained in a Debye sphere, that is, in other terms, provided $n\lambda_D^3\gg1$: the mean interparticle volume, equal to $n^{-1}$, is much smaller than Debye sphere's volume. This huge collection of $N_D$ particles naturally organizes to (quasi)neutralize the plasma, the small charge left giving rise, as suggested by the discussion made in the previous subsection, to potentials at most of the order $k_BT_e$ and, so, to interactions on longer distances. As a consequence, any electromagnetic field applied to a plasma produces a perturbation in all the particles (due to their charge) quite simultaneously, and the rearrangement of one particle affects the redistribution of the others, resulting in a collective response by the system: the basic example is given by electron plasma waves (see subsection \ref{plasmawaves}), namely the small electron oscillations about their equilibrium position. The fact is that plasmas are constituted by particles which are charged (ions and electrons) and quite free to move, so most of them experience, collectively, electromagnetic long-range interactions, rather than collisions: precisely, the relaxation time $\tau$ is long when compared with the time taken by particles to undergo a Coulomb (binary) collision, which is of the order of the inverse of the so-called \textit{plasma frequency} $\omega_p$ (see subsection \ref{plasmawaves}), namely the frequency of oscillation about the equilibrium position. In summary, $\omega_p\tau\gg1$. Ultimately, a plasma shows a collective behavior: this feature makes a distinction with ordinary gases and, also, with systems which, though ionized, are strongly coupled. In fact, being globally neutral and, at most, polar, gas molecules give rise to somewhat weak, and generally binary, interactions: only when two molecules are close enough, they really experience intermolecular forces such as van der Waals forces or hydrogen bonding and, ultimately, the repulsion caused by their respective cohesion forces, which prevents interpenetration. So, gases are dominated by short-range interactions, i.e. collisions, which are essentially binary, random and independent one from the other, while the strong Coulombian and electromagnetic interactions, dominant in plasmas, are only found inside molecules and, in particular, between internal nuclei and electrons.

\section{Plasma dynamics}

Like gaseous systems, also plasmas admit a thermomechanical description both from the macroscopic point of view, namely by means of fluid mechanics and classical thermodynamics, and from the microscopic point of view, namely by means of kinetic theory. Within the microscopic description, the use of modern supercomputers allows us to numerically simulate plasma behavior by following up to $10^{10}$ particle trajectories by the so-called \textit{particle-in-cell} (PIC) approach \cite{Birdsall2004}. Meanwhile, plasma particles carry electric charges that, even in the absence of external interactions, may generate perturbed internal electromagnetic fields. A first approach to study plasma physics is to build a description for ideal systems, i.e. to use the Boltzmann equation for ideal gases or the macroscopic equations of ideal fluids, along with Maxwell's equations in vacuum.

The two Boltzmann equations, ionic and electronic, express the time variation, due to collisions, of the distribution functions $f_{i,e}$ by means of an integral collision term $\mathscr{C}\left(f_{i,e}\right)$, which computes the transition rate from initial to final state and whose expression depends on the system under study and on the adopted mathematical model, \al{\begin{aligned}
\frac{\partial f_{i,e}}{\partial t}+\nabla_{\mathbf{x}}f_{i,e}\cdot\mathbf{v}+\nabla_\mathbf{v}f_{i,e}\cdot\frac{\mathbf{F}}{m}=\mathscr{C}\left(f_{i,e}\right),\end{aligned}
} where $\mathbf{F}$ is the generic force acting on the particle at $(\mathbf{x},\mathbf{v})$. As for the left-hand side, we can restrict $\mathbf{F}$ to the Lorentz force only, describing the aforementioned long-range electromagnetic interactions that regulate the plasma dynamics. Thus, considering, for simplicity, a hydrogen plasma, made of ions and electrons of charge $e$ and $-e$, respectively, the kinetic description is \al{
\frac{\partial f_{i,e}}{\partial t}+\nabla_{\mathbf{x}}f_{i,e}\cdot\mathbf{v}_{i,e}\pm\frac{e}{m_{i,e}}\nabla_\mathbf{v}f_{i,e}\cdot\left(\mathbf{E}+\frac{\mathbf{v}_{i,e}}{c}\times\mathbf{B}\right)=\mathscr{C}(f_{i,e}).\label{boltzmann}
}

The macroscopic counterpart of this description is deduced by taking the appropriate moments of the kinetic equations \cite{Freidberg}, and is represented by the equation of continuity for mass densities $\varrho_{i,e}$ and by the cardinal dynamical equations, which together define a two-fluid model: \begin{gather}
\frac{\partial\varrho_{i,e}}{\partial t}+\nabla\cdot\varrho_{i,e}\mathbf{v}_{i,e}=0,\label{continuitytwo}\\
\varrho_{i,e}\frac{\textrm{d}\mathbf{v}_{i,e}}{\textrm{d}t}=\sigma_{i,e}\left(\mathbf{E}+\frac{\mathbf{v}_{i,e}}{c}\times\mathbf{B}\right)-\nabla p_{i,e},\label{motiontwo}
\end{gather} where $\varrho_{i,e}=n_{i,e}m_{i,e}$ and $\sigma_{i,e}=\pm n_{i,e}e$. However, due to quasineutrality, ionic and electronic densities are the same (plasma density $n$): this further implies that ionic and electronic charge densities are opposite, $\sigma_i=ne=-\sigma_e$, and, then, that the electric part of the total Lorentz force density, $\mathbf{F}_E=(\sigma_i+\sigma_e)\mathbf{E}$, vanishes; moreover, considering that $m_i\gg m_e$, mass density is essentially ionic mass density, $\varrho=\varrho_i$, and as a consequence, although electronic velocity $\mathbf{v}_e$ is higher than $\mathbf{v}_i$, the inertial force density on electrons is negligible with respect to the force density on ions, $\varrho\,\textrm{d}\mathbf{v}_i/\textrm{d}t$. What's more, assuming that the frequency of collisions is enough to randomize all individual motions, so that positively and negatively charged particles have the same, uniform temperature and, as a consequence, exhert the same, uniform pressure, we can also introduce an equation of state. Ultimately, we are led to the single-fluid model, described by only one set of equations, \begin{gather}
\frac{\partial\varrho}{\partial t}+\nabla\cdot\varrho\mathbf{v}_i=0,\label{continuity}\\
\varrho\frac{\textrm{d}\mathbf{v}_i}{\textrm{d}t}=\frac{\mathbf{J}\times\mathbf{B}}{c}-\nabla p,\label{motion}\\
\frac{\textrm{d}}{\textrm{d}t}\left(\frac{p}{\varrho^\gamma}\right)=0,\label{energy}
\end{gather} where $\mathbf{J}=ne(\mathbf{v}_i-\mathbf{v}_e)$ is the current density, $p=p_i+p_e$ and, as a rule, $\gamma=5/3$ is the ratio of specific heats. Writing the derivative in \eqref{energy} explicitly and, then, using \eqref{continuity}, we also obtain the useful identities \al{
\frac{\textrm{d}p}{\textrm{d}t}=\gamma\frac{p}{\varrho}\frac{\textrm{d}\varrho}{\textrm{d}t}=-\gamma p\nabla\cdot\mathbf{v}_i.\label{energycont}
} This one fluid approach to (ideal) plasmas is also called (ideal) MHD, an acronym which stands for MagnetoHydroDynamics.

Both kinetic model and ideal-MHD single-fluid model are closed by Maxwell's equations for the self-consistent evolution of the plasma electromagnetic fields, \begin{gather}
\nabla\times\mathbf{E}=-\frac{1}{c}\frac{\partial\mathbf{B}}{\partial t},\label{Max1}\\
\nabla\times\mathbf{B}=\frac{4\pi}{c}\mathbf{J}.\label{Max2}
\end{gather} Note that Poisson's law is replaced by the quasineutrality condition, as discussed above. Meanwhile, the displacement current is neglected in \eqref{Max2}, which holds for low-frequency fluctuations, i.e., for $|\partial_t|^2 \ll |c \nabla|^2$.

In section \ref{collective}, we have discussed the property of low collisionality in plasmas. In most cases, collisionality is so low that the electrical resistivity $\eta$ becomes negligible, or, the conductivity is almost infinite: from Ohm's formula, \al{
\eta=\frac{|\mathbf{E}+\left(\mathbf{v}/c\right)\times\mathbf{B}|}{|\mathbf{J}|}\approx0
} means that, because the current density is finite, the effective electric field in the plasma moving frame can be taken as zero, \al{
\mathbf{E}+\frac{\mathbf{v}}{c}\times\mathbf{B}=\mathbf{0};\label{lorentzero}
} see \cite{Freidberg} for more detailed analyses. In the collisionless limit, the collision integral itself can be neglected in the kinetic equation: then, \eqref{boltzmann} reduces to \al{
\frac{\partial f_{i,e}}{\partial t}+\nabla_{\mathbf{x}}f_{i,e}\cdot\mathbf{v}_{i,e}\pm\frac{e}{m_{i,e}}\nabla_\mathbf{v}f_{i,e}\cdot\left(\mathbf{E}+\frac{\mathbf{v}_{i,e}}{c}\times\mathbf{B}\right)=0;\label{vlasov}
} this equation, i.e. Boltzmann's equation with null right-hand side and with the Lorentz force as only force, and where, as a rule, $\mathbf{E}$ and $\mathbf{B}$ are defined in a self-consistent way by means of \eqref{Max1} and \eqref{Max2} (with charge quasineutrality and current density $\mathbf{J}$ defined through the distribution functions $f_{i,e}$), is known as the \emph{Vlasov equation}. Indeed, one could wonder how both the use of a Maxwell-Boltzmann distribution and of the MHD description, as introduced before, can work with low-collisional systems like plasmas: a partial answer \cite{Chenint} is that a strong equilibrium magnetic field, when present, causes the plasma response to be rapidly isotropized in the orthogonal plane and somehow mimics the effects of collisions. For example, the proportionality relation $\mathbf{v}=\mu\mathbf{E}$ for the drift velocity of electrons in metals may be thought of as counterpart of the drift velocity $\mathbf{v}_E=c\,\mathbf{E}\times\mathbf{B}/B^2$ in strongly magnetized plasmas. Moreover, electrons in plasmas experimentally show to distribute in a way which is essentially Maxwellian within good approximations (a situation known as \textit{Langmuir paradox} \cite{Langmuir1925}) and, however, the fluid theory is often not so sensitive to deviations from the Maxwell-Boltzmann distribution. When this is not the case, for example when studying the motion along the magnetic field lines or when small scales play an important role, fluid theory may fail but kinetic theory still applies in forms such as the one based on \eqref{vlasov}. More in depth discussion of the deviation of the plasma equilibrium from local thermodynamic equilibrium can be found in the recent work \cite{Falessi2019}.

\subsection{The gyrokinetic model}\label{gyromodsub}

A kinetic equation such as the Vlasov equation \eqref{vlasov} takes account of the particle motion in long-range mean electromagnetic field and so, in principle, also of the helical motion around the magnetic field lines, meaning that one is forced to handle a quite complicated dynamics and to operate at different scales: the spatiotemporal scales of the particles moving along equilibrium magnetic field lines and slowly drifting across it, the spatiotemporal scale of ion gyromotion (ion gyrofrequency $\Omega_i$ and ion Larmor radius $\rho_i$) and the spatiotemporal scale of electron gyromotion (electron gyrofrequency $\Omega_e\gg\Omega_i$ and electron Larmor radius $\rho_e\ll\rho_i$). However, predominant perturbations that are responsible for fluctuation induced transport in strongly magnetized plasmas are characterized by low frequencies compared to $\Omega_i$. This often motivates the use of the gyrokinetic model, where the trajectory of charged particles in a magnetic field, a helix winding around the field line, is decomposed into a fast parallel motion along the magnetic field line plus a relatively slow perpendicular drift motion of the guiding center, and a fast circular motion, the gyromotion, which for most plasma behavior can be averaged over. Averaging over this gyromotion (that is, neglecting the dependence on the gyrophase angle) eliminates one velocity coordinate from the original 6-dimensional phase space (3 spatial coordinates and 3 velocity coordinates), so formally describing the dynamics of charged rings with a guiding center position, instead of the dynamics of gyrating charged particles. For this model to work, it is necessary that $\rho_i\ll L$ (strong magnetization), where $L$ is the characteristic length-scale of the system; meanwhile, the frequencies involved are assumed to be much lower than the particles gyrofrequency ($\omega\ll\Omega_i$), while the plasma behavior on perpendicular spatial scales is taken to be comparable to the ion gyroradius ($k_\perp\rho_i\sim 1$). Gyrokinetic theory and correponding governing equations are then obtained as asymptotic description of the plasma response in terms of the small asymptotic expansion parameter $\varepsilon = \rho_i/L \sim |\omega/\Omega_i|$: in typical magnetized fusion plasmas, $\varepsilon \sim \mathcal{O}(10^{-2})$. In order to account for Landau resonance, $\omega \sim k_\parallel v_\parallel$, gyrokinetic theory assumes $k_\parallel \rho_i \sim \varepsilon$ as well as $k_\parallel/k_\perp \sim \varepsilon$; finally, perturbations are assumed to be of small amplitude, $e\delta\phi\ll k_BT_e$. Gyrokinetics has been experimentally shown to be particularly appropriate for modeling, for example, plasma turbulence.

In the gyrokinetic framework, fluctuating electromagnetic fields are assumed small, \al{
\left|\frac{\delta B}{B_0}\right| \sim \varepsilon,\qquad \left|\frac{\delta E}{B_0}\right| \sim \varepsilon\frac{\rho_i\Omega_i}{c},
} with $B_0$ the modulus of the equilibrium magnetic field $\mathbf{B}_0$. As a consequence, the distribution function, $f_{i,e}=F_M+\delta f+o(\delta f)$, is also expressed as a perturbation of the Maxwell-Boltzmann equilibrium distribution $F_M$, with $\delta f/F_M\sim\mathcal{O}(\varepsilon)$. The gyrophase independent particle response is most conveniently represented when the spatial variable is changed from the particle position $\mathbf{x}$ to the guiding center position $\mathbf{X}=\mathbf{X}_\perp+X_\parallel\mathbf{B}_0/B_0$, being \al{
\mathbf{X}_\perp=\mathbf{x}_\perp+\boldsymbol{\rho},\qquad \boldsymbol{\rho}:=\frac{\mathbf{v}\times\mathbf{B}_0}{\Omega_{i,e}B_0},
} and the three components of velocity are expressed through \al{
\mathscr{E}:=\frac{v^2}{2},\qquad \mu:=\frac{v_\perp^2}{2B_0},\qquad \sigma:=\text{sgn}\left(v_\parallel\right).
} Here, $v_\parallel:=\mathbf{v}\cdot\mathbf{B}_0/B_0$ is the parallel (to the magnetic field) velocity and $\mu$ is the magnetic moment adiabatic invariant. The Maxwell-Boltzmann distribution reads \al{
F_M=\frac{n_0}{\pi^{3/2}v_T^3}e^{-v^2/v_T^2},
} with $n_0$ equilibrium particle density and $v_T$ is the thermal speed in the sense of root mean square. The gyrokinetic equation \cite{Catto1978,Antonsen1980,Frieman1982} is derived from Vlasov's equation and describes the change in time of the perturbed distribution function \al{
\delta g=e^{\boldsymbol{\rho}\cdot\nabla_\mathbf{X}}\left(\frac{e\delta\phi}{T}F_M+\delta f\right),\label{deltagresponse}
} where $\delta\phi$ is the perturbed scalar potential and $T$, from now on, is expressed in energy units in order to be consistent with the standard notation adopted in magnetized fusion plasma literature (so, in particular, from now on, Boltzmann's constant $k_B$ will not appear anymore). Introducing the vector potential perturbation $\delta\mathbf{A}$, so to define \al{
\delta L_g:=e^{\boldsymbol{\rho}\cdot\nabla_\mathbf{X}}\delta L:=e^{\boldsymbol{\rho}\cdot\nabla_\mathbf{X}}\left(\delta\phi-\frac{v_\parallel}{c}\delta A_\parallel\right),\qquad \delta A_\parallel:=\delta\mathbf{A}\cdot\mathbf{B}_0/B_0,
} and denoting by $\langle...\rangle_\alpha$ the averaging over the gyrophase angle $\alpha$ at fixed guiding center position, this equation can be expressed in a quite general form as \cite{Zoncarev} \al{\begin{aligned}
\DP{\delta g}{t}+\left(v_\parallel\frac{\mathbf{B}_0}{B_0}+\mathbf{V}_{d}+\frac{c}{B_0^2} \mathbf{B}_0 \times \nabla \left \langle \delta L_g \right \rangle\right)\cdot\nabla_\mathbf{X}\delta g&=\langle\mathscr{C}\left(\delta g\right)\rangle_\alpha+\frac{eF_M}{T}\DP{\langle\delta L_g\rangle_\alpha}{t}\\&-\frac{c}{B_0^2}\mathbf{B}_0\times\nabla\langle\delta L_g\rangle\cdot\nabla F_M.\end{aligned}\label{gyroeq}
} Here, we have adopted the Coulomb gauge ($\nabla \cdot \delta\mathbf{A}=0$), consistent with the usual gyrokinetic approach, and neglected the compressional magnetic field component $\delta B_\parallel$. In fact \cite{Chen2016,Zoncarev}, introducing the Alfvén speed $v_A:=B_0/(4\pi \varrho)^{1/2}$ (see subsection \ref{uniformplasmaslab}), we can estimate \al{
\frac{\omega}{k_\perp v_A} \sim \frac{\omega}{\Omega_i}\frac{v_{Ti}}{k_\perp \rho_iv_A}<\mathcal{O}(\varepsilon)
} and the compressional Alfvén wave (see subsection \ref{uniformplasmaslab}), responsible for small but finite $\delta B_\parallel$ fluctuations, can be neglected: this important point is further discussed in chapter \ref{alfvenchapter}. The left-hand side of \eqref{gyroeq} is the total time derivative of the non-adiabatic particle response, $\delta g$: in particular, the first, second and third gradient term represent, respectively, the particle streaming along the magnetic field line, the effects of cross-field particle drifts (including the curvature drift and the grad-$\mathbf{B}$ drift) accounted for by the drift velocity $\mathbf{V}_d$, and the nonlinear particle drift effect due to the perturbed $\mathbf{E}\times\mathbf{B}$ drift and the perpendicular magnetic field fluctuation $\delta\mathbf{B}_\perp$. On the right-hand side, the first term is the operator accounting for particle collisions, the second one represents the Maxwell–Boltzmann response to the perturbed scalar potential in the guiding center moving frame, and the last term accounts for the $\mathbf{E}\times\mathbf{B}$ response in the guiding center moving frame. In particular, this last term includes effects from the temperature and density gradients of the background distribution function, which may drive the perturbation. These gradients are only significant in the direction across flux surfaces, labeled by the magnetic flux function. So, once the gyrokinetic equation is solved for $\delta g$ and the corresponding particle response is determined by inversion of \eqref{deltagresponse}, the perturbed plasma densities and current can be consistently computed, while electromagnetic fluctuations are given by \eqref{Max1} and \eqref{Max2}.

\section{Waves and oscillations in plasmas}

\subsection{Electron plasma waves}\label{plasmawaves}
As in other physical systems, oscillations and wave propagation also occur in plasmas \cite{Stix1992,Chenint}. First of all, perturbations cause disturbances in the configuration of electrons, to which they respond with small oscillations at a particular frequency, known as plasma frequency, already introduced in section \ref{collective}. This frequency can be quite easily computed by using a simple but significant model \cite{Chenint}. Let's consider an infinite plasma  with no equilibrium electric or magnetic fields ($\mathbf{E}_0=\mathbf{B}_0=\mathbf{0}$), whose thermal motion is neglected ($T\approx0$) and electrons are at rest ($\mathbf{v}_0=0$), with ions fixed in space and electrons free to move only in the $x$ direction. Due to a perturbation, the quantities of interest are \al{
n=n_0+\delta n,\qquad v=\delta v,\qquad E=\delta E.
} Linearizing the plasma response, and supposing the plasma is uniform ($\nabla n_0=\mathbf{0}$), the continuity equation for the perturbed density is derived from \eqref{continuitytwo} for the electrons, \al{
\frac{\partial n}{\partial t}+\frac{\partial(nv)}{\partial x}\approx\frac{\partial\,\delta n}{\partial t}+n_0\frac{\partial v}{\partial x}=0,
} while the force balance equation is \eqref{motiontwo} for electrons, \al{
m_e\frac{\textrm{d}v}{\textrm{d}t}+eE+\frac{\partial p}{\partial x}=m_e\left(\frac{\partial v}{\partial t}+v\frac{\partial v}{\partial x}\right)+eE+\frac{\partial p}{\partial x}\approx m_e\frac{\partial\,\delta v}{\partial t}+e\,\delta E=0,\label{motion2}
} where we have used $p=3nT\approx0$ since $T\approx0$. Finally, Poisson's equation, keeping in mind that ions are fixed and, thus, no perturbation in their density occurs, is \al{
\frac{\partial E}{\partial x}-4\pi\sigma\approx\frac{\partial\,\delta E }{\partial x}+4\pi e\,\delta n=0.
} In this case, the divergence of the perturbed electric field is not zero, because electronic density is changed, while ions are fixed, and, thus, quasineutrality doesn't hold. Writing the perturbed quantities in the form of plane waves \al{
\delta n=\delta n\,e^{\textrm{i}(kx-\omega t)},\qquad \delta v=\delta v\,e^{\textrm{i}(kx-\omega t)},\qquad \delta E=\delta E\,e^{\textrm{i}(kx-\omega t)},
} so that, in particular, $\partial/\partial t=-\textrm{i}\omega$ and $\partial/\partial x=\textrm{i}k$, the above equations become \al{
-\textrm{i}\omega\delta n+\textrm{i}kn_0\delta v&=0,\\
-\textrm{i}m_e\omega\delta v+e\delta E&=0,\\
\textrm{i}k\delta E+4\pi e\delta n&=0.
} This readily yields the plasma frequency in the form \al{
\omega=\omega_p=\sqrt{\frac{4\pi e^2n_0}{m_e}}.
} Because $\omega_p$ does not depend on $k$, the group velocity $v_g:=\text{d}\omega_p/\text{d}k$ is zero. Actually, this is due to the fact that we have neglected electron thermal motion: if we return to \eqref{motion2} but with \al{
\frac{\partial p}{\partial x}=3T\frac{\partial\,\delta n}{\partial x},
} we easily obtain the dispersion relation \al{
\omega_p^2(k)=\omega_p^2+\frac{3T}{m_e}k^2,
} and, therefore, the group velocity \al{
v_g=\frac{\textrm{d}\omega_p(k)}{\textrm{d}k}=\frac{3Tk}{m_e\omega_p(k)}.
} Finally, when the plasma is magnetized, spiral motions around the direction of magnetic field superimpose, and we could easily demonstrate that the frequency becomes \al{
\omega_h=\sqrt{\omega_p^2(k)+\Omega_e^2},
} being \al{
\Omega_e=\frac{eB_0}{m_ec}
} the electron gyrofrequency. Here, $\omega_h$ is known as the \emph{upper hybrid frequency} and the corresponding wave \emph{upper hybrid oscillation}.

\subsection{Ion acoustic waves}
In ordinary gases, due to pressure (and then, ultimately, to temperature), propagation of sound waves occurs, whose description can be derived from the equation of fluid motion (cardinal equation of ideal fluids or Navier-Stokes equation). A similar phenomenon is also typical of plasmas, with some differences mainly related to the presence of electromagnetic fields: not simply kinetic energy but also the repulsion among positive charges causes ions groups to expand, and then to get closer to other groups, which in turn, because of pressure and electric repulsion, move away, so creating a game of compressions and decompressions that is the ion sound wave in a plasma. The linearization of the ion fluid equation \eqref{motiontwo}, with, for the moment, $\mathbf{B}=\mathbf{0}$, along with the equation of continuity \eqref{continuity}, leads, through analogous computations as in the case of electron oscillations, to the dispersion relation for \emph{ion acoustic waves}, \al{
v_s=\frac{\Omega_s}{k}=\sqrt{\frac{\left(\gamma_i T_i+\gamma_e T_e\right)}{m_i}}=\sqrt{\frac{\gamma_ip_i}{\varrho_i}+\frac{\gamma_ep_e}{\varrho}}=\sqrt{v_{si}^2+v_{se}^2},\label{ionwave}
} where we have used $p_{i,e}=nT_{i,e}$, so to recognize the familiar expressions for sound speed, which here are to be considered as the ionic and electronic \emph{adiabatic sound speeds} (in general, $\gamma_e=1$).

While electron plasma oscillations have a thermal correction in their dispersion relation but also exist at zero temperature (though they cannot propagate in that case), ion waves only exist with thermal motions, like ordinary sound waves. However, ion waves don't need thermal ion motion, specifically, when the dominant term in \eqref{ionwave} is $\sim T_e$: we can say that ion waves essentially depend on electron thermal motion and ion mass. As in the case of plasma oscillations, the extension for magnetized plasmas of the ion sound waves relation can be considered, giving the frequency of the \emph{electrostatic ion cyclotron waves}, \al{
\Omega_C=\sqrt{k^2v_s^2+\Omega_i^2},
} being \al{
\Omega_i=\frac{eB_0}{m_ic}
} the ion gyrofrequency.

\chapter{\textsc{alfvén waves}}\label{alfvenchapter}
\vspace{100pt}

Hydromagnetic waves are the waves propagating in an electrically conducting fluid, such as a plasma, and in a magnetic field; among these waves, we focus on the \textit{(shear) Alfvén wave} in a non-uniform plasma. Non-uniformity is a crucial feature since it causes the wavelength in the non-uniformity direction to decrease in time as $1/t$ \cite{Hasegawa1974} and the MHD wave equations to exhibit a singular behavior: the kinetic description, able to investigate shorter scales, shows that the Alfvén wave actually changes its character by evolving to a different type of wave (the \textit{kinetic Alfvén wave} \cite{Hasegawa1975,Hasegawa1976}). This fact represents an example of the process known as wave \textit{mode conversion} \cite{Stix1992}.

The general theory of Alfvén waves starts during the 1940s, when Hannes Alfvén combined \cite{Alfven1942,Alfven1949} fluid mechanics and electromagnetic waves to try to solve the "coronal heating problem", namely the problem of understanding why the solar corona is hotter than the photosphere despite its distance from the core: a magneto-hydrodynamic wave could exist able to carry energy from the photosphere towards the corona and the solar wind. Enrico Fermi readily adopted this thesis for his own theory of cosmic rays \cite{Fermi1949}. Though it is still not clear whether Alfvén waves, and in particular the mode converted short length scale kinetic Alfvén waves, suffice to explain the observed thermal gradient, we now actually know that they are produced by electromagnetic fluctuations and varying pressure gradients in the convective zone beneath the photosphere.

The first descriptions of kinetic Alfvén waves in plasmas, produced by phase mixing (see section \ref{nonuniformslab}) and mode conversion, date back to 1975-76 \cite{Hasegawa1975,Hasegawa1976}. For some time, the strong (linear) absorption of these waves has been investigated as a possible, convenient way for plasma heating in magnetic fusion devices \cite{Itoh1982,Itoh1983}, requesting quite simple antenna design and low-frequency radio instead of higher-frequency waves; eventually, these attempts have been abandoned due to the evidence that power coupling and efficiency for a kinetic Alfvén wave aren't adequate. In the present work, our main interest is in the nonlinear physics connected to the opposite case, that is weak absorption \cite{Zonca2015,Zoncarev}, because the energy not absorbed by plasma can lead to the formation of nonlinear structures such as convective cells, with interesting novel underlying physics and applications.

\section{Hydromagnetic waves. Shear Alfvén wave (SAW)}

At the end of the first chapter, basic examples of waves originating in a plasma have been introduced. In this chapter, we shall switch our attention to the quite general case of electromagnetic waves in a magnetized ($\mathbf{B}_0=B_0\hat{\mathbf{z}}\neq\mathbf{0}$), ideally infinite, plasma slab at rest ($\mathbf{v}_0=\mathbf{J}_0=\mathbf{0}$), initially taking a uniform system, then studying some important consequences of non-uniformity.

\subsection{Uniform plasma slab}\label{uniformplasmaslab}
For a homogeneous ($\nabla p=\nabla\varrho=0$) plasma \cite{Freidberg}, the equation of continuity, after linearization, is \al{
\omega\delta\varrho=\varrho_0\mathbf{k}\cdot\delta\mathbf{v},\label{freidberg1}
} while \eqref{energycont} is \al{
\omega\delta p=\gamma p_0\mathbf{k}\cdot\delta\mathbf{v}.\label{freidberg2}
} Meanwhile, the Lenz's law can be written as, using Ohm's law in the form of \eqref{lorentzero}, \al{
\omega\delta\mathbf{B}=-\mathbf{k}\times\left(\delta\mathbf{v}\times\mathbf{B}_0\right),\label{max1lin}
} while Ampère's law (in the low-frequency limit $\textrm{d}\mathbf{E}/\textrm{d}t=\mathbf{0}$, as usual) is, taking into account \eqref{max1lin}, \al{
\frac{4\pi}{c}\omega\delta\mathbf{J}=-\textrm{i}\mathbf{k}\times\left(\mathbf{k}\times\left(\delta\mathbf{v}\times\mathbf{B}_0\right)\right).\label{max2lin}
} Maxwell's equations for the divergence of the fields are redundant: in particular, the first of \eqref{Max2}, namely $\mathbf{k}\cdot\delta\mathbf{B}=0$, is a consequence of \eqref{max1lin}. For convenience, we rotate the $(x,y)$ plane so that $k_x=0$, and, then, set $k_\perp:=k_y$ and $k_\parallel:=k_z$. The linearized cardinal equation \eqref{motion} is \al{
-\textrm{i}\omega\varrho_0\delta\mathbf{v}=\frac{\delta\mathbf{J}}{c}\times\mathbf{B}_0-\textrm{i}\delta p\mathbf{k},
} which, by substitution of \eqref{max2lin}, in components reads \al{
\omega\varrho_0\delta v_x&=\frac{B_0^2k_\parallel^2}{4\pi\omega}\delta v_x,\\
\omega\varrho_0\delta v_y&=\frac{B_0^2k^2}{4\pi\omega}\delta v_y+\delta pk_\perp,\\
\omega\varrho_0\delta v_z&=\delta pk_\parallel.
} Further substitution of \eqref{freidberg2} in the above equations yields \al{
\left(\omega^2-k_\parallel^2v_A^2\right)\delta v_x&=0,\\
\left(\omega^2-k_\perp^2v_s^2-k^2v_A^2\right)\delta v_y-k_\perp k_\parallel v_s^2\,\delta v_z&=0,\\
k_\perp k_\parallel v_s^2\,\delta v_y-\left(\omega^2-k_\parallel^2v_s^2\right)\delta v_z&=0,
} where \al{
v_A=\frac{B_0}{\sqrt{4\pi\varrho_0}}
} is the \emph{Alfvén speed}, while $v_s=v_{si}=\sqrt{\gamma p_0/\varrho_0}$ is the ion adiabatic sound speed, introduced in \eqref{ionwave}. In order to solve the above linear system, we first set to zero the determinant of its coefficient matrix, getting the dispersion relation in three branches, i.e. three solutions, \al{
\omega_A^2=k_\parallel^2v_A^2\qquad&\textrm{(shear Alfvén wave)},\label{saw}\\
\omega_{FM}^2=\frac{k^2\left(v_A^2+v_s^2\right)\left(1+\sqrt{1-\alpha^2}\right)}{2}\qquad&\textrm{(fast magnetosonic wave)},\label{fmw}\\
\omega_{SM}^2=\frac{k^2\left(v_A^2+v_s^2\right)\left(1-\sqrt{1-\alpha^2}\right)}{2}\qquad&\textrm{(slow magnetosonic wave)},\label{smw}
} with \al{
\alpha^2:=\frac{4k_\parallel^2v_s^2v_A^2}{k^2\left(v_s^2+v_A^2\right)^2};\label{alpha}
} then, we substitute these solutions in the system, obtaining the corresponding eigenvectors, namely the components of the perturbed velocity, for each case. In particular, for shear Alfvén waves (SAWs), the velocity is constrained to the $x$ direction, $\delta\mathbf{v}=(\delta v_x,0,0)$, from which the name of shear waves. They only depend on $k_\parallel$ and not on $k_\perp$, and, because $\mathbf{k}\cdot\delta\mathbf{v}_x=0$, we see, from \eqref{freidberg1} and \eqref{freidberg2} and from $\nabla\cdot\delta\mathbf{v}\propto\mathbf{k}\cdot\delta\mathbf{v}$, that no perturbation in density and pressure arises: i.e., they are incompressible. The perturbation of the magnetic field is, according to \eqref{max1lin} and \eqref{saw} and, in the last step, to \eqref{lorentzero}, \eq{
\delta\mathbf{B}=\delta B_x=-\frac{k_\parallel B_0\delta v_x}{\omega_A}=\mp\frac{B_0\delta v_x}{v_A}=\mp\frac{c\delta E}{v_A},\label{deltaB=deltaBx}
} where $\delta E=\delta E_y$ and $\mp$ signs refer to $\pm$ roots of the shear Alfvén dispersion relation $\omega_A=\pm k_\parallel v_A$. Equation \eqref{deltaB=deltaBx} is known as Walén relation \cite{Walen1944}. Still from \eqref{saw}, we see that $v_A\hat{\mathbf{z}}=(\partial\omega_A/\partial k_\parallel)\hat{\mathbf{z}}$ is the group velocity, which is directed along the ambient magnetic field.

For the magnetosonic waves, the fluctuation of velocity lies in the $(y,z)$ plane: it holds that \al{
\omega_{SM}\leq\omega_A\leq\omega_{FM}.
} In strongly magnetized plasmas, where ion Larmor radius is much lower than the system size and, typically, thermal plasma pressure, $p\propto nT_i$, is much lower than the magnetic pressure, $p_m=B^2/8\pi$, one has \al{
\beta:=\frac{4\pi n_0T}{B_0^2}=\frac{v_{si}^2}{\gamma_i v_A^2}\ll 1,
} and, therefore, $v_s\ll v_A$. By looking at \eqref{smw} and \eqref{alpha}, in this case slow magnetosonic waves reduce to the ion sound waves, $\omega_{SM}^2=k_\parallel^2v_s^2$, while fast waves simplify their dispersion relation to $\omega_{FM}^2=k^2v_A^2$ and become the so-called \emph{compressional Alfvén waves}.

\subsection{Non-uniform plasma slab}\label{nonuniformslab}

\begin{figure}[h]
	\centering
	\includegraphics[scale=0.22]{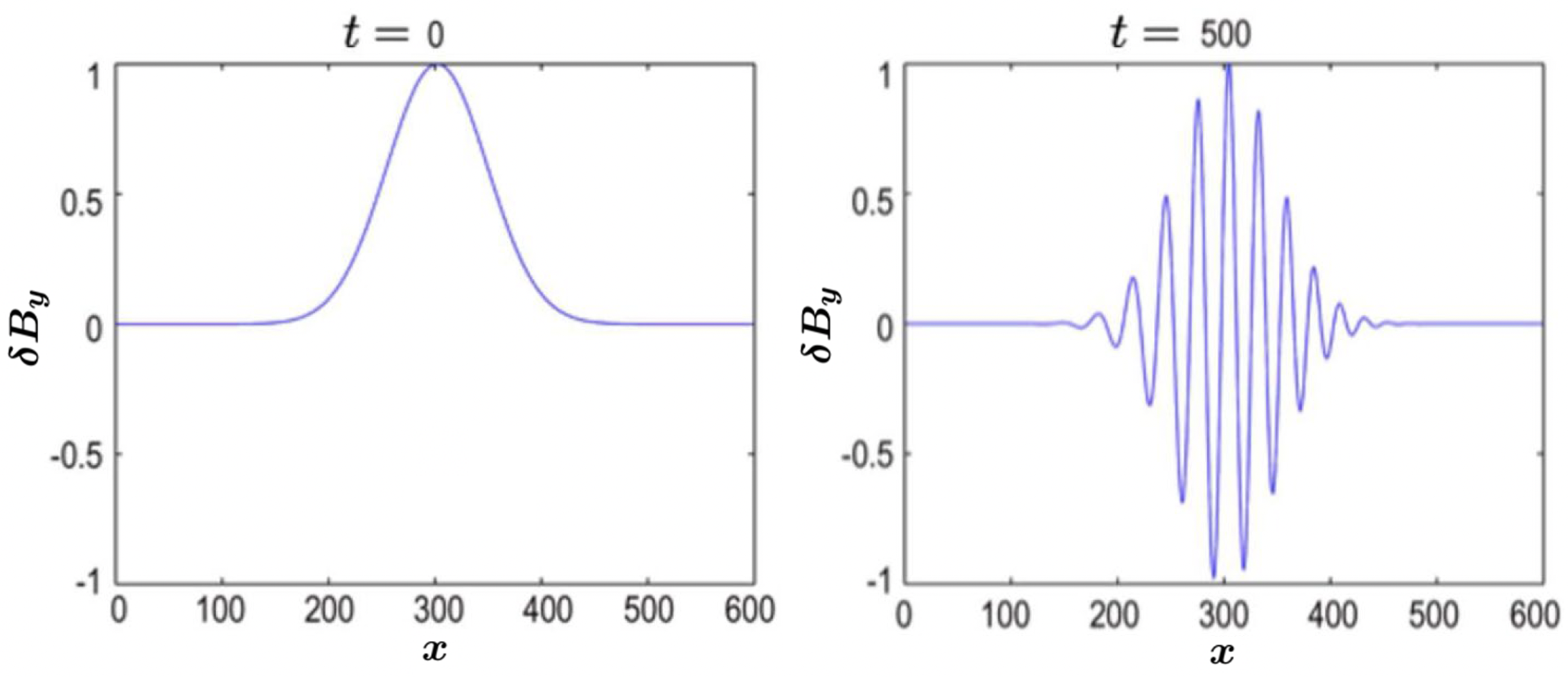}
	\caption{\emph{Fig.\ref{Qiu}} Snapshots of the spatial structure of $\delta B_y(x,t)$ with respect to $x$ at different times, illustrating the formation of shorter scales at later times, from original figure in \cite{Qiu2010,Qiu2011}.}
	\label{Qiu}
\end{figure}

In a non-uniform plasma, whose density, and pressure, vary along a given direction, say the $x$ direction, orthogonal to the magnetic field $\mathbf{B}_0(x)=B_0(x)\hat{\mathbf{z}}$, the perturbation of a generic physical quantity $\mathcal{F}$ will be expressed as \al{
\delta\mathcal{F}(x,y,z,t)=\delta\mathcal{F}(x)e^{\text{i}\left(k_yy+k_\parallel z-\omega t\right)}.
} So, the equation of continuity is \al{
\text{i}\omega\delta\varrho=\D{\varrho_0}{x}\delta v_x+\varrho_0\left(\D{\delta v_x}{x}+\text{i}k_y\delta v_y+\text{i}k_z\delta v_z\right),
} while \eqref{energycont} is \al{
\text{i}\omega\delta p=\gamma p_0\left(\D{\delta v_x}{x}+\text{i}k_y\delta v_y+\text{i}k_z\delta v_z\right).
} Ampère's law \eqref{Max2}, taking Ohm's law \eqref{lorentzero} into account, generally reads \al{
\text{i}\omega\delta\mathbf{B}=-\nabla\times\left(\delta\mathbf{v}\times\mathbf{B}_0\right),
} or, \al{
\delta\mathbf{B}=\nabla\times\left(\delta\boldsymbol{\xi}\times\mathbf{B}_0\right),\label{deltaBdeltaxi}
} being $\delta\boldsymbol{\xi}$ the plasma displacement. Meanwhile, the momentum equation becomes \al{
\varrho_{0}\DP{^2}{t^2}\delta\boldsymbol{\xi}=-\nabla\delta\tilde{p}+\frac{1}{4\pi}\left(\mathbf{B}_0\cdot\nabla\right)^2\delta\boldsymbol{\xi}-\frac{1}{4\pi}\mathbf{B}_0\left(\mathbf{B}_0\cdot\nabla\right)\left(\nabla\cdot\delta\boldsymbol{\xi}\right),\label{momkin}
} where $\tilde{p}$ indicates the total, kinetic plus magnetic, pressure perturbation, \al{
\delta\tilde{p}=\delta p+\frac{B_0\delta B_\parallel}{4\pi}.\label{presskin}
} In particular, for the parallel, $z$-component of \eqref{momkin}, we have \al{
\left(\varrho_{0}\omega^2-\frac{B_0^2}{4\pi}k_\parallel^2\right)\delta\xi_\parallel=\text{i}k_\parallel\delta\tilde{p}+\frac{B_0^2}{4\pi}\text{i}k_\parallel\left(\text{i}k_\parallel\delta\xi_\parallel+\text{i}k_y\delta\xi_y+\D{\delta\xi_x}{x}\right),
} or, multiplying by $4\pi/B_0^2$, \al{
D_A\delta\xi_\parallel=\frac{4\pi}{B_0^2}\text{i}k_\parallel\delta\tilde{p}+\text{i}k_\parallel\left(\text{i}k_\parallel\delta\xi_\parallel+\text{i}k_y\delta\xi_y+\D{\delta\xi_x}{x}\right),\label{comp1}
} where $D_A(x):=\frac{\omega^2}{v_A^2(x)}-k_\parallel^2$. Likewise, we obtain \al{
D_A\delta\xi_x&=\frac{4\pi}{B_0^2}\D{\delta\tilde{p}}{x},\label{comp3}\\
D_A\delta\xi_y&=\frac{4\pi}{B_0^2}\text{i}k_y\delta\tilde{p}.\label{comp2}
} In all three components, then, the left-hand side vanishes when the shear Alfvén wave dispersion relation is locally satisfied. Using \eqref{energycont} in the first step below, and \eqref{momkin} in the second step, \eqref{presskin} explicitly reads \al{\begin{aligned}
\delta\tilde{p}&=-\delta\xi_x\D{p_0}{x}-\gamma p_0\left(\nabla\cdot\delta\boldsymbol{\xi}\right)+\frac{B_0\delta B_\parallel}{4\pi}=\\
&=-\text{i}k_\parallel\gamma p_0\delta\xi_\parallel-\left(\gamma p_0+\frac{B_0^2}{4\pi}\right)\left(\text{i}k_y\delta\xi_y+\D{}{x}\delta\xi_x\right).\end{aligned}\label{deltaptilde}
} Substituting \eqref{comp1} in this equation, and taking into account \eqref{comp2}, we have \al{
\delta\xi_\parallel=-\text{i}\frac{\gamma\beta k_\parallel v_A^2}{2\omega^2-\gamma\beta k_\parallel^2v_A^2}\nabla\cdot\delta\boldsymbol{\xi}_\perp;
} with the help of the same three equations, we also get \al{
\delta\xi_y=\frac{\text{i}\overline{\alpha}k_y}{\overline{\alpha}k_y^2-D_A}\D{\delta\xi_x}{x},\label{deltaxiy}
} with \al{
\overline{\alpha}:=1+\frac{\gamma\beta\omega^2}{2\omega^2-\gamma\beta k_\parallel^2v_A^2}.
} Combining \eqref{deltaxiy} with \eqref{comp3} and \eqref{comp2}, we finally obtain the equation \al{
\D{}{x}\left(\frac{B_0^2D_A\overline{\alpha}}{\overline{\alpha}k_y^2-D_A}\D{\delta\xi_x}{x}\right)-B_0^2D_A\delta\xi_x=0.\label{kaw1}
} This equation, and its solutions, are singular when $D_A\overline{\alpha}=0$, namely when either $\overline{\alpha}=0$, corresponding to the dispersion relation of ion sound waves, \al{
\omega^2=\omega^2_s(x):=\frac{k_\parallel^2 v_s^2(x)}{1+\frac{v_s^2(x)}{v_A^2(x)}},
} or $D_A=0$, corresponding to \al{
\omega^2=\omega^2_A(x):=k_\parallel^2 v_A^2(x),
} namely the SAW dispersion relation for a non-uniform plasma: in both cases, the spectrum is naturally non-uniform, continuous. In a cold plasma, $\gamma\beta\ll1$, so $\overline{\alpha}\approx1$ and ion sound waves are suppressed, as we expected; furthermore, \al{
\delta\xi_\parallel=0,\label{deltaxiparallel}
} so only the perpendicular force balance is relevant and, according to \eqref{momkin} and \eqref{presskin}, is given by \al{
D_A\delta\boldsymbol{\xi}_\perp=\nabla_\perp\left(\frac{\delta B_\parallel}{B_0}\right).\label{deltaxiperp}
} Note that $k_y^2\sim a^{-2}$, where $a$ is the perpendicular (to the ambient magnetic field) macroscopic scale-length, whereas $D_A$ can be taken of the order $k_\parallel^2$, scaling with the longitudinal length as $L^{-2}$: since generally $a<L$ (this is definitely the case for the cylindrical plasma model studied in the next chapters), then $|D_A|\ll k_y^2$, and \eqref{kaw1} can be simplified to \al{
\D{}{x}\left(\frac{B_0^2D_A}{k_y^2}\D{\delta\xi_x}{x}\right)-B_0^2D_A\delta\xi_x=0,
} that is, \al{
\left(\DP{}{x}\left(\omega^2-\omega_A^2\right)\DP{}{x}-k_y^2\left(\omega^2-\omega_A^2\right)\right)\delta\xi_x=0.\label{dxkyxix}
} By identifying $\omega^2$ with $-\partial_t^2$, and neglecting $k_y$ with respect to $\partial_x$ (an assumption that will be justified a-posteriori), this equation becomes \al{
\DP{}{x}\left(\DP{^2}{t^2}+\omega_A^2\right)\DP{}{x}\delta\xi_x=0.\label{xixt}
} Integration in $t$ yields \al{
\DP{}{x}\delta\xi_x(x,t)=C(x)e^{\pm\text{i}\omega_A(x)t}\label{xix}
} and it is easy to verify that, for $t\rightarrow\infty$, the right-hand side is the derivative of \al{
\delta\xi_x(x,t)\approx\pm\frac{C(x)}{\text{i}\omega_A'(x)t}e^{\pm\text{i}\omega_A(x)t},\label{xix2}
} namely $\delta\xi_x$ decays with time as $1/t$. This result implies that, asymptotically, \al{
\DP{}{x}\approx k_x=\pm\text{i}\omega_A'(x)t,\label{kincrease}
} meaning, in particular, that the modulus $|k_x|$ of the wavenumber increases in time, that is, the wavelength decreases: any initially long-scale perturbation evolves into short scales perturbations. Equation \eqref{kincrease} justifies having neglected $|k_y|$ with respect to $|k_x|$ time asymptotically. Close to the resonant point $x_0$, where the resonance condition $\omega_A^2(x_0)=\omega^2$ is satisfied, we can write the Alfvén frequency as \al{
\omega_A^2(x)\approx\omega^2+\left(\omega_A^2\right)'(x_0)\left(x-x_0\right)
} and, then, \eqref{xixt} as \al{
\left(x-x_0\right)\DP{^2}{x^2}\delta\xi_x+\DP{}{x}\delta\xi_x=0,
} which, for $x\neq x_0$, has solution \al{
\delta\xi_x=C\ln\left(x-x_0\right)+D,\label{xixres}
} where $C$ and $D$ are constants, that is evidently singular in $x_0$. In other words, the singularity is logarithmic and, consistently with \eqref{kincrease}, it indicates that arbitrarily short scales are generated in the time asymptotic limit. Using \eqref{deltaxiy} still in the limit $|D_A|\ll k_y^2$, \eqref{xixt} and \eqref{xix} yield \al{
\left(\DP{^2}{t^2}+\omega_A^2\right)\delta\xi_y=0
} and, so, \al{
\delta\xi_y(x,t)=\frac{\text{i}}{k_y}C(x)e^{\pm\text{i}\omega_A(x)t}=\delta\xi_y(x,0)e^{\pm\text{i}\omega_A(x)t}:\label{deltaxiyeq}
} the $y$ component of the displacement shows undamped oscillations at SAW continuum frequencies. Equation \eqref{deltaxiyeq} also allows us to rewrite \eqref{xix2} as \al{
\delta\xi_x(x,t)\approx\mp\frac{k_y}{\omega_A'(x)t}\delta\xi_y(x,0)e^{\pm\text{i}\omega_A(x)t}.
} Finally, from \eqref{deltaBdeltaxi}, magnetic field and displacement perturbations are proportional (and "in quadrature"), \al{
\delta B_{x,y}=\text{i}B_0k_\parallel\delta\xi_{x,y},\label{deltaBdeltaxiprop}
} so, also for $\delta B_x$, the SAW equation is \al{
\DP{}{x}\left(\DP{^2}{t^2}+\omega_A^2\right)\DP{}{x}\delta B_x=0,
} with time decaying solution \al{
\delta B_x(x,t)\approx\mp\frac{k_y}{\omega_A'(x)t}\delta B_y(x,0)e^{\pm\text{i}\omega_A(x)t},
} and, near the resonance, \al{
\delta B_x\approx C'\ln\left(x-x_0\right)+D'.\label{Bxres}
} That every point in $x$ oscillates at a different frequency $\omega_A(x)$ is a manifestation of \textit{phase mixing} \cite{Hasegawa1974,Grad1969} and of \eqref{kincrease}. The (spatial) phase mixing is the fast spatial variation of the phase $\text{i}(k_yy+k_\parallel z-\omega t)$ in the $x$ direction close to the SAW resonance, that's to say, the fact that nearby points in the $x$ direction decorrelate in phase, corresponding to $k_x \sim \omega_A' t$ and to the singular structures decaying in time (as $1/t$). And, as we shall see, to the mode conversion to the kinetic Alfvén wave. Meanwhile, as to the other perpendicular magnetic field fluctuation component, \al{
\left(\DP{^2}{t^2}+\omega_A^2\right)\delta B_y=0,\label{waveeq}
} that is, \al{
\delta B_y(x,t)=\delta B_y(x,0)e^{\pm\text{i}\omega_A(x)t},
} with, time asymptotically, $\partial_x\ln\delta B_y=k_x$, as given by \eqref{kincrease}: see also fig.\ref{Qiu} \cite{Qiu2010,Qiu2011}. In fact, the equations for the $y$ components of $\delta\boldsymbol{\xi}$ and $\delta\mathbf{B}$ can also be obtained from \eqref{dxkyxix} by neglecting $k_y^2$ with respect to $\partial_x^2$, integrating in $x$ and, then, using $\nabla\cdot \delta\xi\approx 0$ by analogy with $\nabla\cdot\delta\mathbf{B}=\partial_x\delta B_x+k_y\delta B_y=0$.

The locality of the SAW spectrum in the case of non-uniformity and the fact that the wavelength decreases as $t^{-1}$ are responsible for the singularity of the wave equation and imply the emergence of short scales. In particular, \eqref{xixres} and \eqref{Bxres} mean that the perturbations in the non-uniform direction develop a logarithmic singularity at the Alfvén resonant layer $x=x_0$, namely the layer such that $\omega^2(x_0)=\omega^2$, with a finite resonant wave-energy absorption rate. The existence of such a singularity naturally suggests that the microscopic length-scale physics neglected in the ideal MHD fluid description should be included in the long-time-scale dynamics of SAWs, and that the fluid, macroscopic description must be replaced by a kinetic description. For low-frequency SAWs, the relevant, and perpendicular to $\mathbf{B}_0$, microscopic scales are \cite{Zoncarev} either the ion Larmor radius, $\rho_i=v_{Ti}/\Omega_i$, with $v_{Ti}$ the ion thermal speed, and/or $\rho_s=v_s/\Omega_i$, with $v_s^2\sim T_e/m_i$ the ion-sound velocity \eqref{ionwave}. Including the effects of finite $\rho_i$ and/or $\rho_s$ in the SAW dynamics led to the discovery of the so-called kinetic Alfvén wave (KAW) \cite{Hasegawa1975,Hasegawa1976}.

\section{Kinetic Alfvén wave (KAW)}\label{kawsection}

In the previous section, we demonstrated that the existence of a continuous spectrum with oscillation frequency varying in the non-uniformity direction implies a phase-mixing process \cite{Hasegawa1974,Grad1969}, according to which any initial perturbation of SAW structures eventually evolves into short-wavelength structures, while the singularity at the resonant surface is a manifestation of the impossibility for a fluid model to provide further information about such scales. A kinetic theory approach is needed to properly describe the dynamics of Alfvén waves and, by including effects such as finite ion Larmor radius (FILR) and/or wave-particle interactions, allows the new short-wavelength structures involved, named kinetic Alfvén waves (KAWs) \cite{Hasegawa1975,Hasegawa1976,Wu2012,Wu2020}, to be studied. The discovery and first discussions of KAWs (since 1975-76 \cite{Hasegawa1975,Hasegawa1976}), having come a little before the introduction of the linear (in 1978 \cite{Catto1978,Antonsen1980}) and of the nonlinear (in 1982 \cite{Frieman1982}) electromagnetic gyrokinetic theories, have often been characterized by the complicated procedures of taking the low-frequency ($\omega\ll\Omega_i$) limit of the Vlasov kinetic theory and/or of employing the drift-kinetic theory approach, making the theoretical analysis of KAW dynamics in non-uniform plasmas with realistic $\mathbf{B}_0(x)$ intractable, especially when dealing with the nonlinear physics, where either the drift-kinetic or the two-fluid description have been adopted \cite{Mikhailovskii2007,Onishchenko2004a,Onishchenko2004b,Pokhotelov2003,Pokhotelov2004,Zhao2011}. Also, such approximations, apart from being inadequate for treating realistic plasma regimes, often leave out important physics, forsaking for example the FILR effects, so finally motivating, in recent years \cite{Chen2011,Chen2013,Zonca2015,Zoncarev}, the use of the powerful nonlinear gyrokinetic theory for re-examining both the linear and the nonlinear physics of KAWs.

Neglecting nonlinear terms (which will be reintroduced in chapter \ref{nonlinearGK}) and drift terms (since the ratio of drift velocity to thermal velocity scales as $\rho_i/L\ll1$; however, the effect of neglecting drift terms in cylindrical plasmas is discussed in detail in section \ref{neglectingdrift}), as well as collisions, the gyrokinetic equation \eqref{gyroeq} becomes \al{
\left(\DP{}{t}+v_\parallel\frac{\mathbf{B}_0}{B_0}\cdot\nabla\right)\delta g=\frac{q}{T}F_M(\mathscr{E})\DP{}{t}\langle\delta L_g\rangle_\alpha,\label{gyroeqlin}
} where $\nabla=\nabla_\mathbf{X}$ is the gradient operator with respect to the gyrokinetic coordinates. For a plane-wave perturbation of frequency $\omega$ and parallel wavenumber $\mathbf{k}_\parallel$, this equation has solution \al{
\delta g_{{k}}=-\frac{q}{T}F_MJ_0(k_\perp\rho)\frac{\omega}{k_\parallel v_\parallel-\omega}\delta L_{{k}},
} where $J_0(k_\perp\rho)=\langle e^{-\boldsymbol{\rho}\cdot{\nabla}}\rangle$, namely the first-kind Bessel function $J_0$ accounts for the gyro-averaging of the coordinate transformation. Since we are dealing with potential functions, the basic equations are Poisson's equation for $\delta\phi_{k}$, which in the regime $|k\lambda_D|\ll1$ is the quasineutrality condition \al{
\sum_{j=i,e}n_{0j}q_j\int\delta f_{j{k}}\,\text{d}\mathbf{v}\approx0,\label{quasi0}
} where $\mathbf{v}$ is the particle velocity, and Ampère's equation for the vector potential, \al{
\nabla^2\delta A_{\parallel{k}}=-\frac{4\pi}{c}\delta J_{\parallel{k}}=-\frac{4\pi}{c} \sum_{j=i,e} n_{0j}q_j \int v_\parallel \delta f_{j{k}}.\label{ampere0}
} Introducing the rescaled parallel vector potential \al{
\delta\psi_{k}=\frac{\omega\delta A_{\parallel{k}}}{ck_\parallel},
} that is, $\nabla\delta\psi\cdot\mathbf{B}_0/B_0:=-\partial_t\delta A_\parallel/c$, and such that the net parallel electric field is given by \al{
\delta E_{\parallel{k}}=-\frac{\mathbf{B}_0}{B_0}\cdot\nabla\left(\delta\phi-\delta\psi\right)=-\text{i}k_\parallel\left(\delta\phi-\delta\psi\right)_{k},\label{deltaEpar}
} the quasineutrality or Poisson equation \eqref{quasi0} becomes \al{
\sum_j\left(\frac{n_0q^2}{T_0}\right)_j\left\{\delta\phi_{{k}}+\Gamma_{0kj}\left[\xi_{kj}Z_{kj}\delta\phi_{{k}}-\left(1+\xi_{kj}Z_{kj}\right)\delta\psi_{{k}}\right]\right\}=0,
} where $\Gamma_{0kj}=I_0(b_{kj})e^{-b_{kj}}$, with $I_0$ the modified Bessel function and $b_{kj}:=k_\perp^2\rho_j^2=k_\perp^2T_j/(m_j\Omega_j^2)$, $\xi_{kj}=\omega/(|k_\parallel|v_{Tj})$ and \al{
Z_{kj}:=Z\left(\xi_{kj}\right)=\frac{1}{\sqrt{\pi}}\int_{-\infty}^{+\infty}{\frac{e^{-\zeta^2}}{\zeta-\xi_{kj}}\,\text{d}\zeta}\label{plasmadispersionf}
} is the \textit{plasma dispersion function} \cite{Chenint,Freidberg}. Meanwhile, substitution of the Ampère law \eqref{ampere0} in the relation $\nabla\cdot\delta\mathbf{J}=0$ gives the linear gyrokinetic \textit{vorticity equation} \al{
\frac{c^2}{4\pi\omega}k_\parallel^2k_\perp^2\delta\psi_{{k}}-\sum_j\left(\frac{n_0q^2}{T_0}\right)_j\left(1-\Gamma_{0kj}\right)\omega\delta\phi_{{k}}=0.
} Note that in the ideal MHD limit $\delta E_\parallel=0$ (as it can be seen from \eqref{lorentzero}), that is, $\delta\phi=\delta\psi$. With a wavelength comparable to the ion Larmor radius, namely $\left|k_\perp\rho_i\right|\sim1$, $\left|k_\perp\rho_e\right|\ll1$ and, as a consequence, $\Gamma_{0ke}\approx1$, the previous equations become \al{\begin{split}
\epsilon_{s{k}}\delta\phi_{{k}}&:=\left[1+\xi_{ke}Z_{ke}+\tau\left(1+\Gamma_k\xi_{ki}Z_{ki}\right)\right]\delta\phi_{{k}}=\\&=\left[1+\xi_{ke}Z_{ke}+\tau\Gamma_k\left(1+\xi_{ki}Z_{ki}\right)\right]\delta\psi_{{k}},\end{split}\\
\begin{split}\omega^2\delta\phi_{{k}}&=k_\parallel^2v_A^2\frac{b_k}{1-\Gamma_k}\delta\psi_{{k}}\label{deltaphideltapsi0}\end{split}
} where \al{
\epsilon_{s{k}}=1+\xi_{ke}Z_{ke}+\tau\left(1+\Gamma_k\xi_{ki}Z_{ki}\right)\label{dielectricconstant}
} is the dielectric constant for slow-sound ion-acoustic waves, $\tau:=T_{0e}/T_{0i}$ and $\Gamma_k:=\Gamma_{0ki}$. Introducing $\delta\phi_{\parallel{k}}=\delta\phi_{k}-\delta\psi_{k}$ as the effective parallel potential, we finally get \al{
\epsilon_{s{k}}\delta\phi_{\parallel{k}}&=-\tau\left(1-\Gamma_k\right)\delta\psi_{k},\label{quasi}\\
\omega^2\delta\phi_{\parallel{k}}&=-\left(\omega^2-k_\parallel^2v_A^2\frac{b_k}{1-\Gamma_k}\right)\delta\psi_{k}.\label{vorticity}
} Substituting \eqref{quasi} in \eqref{vorticity} we obtain \al{
\left(\omega^2-k_\parallel^2v_A^2\frac{b_k}{1-\Gamma_k}\right)\delta\psi_{k}=\omega^2\frac{\tau\left(1-\Gamma_k\right)}{\epsilon_{s{k}}}\delta\psi_{k},
} from which the KAW dispersion relation can be cast as \al{
\omega^2\left(1-\frac{\tau\left(1-\Gamma_k\right)}{\epsilon_{s{k}}}\right)=k_\parallel^2v_A^2\frac{b_k}{1-\Gamma_k}.
} At the lowest order, and neglecting the algebraically small electron Landau damping and the exponentially small ion Landau damping \cite{Zoncarev,Hasegawa1974,Hasegawa1975}, we can write $\epsilon_{s{k}}\approx1+\tau(1-\Gamma_k)=:\sigma_k$, so the KAW real frequency is given by \eq{
\omega_{{k}r}^2\approx k_\parallel^2v_A^2b_k\frac{1+\tau\left(1-\Gamma_k\right)}{1-\Gamma_k}=k_\parallel^2v_A^2\frac{\sigma_kb_k}{1-\Gamma_k}.
} Assuming $|k_\perp\rho_i|^2\ll1$ and defining $\hat{\rho}^2:=(3/4+\tau)\rho_i^2$, we finally reduce the KAW dispersion relation to \eq{
\omega_{k}^2\approx k_\parallel^2v_A^2\left(1+k_\perp^2\hat{\rho}^2\right),\label{kawdispersion}
} which is a minimal extension of the SAW dispersion relation \eqref{saw}, exactly recovered in the limit $k_\perp^2\rho_i^2\rightarrow0$, that's to say when the finite ion Larmor radius effects become negligible (long-wavelength limit).

Letting, in \eqref{kawdispersion}, $\omega_{k}^2=-\partial_t^2$ and $k_\perp^2=k_x^2+k_y^2=-\partial_x^2-\partial_y^2$, we can, in the case of a non-uniformity in the $x$ direction, straightforwardly extend the SAW equation \eqref{waveeq} as \eq{
\left(\DP{^2}{t^2}+\omega^2_{k}\right)\delta B_y(x,t)=0,
} i.e., formally using, as definition, the SAW dispersion relation $\omega_A:=k_\parallel v_A$, \eq{
\left(\frac{\partial^2}{\partial t^2}+\omega_A^2(x)-\hat{\rho}^2\omega_A^2(x)\frac{\partial^2}{\partial x^2}\right)\delta B_y(x,t)=0.\label{deltab}
} These equations show that one effect of FILR is the removal of the singularity for $\omega^2=\omega_A^2$. According to \eqref{kawdispersion}, the kinetic wave propagates in the $\omega_{k}^2>\omega_A^2$ region, and exponentially decays for $\omega_{k}^2<\omega_A^2$, where $k_x^2$ is evidently negative (cutoff). Furthermore, \eqref{kawdispersion} also shows that the group velocity has, in contrast to the MHD case, non-zero perpendicular component, \al{
\mathbf{v}_{g\perp}\approx\frac{\omega_A^2\hat{\rho}^2}{\omega_{k}}\mathbf{k}_\perp.
}

In conclusion, the kinetic description goes beyond the singular response of MHD introducing mode conversion of the shear to the kinetic Alfvén wave. In the long-wavelength limit, the kinetic correction $k_\perp^2\hat{\rho}^2$ in \eqref{kawdispersion} provides the simplest model to quantitatively describe conversion from long-wavelength MHD perturbations to short-wavelength gyrokinetic fluctuations, which will be further analyzed in chapter \ref{kawlinchap}.

\subsection{Observation of KAWs}

The efficient energization, including heating and acceleration, of plasma particles by kinetic Alfvén waves, pointed out by Hasegawa and Chen \cite{Hasegawa1976}, is an ubiquitous phenomenon in laboratory, space, and astrophysical magnetized plasmas \cite{Wu2020}, and equally important are its implications on particle transport.

In laboratory fusion plasmas, realistic plasma non-uniformities and magnetic field geometries often play crucially important roles in determining SAW/KAW mode structures and stability properties \cite{Chen2016}; for example, in toroidal fusion plasmas, the Kinetic Toroidal Alfvén Eigenmodes (KTAEs) \cite{Mett1992} may exist within the SAW continuum and their dynamics are intrinsically related to those of KAWs. Furthermore, laboratory plasma experiments have shown evidence of coupling between SAW eigenmodes and KAWs \cite{Wong1996}, which may also be externally driven by mode conversion of fast modes \cite{Fasoli1996}.

\begin{figure}[h]
	\centering
	\includegraphics[scale=0.5]{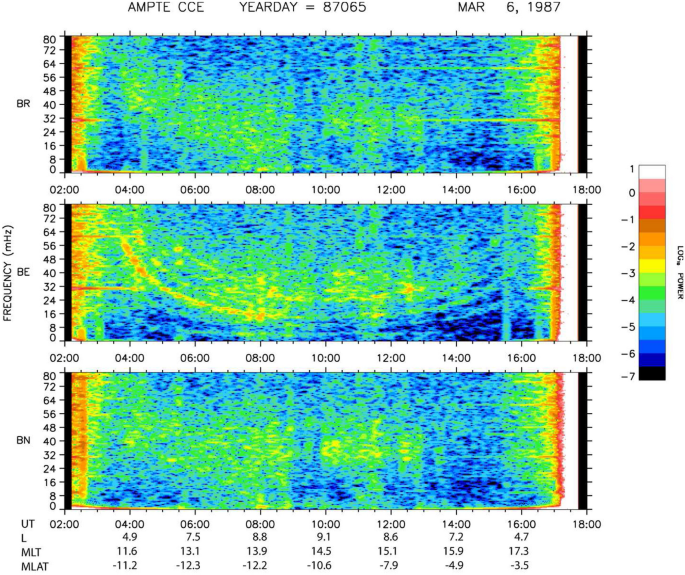}
	\caption{\emph{Fig.\ref{powerspectrum}} Three-component dynamic power spectrum of magnetic field data from AMPTE CCE satellite, from original figure in \cite{Engebretson1987}. Apogee is at the center of the figure. \emph{L}, \emph{MLT}, and \emph{MLAT} correspond, respectively, to the equatorial distance of the magnetic field line (in unit of the Earth radius), magnetic local time, and magnetic latitude.}
	\label{powerspectrum}
\end{figure}

However, due to diagnostics constraints in laboratory plasmas, most of the KAW observations have been made by satellites in the Sun-Earth space plasma environments, which, as mentioned at the beginning of this chapter, were a motivation for the very introduction of Alfvén waves in physics. A good example is geomagnetic pulsations in the Earth's magnetosphere: fig.\ref{powerspectrum} shows the oscillations in the Earth's magnetic field as observed by the satellite AMPTE CCE \cite{Engebretson1987}, illustrating the three-component dynamic power spectrum of magnetic field data from a full orbit from 02:30 to 17:30 UT March 6, 1987. The geomagnetic $B_R$, radially outward from the center of the Earth, $B_E$, magnetically Eastward, and $B_N$, approximately along local magnetic field lines, correspond to, respectively, $B_x$, $B_y$, and $B_z$. As the satellite moved outward from the morning side, $\omega_A$ should decrease due to the decreasing $|\mathbf{B}_0|$ and $|k_\parallel|$ (increasing field-line length), and this was clearly exhibited in the wave frequency of $B_E$, the azimuthal (East-West) component of $\delta\mathbf{B}$ (i.e., the effective $\delta B_y$). $B_E$ also shows that the wave frequency increases as the satellite moved inward toward the dusk side, consistently, again, with $\omega_A$. Furthermore, the observed wave frequency consisted of several bands, which could be understood as harmonics of standing waves along the field line, i.e., different $k_\parallel$.

\begin{figure}[h]
	\centering
	\includegraphics[scale=0.2]{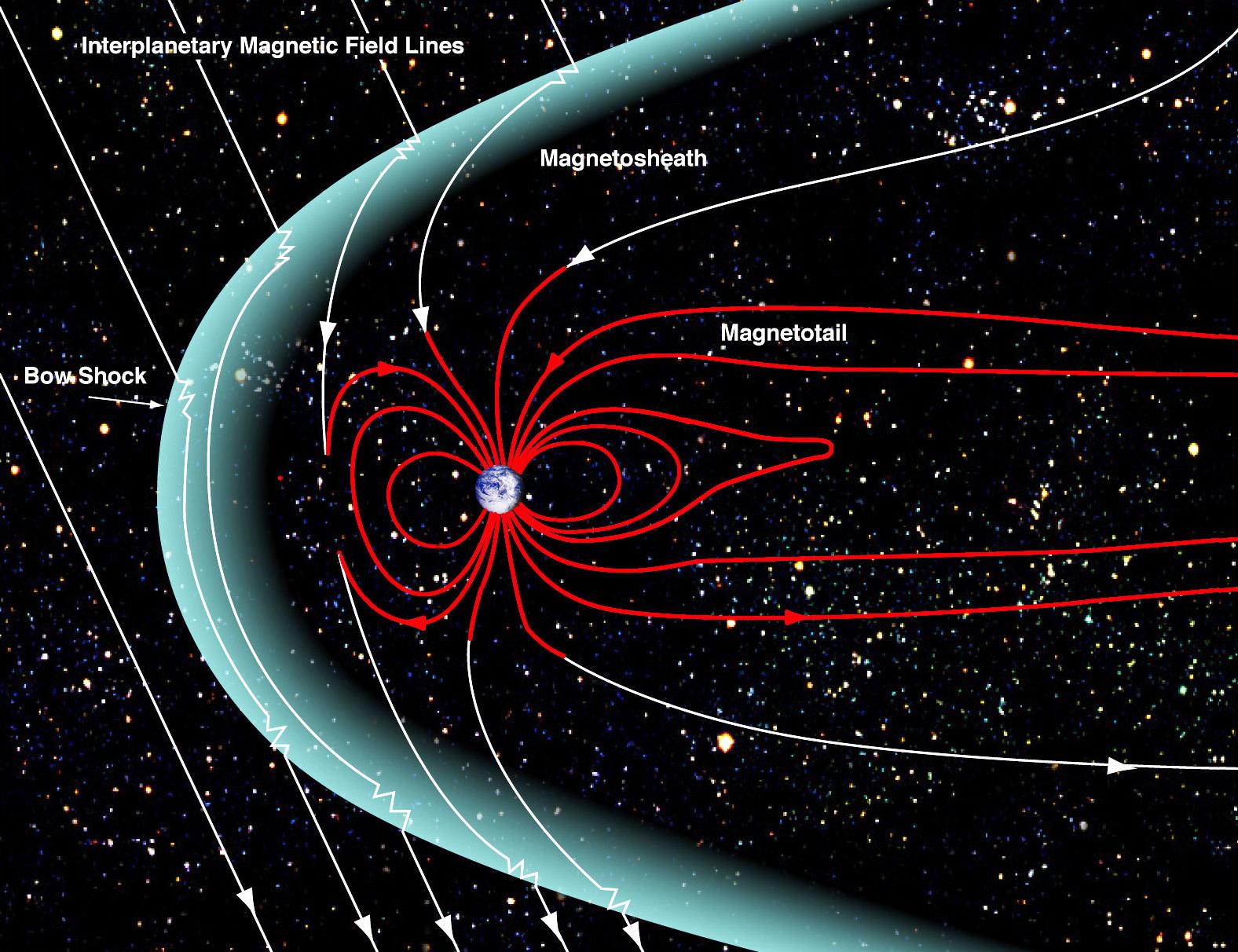}
	\caption{\emph{Fig.\ref{bowshock}} A supersonic shock wave is created sunward of Earth: the bow shock. Most of the solar wind particles are heated and slowed at the bow shock and detour around the Earth in the magnetosheath. The solar wind drags out the night-side magnetosphere to possibly 1000 times Earth's radius; its exact length is not known. This extension of the magnetosphere is known as the magnetotail. The outer boundary of Earth's confined geomagnetic field is called the magnetopause.}\label{bowshock} (Credit: NASA/Goddard/Aaron Kaase\footnotemark)
\end{figure}

Shear Alfvénic oscillations in the magnetosphere have been linked to excitations from the upstream solar wind: that is, kinetic Alfvén waves are produced by the interaction between solar wind and Earth's magnetosphere. Due to the collisionless nature of space plasmas, kinetic effects create large-amplitude waves and pressure pulses in the foreshock region upstream from the quasi-parallel so-called bow shock (fig.\ref{bowshock}), which occurs when the magnetosphere of an astrophysical object interacts with the nearby flowing ambient plasma such as the solar wind: for Earth and other magnetized planets, it is the boundary at which the speed of the stellar wind abruptly drops as a result of its approach to the magnetopause, while, for stars, this boundary is typically the edge of the astrosphere, where the stellar wind meets the interstellar medium. The foreshock is found to be an important source of (magnetic) pulsating continuous (Pc) magnetospheric waves \cite{Fairfield1990,Engebretson1991,Chi1994,Clausen2009,Wang2019}. The mode conversion process associated with the compressional modes of the foreshock waves has been suggested as a directly driven mechanism for the generation of the frequently observed discrete harmonic frequencies of shear Alfvénic field-line resonances \cite{Hasegawa1979,Hasegawa1975,Lee1994}. Indeed, near the magnetopause boundary, a sharp transition is frequently found in wave polarization from predominantly compressional waves in the magnetosheath to transverse in the boundary layer \cite{Song1993,Rezeau1989,Chaston2008}. THEMIS observations \cite{Chaston2008} show a direct evidence of a turbulent spectrum of KAWs at the magnetopause with sufficient power to provide massive particle transport; the use of coordinated observations in the foreshock and the magnetosphere \cite{Wang2019} found direct evidence of pulsating continuous field line resonances driven by the foreshock perturbations. The main mode identification method for KAWs is based on the measurement of the wave polarization, which has to be \al{
\frac{c\delta\mathbf{E}_\perp}{\delta\mathbf{B}_\perp}=v_A\sqrt{\frac{b_k}{\sigma_k\left(1-\Gamma_k\right)}},\label{KAWpol}
} reducing to \al{
\frac{c\delta\mathbf{E}_\perp}{\delta\mathbf{B}_\perp}=v_A
} for SAWs; this method is used in several kind of measurements, for example with Van Allen Probes in the Earth's inner magnetosphere \cite{Chaston2014} and Cluster satellites in the solar wind \cite{Salem2012}, the two of them showing qualitative and/or quantitative agreement with the KAW value given by \eqref{KAWpol}.

\footnotetext{https://www.nasa.gov/mission\_pages/sunearth/multimedia/magnetosphere.html}

The role of KAWs in energy transport and particle acceleration in the magnetotail with a strong parallel disturbed electric field \cite{Zoncarev}, \al{
\left|\frac{c\delta E_\parallel}{\delta\mathbf{B}_\perp}\right|=v_A\left|\frac{k_\parallel}{k_\perp}\right|\tau\sqrt{\frac{b_k\left(1-\Gamma_k\right)}{\sigma_k}},
} has been investigated until very recently \cite{Zhang2022}, suggesting that the energy enhancement of electron beams is caused by KAWs.

The phenomena observed in the laboratory experiments show striking similarities to what has been observed by satellites in space plasmas, motivating an increasing interest in KAWs.

\chapter{\textsc{shear alfvén waves in a cylindrical plasma}}\label{mhdchapter}
\vspace{100pt}

In the present and in the next chapters, our focus will be on the study of the interaction (propagation and absorption) of Alfvén waves, whose frequency is such that $|\omega/\Omega_i|\ll1$, in a magnetized periodic cylindrical plasma with a non-uniformity in the radial direction.

As a first step, in this chapter we investigate the problem of the shear Alfvén wave by using the ideal, linear MHD model. As seen in the previous chapter, the existence of a spatial inhomogeneity, corresponding, in general, to realistic conditions, modifies the SAW dispersion relation from a constant value to a continuous spectrum, with oscillation frequency varying in the non-uniform (radial) direction.

\section{Model}\label{Model}

\begin{figure}[h]
	\centering
	\includegraphics[scale=0.4]{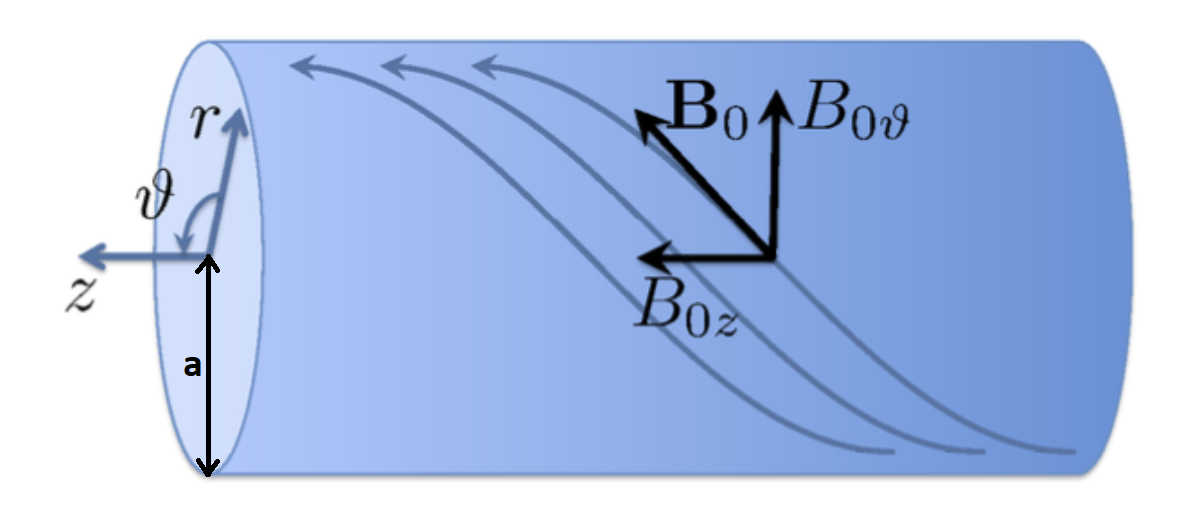}
	\caption{\emph{Fig.\ref{cyl}} Magnetized cylindrical plasma.}
	\label{cyl}
\end{figure}

Let's consider a periodic, cylindrical plasma (fig.\ref{cyl}), with periodic length $R$ and with radius $a$, and whose axis is directed along $\vec{z}$. The non-uniformity is in the radial direction $r$, so there is axial symmetry, and the magnetic equilibrium configuration is the \textit{screw pinch} \cite{Freidberg}, namely $\mathbf{B}_0(r)=(0,B_{0\theta}(r),B_{0z}(r))$. The components of the magnetic field are related to the components of the plasma current density through the Ampère-Maxwell equation (second of \eqref{Max2}), \begin{gather}
\D{B_{0\theta}}{r}+\frac{B_{0\theta}}{r}=\frac{4\pi J_{0z}}{c},\label{B0phi}\\
\D{B_{0z}}{r}=-\frac{4\pi J_{0\theta}}{c},\label{ampz}
\end{gather} while, being the non-uniformity purely radial, $J_{0r}=0$. Since $B_{0r}=0$, the radial condition for the momentum equilibrium is, from \eqref{motion}, \al{
\D{p_0}{r}=J_{0\theta}B_{0z}-J_{0z}B_{0\theta},\label{radmomeq}
} which, by substitution of \eqref{B0phi} and \eqref{ampz}, reads \al{
\D{}{r}\left(4\pi p_0+\frac{B_0^2}{2}\right)=-\frac{B_{0\theta}^2}{r}.\label{PBequilibrium}
}

\begin{figure}[h]
	\centering
	\includegraphics[scale=1.1]{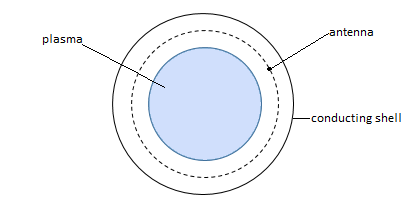}
	\caption{\emph{Fig.\ref{cylsect}} Cross-sectional view.}
	\label{cylsect}
\end{figure}

The plasma cylinder is surrounded (fig.\ref{cylsect}) by a conducting, coaxial cylindrical wall of radius $b$ and, in the vacuum region, at a distance $d$ such that $a<d<b$, an antenna carries an oscillating helical current density $\delta\mathbf{J}_h$, of components \al{\begin{gathered}
\delta J_{rh}=0,\\ \delta J_{\theta h}=\delta \tilde{J}_{\theta h}e^{\text{i}\left(-m\theta+kz-\omega_0 t\right)}\,\updelta\left(r-d\right),\qquad
\delta J_{zh}=-\frac{m}{kd}\delta\tilde{J}_{\theta h}e^{\text{i}\left(-m\theta+kz-\omega_0 t\right)}\,\updelta\left(r-d\right),\end{gathered}\label{jz}
} where $\omega_0$ is the externally imposed antenna frequency, $k$ is the wave number in the $z$-direction and will be conventionally denoted as $k=n/R$ to account for the periodicity of the cylindrical equilibrium, and $m$ is the azimuthal wave number. Meanwhile, the expression for $\delta J_{zh}$ is a consequence of the condition on the radial component, $\delta J_{rh}=0$, and of $\nabla\cdot\delta\mathbf{J}_h=0$. The sheet current density is \al{
\delta I_h=2\pi d\delta J_{zh}.
} This toy-model antenna launches a low-frequency wave that, for convenience, can be represented by the radial component of the fluctuating magnetic field $\delta B_{rh}$ \cite{Itoh1982,Itoh1983}. The conducting wall ensures the perfect reflection of the wave at the boundary, $\delta B_{rh}(r=b)=\delta E_{zh}(r=b)=0$. In general, the antenna current perturbations as well as the perturbations of the other physical quantities $\mathcal{F}$ characterizing the plasma can be expressed (using a generic $\omega$) as Fourier series \al{
\sum_{m,n}\delta\mathcal{F}_{mn}(r,\theta,z,t)=\sum_{m,n}\delta\mathcal{F}_{mn}(r)e^{\text{i}\left(-m\theta+\frac{n}{R}z-\omega t\right)}.
} Since all the $(m,n)$ components are linearly decoupled in the considered cylindrical plasma equilibrium, we can treat them separately and, suppressing redundant $m,n$ subscripts, write any generic perturbation as \al{
\delta\mathcal{F}(r,\theta,z,t)=\delta\mathcal{F}(r)e^{\text{i}\left(-m\theta+\frac{n}{R}z-\omega t\right)}=\delta\mathcal{F}(r)e^{\text{i}\left(rk_\theta(r)\theta+k_zz-\omega t\right)}.
} The wavenumber vector, then, is $\mathbf{k}=k_r\hat{\mathbf{r}}+k_\theta\hat{\boldsymbol{\theta}}+k_z\hat{\mathbf{z}}=k_r\hat{\mathbf{r}}-\hat{\boldsymbol{\theta}}m/r+\hat{\mathbf{z}}n/R$, and the gradient operator for perturbed quantities is \al{
\nabla=\hat{\mathbf{r}}\partial_r+\text{i}\hat{\boldsymbol{\theta}}k_\theta+\text{i}\hat{\mathbf{z}}k_z.
} For more clearly separating the parallel and perpendicular behaviors (with respect to $\mathbf{B}_0$) of the perturbation of vector fields, let us introduce an orthonormal projection based on the magnetic field lines with unit vectors \al{\begin{aligned}
\mathbf{e}_r&:=\hat{\mathbf{r}},\\\mathbf{e}_\perp(r):=\frac{1}{B_0(r)}\left(0,B_{0z}(r),-B_{0\theta}(r)\right),\quad &\mathbf{e}_\parallel(r):=\frac{\mathbf{B}_0}{B_0}=\frac{1}{B_0(r)}\left(0,B_{0\theta}(r),B_{0z}(r)\right),\end{aligned}
} according to which the wavenumber vector decomposes as \al{\begin{split}
k_\perp(r)&=\mathbf{k}\cdot\mathbf{e}_\perp(r)=-\frac{1}{B_0(r)}\left(B_{0z}(r)\frac{m}{r}+B_{0\theta}(r)\frac{n}{R}\right),\end{split} \\\begin{split} k_\parallel(r)&=\mathbf{k}\cdot\mathbf{e}_\parallel(r)=\frac{1}{B_0(r)}\left(B_{0z}(r)\frac{n}{R}-B_{0\theta}(r)\frac{m}{r}\right)=\\&=\left(n-\frac{m}{q(r)}\right)\frac{B_{0z}(r)}{RB_0(r)},\end{split}\label{kcyl}
} being \al{
q(r)=\frac{rB_{0z}(r)}{RB_{0\theta}(r)}
} the so-called \textit{safety factor} of the plasma, and such that the gradient operator for perturbed quantities becomes \al{
\nabla=\mathbf{e}_r\partial_r+\text{i}\mathbf{e}_\perp k_\perp+\text{i}\mathbf{e}_\parallel k_\parallel.
} The SAW dispersion relation is, then, \al{
\omega_A^2(r)=k_\parallel^2(r)v_A^2(r)
} with $k_\parallel$ as defined in \eqref{kcyl}.

With the sole exception of the current-sheet antenna, in the vacuum region between the plasma and the wall ($a<r<b$) the basic equations are Maxwell's equations; nonetheless, in order to avoid problems with divergent Alfvén speed, as they will be clearer in section \ref{equilibriumprofile}, we assume, in this region, the existence of a nearly zero-density cold plasma.

\subsection{Relations between fields and potentials}

Equation \eqref{deltaBdeltaxi} gives, at the lowest order, \al{
\delta B_r=\text{i}k_\parallel B_0\delta\xi_r,\label{deltaBdeltaxiprop2}
} which is the analog of \eqref{deltaBdeltaxiprop}, and \al{
\delta B_\theta=\text{i}k_\parallel B_0\delta\xi_\theta-r\DP{}{r}\left(\frac{B_{0\theta}}{r}\right)\delta\xi_r.
} When, as it shall be for our model, $B_{0\theta}\ll B_{0z}$ and $B_{0z}\approx B_0$ is almost uniform, $\partial_r(B_{0\theta}/r)\approx-(B_0/m)\partial_rk_\parallel$ and so \al{
\delta B_\theta\approx\text{i}k_\parallel B_0\delta\xi_\theta-\frac{B_0r}{m}\DP{k_\parallel}{r}\delta\xi_r,
} and finally, since $\nabla_\perp\cdot\boldsymbol{\xi}_\perp\approx0$, \al{\begin{aligned}
\delta B_\theta&=\frac{B_0r}{m}\left(\frac{k_\parallel}{r}\DP{}{r}\left(r\delta\xi_r\right)+\DP{k_\parallel}{r}\delta\xi_r\right)=\\
&=\frac{B_0}{m}\DP{}{r}\left(rk_\parallel\delta\xi_r\right).\end{aligned}
} The relations between (displacement and magnetic) fields and potentials are given by first noting that, according to \eqref{lorentzero}, $\nabla\delta\phi=-\delta\mathbf{E}=\delta\mathbf{v}\times\mathbf{B}_0/c$, and so $\mathbf{e}_\parallel\times\nabla\delta\phi=B_0\partial_t\delta\boldsymbol{\xi}_\perp/c$, from which \al{
\delta\boldsymbol{\xi}_\perp=\frac{\text{i}c}{\omega B_0}\mathbf{e}_\parallel\times\nabla\delta\phi,
} that is, in components, \al{
r\delta\xi_r&=-\frac{cm}{\omega B_0}\delta\phi,\label{xirpot}\\
\delta\xi_\theta&=\frac{\text{i}c}{\omega B_0}\DP{\delta\phi}{r},\label{xitpot}
} and, using \eqref{deltaBdeltaxiprop2}, \al{
r\delta B_r&=-\frac{\text{i}cmk_\parallel}{\omega}\delta\phi,\label{brpot}\\
\delta B_\theta &=-\frac{c}{\omega}\DP{\left(k_\parallel\delta\phi\right)}{r}.\label{btpot}
} So, with $B_0$ almost uniform, the scalar potential is essentially given by $r\delta\xi_r$. From the MHD condition \eqref{lorentzero} it also derives that $\delta E_\parallel=0=\mathbf{e}_\parallel\cdot\nabla\phi-(1/c)\partial_tA_\parallel$, and so the relations involving the parallel vector potential $A_\parallel$, \al{
r\delta B_r&=-\text{i}m\delta A_\parallel,\label{brvecpot}\\
\delta B_\theta&=-\DP{}{r}\delta A_\parallel,\label{btvecpot}
} are fully consistent with \eqref{brpot} and \eqref{btpot}.

\section{Field equation and antenna conditions}

In this section we first derive the external antenna field, and subsequently the internal, SAW master equation. The matching conditions between the two fields are the boundary conditions for the internal solution.

\subsection{Antenna equation}\label{antennaequation}

The wave equation in vacuum is obtained, as usual, from Maxwell's equations in this way: \al{
\nabla^2\mathbf{\delta B}=-\nabla\times\nabla\times\delta\mathbf{B}=-\frac{4\pi}{c}\nabla\times\delta\mathbf{J}-\frac{1}{c}\DP{}{t}\nabla\times\delta\mathbf{E}=-\frac{4\pi}{c}\nabla\times\delta\mathbf{J}+\frac{1}{c^2}\DP{^2}{t^2}\delta\mathbf{B};
} since $[\nabla\times(r\delta\mathbf{B})]_r=[\nabla r\times\delta\mathbf{B}]_r+r[\nabla\times\delta\mathbf{B}]_r=r[\nabla\times\delta\mathbf{B}]_r$, we obtain \al{\begin{aligned}
\frac{1}{r}\partial_r\left[r\partial_r\left(r\delta B_r\right)\right]+\frac{\partial^2_\theta (r\delta B_r)}{r^2}+\partial_z^2(r\delta B_r)-\frac{\partial_t^2(r\delta B_r)}{c^2}&=-\frac{4\pi r}{c}\left(\frac{\partial_\theta\delta J_z}{r}-\partial_z\delta J_\theta\right)=\\&=\frac{4\pi\text{i}r}{k_\theta c}\left(\frac{\partial_\theta^2\delta J_z}{r^2}+\partial_z^2\delta J_z\right).\end{aligned}\label{Itohvacuum}
} Here, we have used that $(1/r) \partial_\theta \delta J_\theta + \partial_z \delta J_z = 0$ for the antenna current density. For our model of plasma of radius $a$ with antenna located at $d$ and carrying current $\delta I_h$, and external conducting shell in $b$, and with the definition $\nu:=\sqrt{k_z^2-\omega_0^2/c^2}$, the solution to \eqref{Itohvacuum} is, for $a\leq r<d$ (see also \cite{Itoh1982,Itoh1983}) and introducing the $|m|$-th order modified Bessel functions of the first and of the second kind, $I_{|m|}$ and $K_{|m|}$ respectively, \al{\begin{aligned}
r\delta B_r&=\alpha_0\left[I_{|m|}(\nu r)K_{|m|}(\nu b)-K_{|m|}(\nu r)I_{|m|}(\nu b)\right]\\&-\frac{2\text{i}m\delta I_h}{c\nu}\left(1+\frac{k_z^2d^2}{m^2}\right)\left[I_{|m|}(\nu r)K_{|m|}(\nu d)-K_{|m|}(\nu)I_{|m|}(\nu d)\right],\end{aligned}\label{Itohvacuumsol0}
} where $\alpha_0$ is a constant depending on the antenna frequency, and having used \al{
I_{|m|}(\nu d)K'_{|m|}(\nu d)-K_{|m|}(\nu d)I'_{|m|}(\nu d)=-\frac{1}{\nu d},
} while, for $d<r\leq b$, only the homogeneous solution survives, \al{
r\delta B_r&=\alpha_0\left[I_{|m|}(\nu r)K_{|m|}(\nu b)-K_{|m|}(\nu r)I_{|m|}(\nu b)\right].\label{Itohvacuumsol}
} With the definition \al{
\delta\hat{B}_{rh}:=\frac{2m\delta I_h}{ca\nu}\left(1+\frac{k_z^2d^2}{m^2}\right),
} solution \eqref{Itohvacuumsol0} can be rewritten as \al{\begin{aligned}
\frac{r}{a}\frac{\delta B_r}{B_0}&=-\text{i}\frac{\delta\hat{B}_{rh}}{B_0}\bigg\{\left[I_{|m|}(\nu r)K_{|m|}(\nu d)-K_{|m|}(\nu r)I_{|m|}(\nu d)\right]\\
&+\alpha\left[I_{|m|}(\nu r)K_{|m|}(\nu b)-K_{|m|}(\nu r)I_{|m|}(\nu b)\right]\bigg\}=\\
&=:-\text{i}\frac{\delta\hat{B}_{rh}}{B_0}\left(y_{1\nu}(r)+\alpha y_{2\nu}(r)\right),\end{aligned}\label{antennasol}
} or, also, \al{
y:=\frac{r\delta\xi_r}{a^2}=\frac{x\delta\xi_r}{a}=-\frac{1}{k_\parallel a}\frac{\delta\hat{B}_{rh}}{B_0}\left(y_{1\nu}+\alpha y_{2\nu}\right),\label{solmatch}
} having introduced for convenience the normalized variable $x:=r/a$. The complex parameter $\alpha$, depending on the antenna frequency, is the characteristic ratio of the solution of the homogeneous equation to the particular solution with forcing term, or antenna solution, namely a kind of field/perturbation ratio, and, as we shall examine in section \ref{Poyntingfluxanalysis}, is a measure of the Poynting flux launched by the antenna from outside into the inner region.

\subsection{SAW equation}\label{sawequation}

The curvature of the magnetic field modifies the momentum equation \eqref{momkin} through an additional term, \al{\begin{aligned}
\varrho_0\DP{^2}{t^2}\delta\boldsymbol{\xi}&=-\nabla\delta\tilde{p}+\frac{1}{4\pi}\left(\mathbf{B}_0\cdot\nabla\right)^2\delta\boldsymbol{\xi}-\frac{1}{4\pi}\mathbf{B}_0\left(\mathbf{B}_0\cdot\nabla\right)\left(\nabla\cdot\delta\boldsymbol{\xi}\right)\\&-\left(2\nabla\cdot\delta\boldsymbol{\xi}+\delta\boldsymbol{\xi}\cdot\nabla\right)\left(\frac{\mathbf{B}_0\cdot\nabla\mathbf{B}_0}{4\pi}\right),\end{aligned}\label{momeqcy}
} which readily yields $\delta\xi_\parallel=0$ in the low-$\beta$ limit, $\beta$ being the ratio between kinetic and magnetic pressures (see subsection \ref{nonuniformslab} and equation \eqref{deltaxiparallel}). Meanwhile, \al{\begin{aligned}
\frac{\delta B_\parallel}{B_0}&=-\nabla\cdot\delta\boldsymbol{\xi}_\perp-\left(\boldsymbol{\kappa}+\frac{\nabla_\perp B_0}{B_0}\right)\cdot\delta\boldsymbol{\xi}_\perp=\\&=-\nabla\cdot\delta\boldsymbol{\xi}_\perp+2\frac{B_{0\theta}^2}{B_0^2r}\delta\xi_r,\end{aligned}\label{deltabpareqcy}
} where $\boldsymbol{\kappa}:=\mathbf{B}_0\cdot\nabla{\mathbf{B}}_0/B_0^2$ is the magnetic curvature. The perpendicular force balance equation \eqref{deltaxiperp} is then modified to \al{
D_A\delta\boldsymbol{\xi}_\perp=\nabla_\perp\left(\frac{\delta B_\parallel}{B_0}\right)-\hat{\mathbf{r}}\DP{}{r}\left(\frac{B_{0\theta}^2}{r^2B_0^2}\right)r\delta\xi_r+\frac{2\text{i}k_\parallel B_{0\theta}}{rB_0}\left(\hat{\mathbf{r}}\delta\xi_\theta-\hat{\boldsymbol{\theta}}\delta\xi_r\right).\label{deltaxiperp2}
} In the limit $|D_A|\ll k_\theta^2$ (see subsection \ref{nonuniformslab}) and considering that \al{
\left(1-\frac{D_A}{k_\theta^2}\right)^{-1}\approx\left(1+\frac{D_A}{k_\theta^2}\right),
} the $\theta$ component of \eqref{deltaxiperp2} is \al{
\delta\xi_\theta\approx\left(1+\frac{D_A}{k_\theta^2}\right)\frac{\text{i}}{k_\theta r}\DP{}{r}\left(r\delta\xi_r\right)+\frac{2\text{i}}{k_\theta^2r}\left(k_\parallel\frac{B_{0\theta}}{B_0}-k_\theta\frac{B_{0\theta}^2}{B_0^2}\right)\delta\xi_r.
} Substituting this equation into \eqref{deltabpareqcy} and into the $r$-component of \eqref{momeqcy} one arrives, respectively, at \al{
\frac{\delta B_\parallel}{B_0}=\frac{D_A}{k_\theta^2r}\DP{}{r}\left(r\delta\xi_r\right)+\frac{2k_\parallel\delta\xi_r}{k_\theta r}\frac{B_{0\theta}}{B_0}\label{2.44}
} and at the wave equation \al{
\DP{}{r}\left(rD_A\DP{}{r}\left(r\delta\xi_r\right)\right)-m^2D_A\delta\xi_r+r\D{}{r}\left[\left(2k_\parallel\frac{B_{0\theta}}{B_0}-k_\theta\frac{B_{0\theta}^2}{B_0^2}\right)k_\theta\right]\delta\xi_r=0.\label{2.45}
} 

Note that, in the typical case $B_{0\theta}\ll B_{0z}$, \al{
k_\parallel\approx\frac{n}{R}+k_\theta\frac{B_{0\theta}}{B_{0z}};\label{kpar}
} accordingly, \eqref{2.45} becomes \al{
\DP{}{r}\left(rD_A\DP{}{r}\left(r\delta\xi_r\right)\right)-\left(m^2D_A-r\D{}{r}k_\parallel^2\right)\delta\xi_r=0.\label{2.45simple}
} Near the resonance, and taking $|k_\theta|\ll|\partial_r|$ as the analogue to $|k_y|\ll|\partial_x|$ in subsection \ref{nonuniformslab}, we find that, approaching $r_0$, \al{
\delta\xi_r\approx C\ln\left(r-r_0\right)+D,\label{deltaxircorr}
} for some constants $C$ and $D$. Equation \eqref{deltaxircorr} describes, thus, the radial singular structures of the SAW continuous spectrum similar to what was discussed in the simpler slab case. Equation \eqref{2.45simple} describes SAW oscillation in a nonuniform cylindrical plasma. In the limit of a cold tenuous plasma, such as that occupying the "vacuum" region in the present case, $\omega_0^2/\omega_A^2\ll1$ and, so, $D_A\approx-k_\parallel^2$, yielding \al{
\DP{}{r}\left(rk_\parallel^2\DP{}{r}\left(r\delta\xi_r\right)\right)-\left(m^2k_\parallel^2+r\D{}{r}k_\parallel^2\right)\delta\xi_r=0.
} When $B_{0z}\approx B_0$ is (almost) uniform, and invoking \eqref{deltaBdeltaxiprop2}, we obtain \al{
\DP{}{r}\left(rk_\parallel^2\DP{}{r}\left(\frac{r\delta B_r}{k_\parallel}\right)\right)-\left(\frac{m^2k_\parallel}{r}+\frac{\partial_rk_\parallel^2}{k_\parallel}\right)r\delta B_r=0,
} or, using \eqref{B0phi} and \eqref{kpar}, the equation \cite{Itoh1982,Itoh1983} \al{
\frac{1}{r}\DP{}{r}\left(r\DP{}{r}\left(r\delta B_r\right)\right)-\frac{m^2}{r^2}r\delta B_r-\frac{4\pi k_\theta\partial_rJ_{0z}}{cB_0k_\parallel}r\delta B_r=0. \label{Itoh}
} Equations \eqref{2.45} or \eqref{2.45simple} may also be extended, assuming that there is an external force adding to the right-hand side, in the form \al{
-\frac{4\pi}{B_0^2c}k_\theta^2r^2\left(\delta\mathbf{J}\times\mathbf{B}_0\right)_r.
} Since $\delta J_r=0$ and $0=\nabla\cdot\delta\mathbf{J}=k_\theta\delta J_\theta+k_z\delta J_z$, \al{
\frac{4\pi k_\theta^2r^2}{B_0^2c}\left(B_{0\theta}\delta J_z-B_0\delta J_\theta\right)=-\frac{4\pi k_\theta^2r^2\delta J_\theta B_0}{B_0^2ck_z}\left(k_z+k_\theta\frac{B_{0\theta}}{B_0}\right)=\frac{4\pi k_\theta r^2}{B_0^2c}\delta J_z B_0k_\parallel.
} Correspondingly, \eqref{Itoh} can be extended as \al{
\frac{1}{r}\DP{}{r}\left(r\DP{}{r}\left(r\delta B_r\right)\right)-\frac{m^2}{r^2}r\delta B_r-\frac{4\pi k_\theta\partial_rJ_{0z}}{cB_0k_\parallel}r\delta B_r=-\frac{4\pi\text{i}k_\theta r}{c}\delta J_z.\label{Itohsource}
} The vacuum limit of \eqref{Itohsource} is taken for $\partial_rJ_{0z}=0$, and corresponds to the limit $\omega_0^2\ll k_z^2c^2\ll k_\perp^2c^2$ of \eqref{Itohvacuum}.

\subsection{Matching conditions}\label{matching}

The "vacuum" region solution \eqref{solmatch} must be smoothly connected at the plasma boundary with the internal plasma solution, namely with the solution to \eqref{2.45} or to \eqref{2.45simple}; in both cases, with the solution of a 2nd-order ordinary differential equation, which is a linear combination of two independent solutions. Since, as we have seen, near the resonant layer this linear combination becomes $D+C\ln(r-r_0)$, the two independent solutions $y_1$ and $y_2$ can be taken in general as those which, approaching the resonance, respectively behave as (remember that $x:=r/a$) \al{
y_1(x\rightarrow r_0/a)\approx1,\qquad y_2(x\rightarrow r_0/a)\approx\frac{\ln\left(x-x_0\right)}{R^2\partial_x\left(xD_A\right)}.\label{y1y2res}
} The two solutions are independent since the wronskian is \al{
y_1y_2'-y_1'y_2=\frac{1}{R^2xD_A}\neq0.
} The inner solution expressed as linear combination of $y_1$ and $y_2$ matches, with different constants, each of the two external solutions $y_{1\nu}$ and $y_{2\nu}$ in \eqref{solmatch}, \al{\left\{ \begin{array}{ll}
Y_1(a):=-\frac{1}{k_\parallel a}\frac{\delta\hat{B}_r}{B_0}y_{1\nu}(a)=C_1y_2(a)+D_1y_1(a),\\
Y_2(a):=-\frac{1}{k_\parallel a}\frac{\delta\hat{B}_r}{B_0}y_{2\nu}(a)=C_2y_2(a)+D_2y_1(a),
\end{array} \right.\label{MHDcond1}
} and so it can also be written as \al{\begin{aligned}
y=x\frac{\delta\xi_r}{a}&=D_1y_1+C_1y_2+\alpha\left(D_2y_1+C_2y_2\right)=\\&=\left(C_1+\alpha C_2\right)y_2+\left(D_1+\alpha D_2\right)y_1.\end{aligned}\label{intextmatch}
} The boundary conditions are to be completed by the matching of the derivatives, \al{
\left\{ \begin{array}{ll}
Y'_1(a)=C_1y'_2(a)+D_1y'_1(a),\\
Y'_2(a)=C_2y'_2(a)+D_2y'_1(a),
\end{array} \right.\label{MHDcond2}
} in order to give the four constants as functions of $y_{1,2}(a)$ and $y'_{1,2}(a)$, \al{
C_{1,2}=\frac{Y_{1,2}y_1'(a)-Y'_{1,2}y_1(a)}{y_1y'_2(a)-y'_1y_2(a)},\\
D_{1,2}=\frac{Y_{1,2}y_2'(a)-Y'_{1,2}y_2(a)}{y_1y'_2(a)-y'_1y_2(a)}.
} The last member of \eqref{intextmatch} also defines the matching conditions with the internal region, \begin{gather}
C=C_1+\alpha C_2,\label{MHDcond30}\\
D=D_1+\alpha D_2.\label{MHDcond3}
\end{gather}
In conclusion, after rescaling the linear solution by the strength of the antenna perturbation as in \eqref{solmatch}, the internal region solution approaching the singular layer at $r=r_0$ is determined up to the constant $\alpha$, which, as specified earlier, represents the ratio between the homogeneous solution and the antenna solution.

\section{Poynting flux analysis}\label{Poyntingfluxanalysis}
Poynting's theorem reads \al{
\DP{}{t}\left(\frac{|\delta \mathbf{E}|^2+|\delta \mathbf{B}|^2}{8\pi}\right)=-\frac{c}{4\pi}\nabla\cdot(\delta\mathbf{E}\times\delta\mathbf{B})-\delta\mathbf{J}\cdot\delta\mathbf{E},
} but considering that, in ideal MHD and with negligible diamagnetic current, \al{
\delta\mathbf{J}_\perp=\frac{c^2}{4\pi v_A^2}\DP{}{t}\delta\mathbf{E}_\perp,
} it becomes \al{
\DP{}{t}\left(\frac{\left(1+\frac{c^2}{v_A^2}\right)|\delta \mathbf{E}|^2+|\delta \mathbf{B}|^2}{8\pi}\right)=-\nabla\cdot\frac{c}{4\pi}(\delta\mathbf{E}\times\delta\mathbf{B}).
} Finally, writing the fields as \al{
\delta\mathbf{E}=\frac{1}{2}\left(\delta\mathbf{E}_{mn}(r)e^{-\text{i}\left(-m\theta+\frac{n}{R}z-\omega_0 t\right)}+c.c.\right),\\ \delta\mathbf{B}=\frac{1}{2}\left(\delta\mathbf{B}_{mn}(r)e^{-\text{i}\left(-m\theta+\frac{n}{R}z-\omega_0 t\right)}+c.c.\right),
} and mediating over the wave period and the cylindrical surface flux, we get \al{
\DP{}{t}\left(\frac{\left(1+\frac{c^2}{v_A^2}\right)|\delta \mathbf{E}_{mn}|^2+|\delta \mathbf{B}_{mn}|^2}{16\pi}\right)=-\frac{c}{8\pi r}\partial_r\left[r\text{Re}(\delta\mathbf{E}_{mn}\times\delta\mathbf{B}^*_{mn})_r\right],\label{pow}
} where a sum over the mode numbers $m,n$ is assumed implicitly in the case of multiple Fourier components. Since $\delta E_\parallel=0$, we have $0=\delta\mathbf{E}\cdot\delta\mathbf{B}=\delta E_\theta\delta B_\theta+\delta E_z\delta B_z$ and $(\delta \mathbf{E}_{mn}\times\delta \mathbf{B}_{mn}^*)_r=\delta E_{mn\theta}\delta B_{mn\parallel}^*$, so, using \eqref{2.44}, we also have \al{
\DP{}{t}\left(\frac{\left(1+\frac{c^2}{v_A^2}\right)|\delta \mathbf{E}_{mn}|^2+|\delta \mathbf{B}_{mn}|^2}{16\pi}\right)=\frac{c}{8\pi r}\partial_r\left[r\text{Im}\left(\frac{\omega_0 B_0^2 D_A}{ck_\theta^2r}\delta\xi_r^*\partial_r(r\delta\xi_r)\right)\right],\label{power}
} having dropped the $mn$ subscript to the radial plasma displacement $\delta \xi_r$, consistently with the present notation. This formula implies that the Poynting flux vanishes when $\delta\xi_r$ is a real quantity. Since, due to the discussion in \ref{nonuniformslab} and to \eqref{xixres}, near the resonance $\delta\xi_r=C_m\ln(r-r_0)+D_m$ for $r>r_0$ ($r=r_0$ is the equation of the resonant layer), it holds that (keeping track, from now on, only of the subscript $m$ to $C_m$ to avoid possible confusion) \cite{Hasegawa1974} \al{
\delta\xi_r=C_m\left[\ln(r_0-r)-\text{i}\pi\text{sgn}\left(\frac{\omega_0}{(\omega_A^2)'}\right)\right]+D_m,\quad\text{for}\quad r<r_0.\label{solreson}
} The imaginary part is then connected with the existence of the resonance and of the logarithmic singularity as manifestation of phase mixing \cite{Hasegawa1974,Grad1969}. The imaginary part in \eqref{solreson} is selected according to the "causality constraint" that, in the real space, corresponds to the modification of the Landau contour in the velocity space for the description of Landau damping. In this respect, spatial phase mixing has deep common roots with velocity space phase mixing in wave-particle resonances \cite{Grad1969}. Clearly, the resonance has to exist inside the plasma so that the wave can lose energy to the plasma itself. The coefficients $C_m$ and $D_m$ or, more precisely, their ratio, since the present problem is linear, have to be found from the conditions at the plasma boundary, as illustrated in the previous section. Then, solution \eqref{solreson} must be continued up to the plasma center, where regularity conditions must apply. These will ultimately determine the constant $\alpha$ and the overall radial structure of the fluctuation generated by the external antenna. Integrating \eqref{power} one obtains the absorption rate of the total energy (in the components $m,n$ of the wave spectrum), \al{
W=4\pi^2R\bigintsss_0^ar\,\text{d}r\left[\frac{\left(1+\frac{c^2}{v_A^2}\right)|\delta \mathbf{E}_m|^2+|\delta \mathbf{B}_m|^2}{16\pi}\right].
} In fact, if we integrate over the plasma cylinder and apply Stokes' theorem, we obtain \al{
\D{W}{t}=\frac{\pi^2}{2}r_0R\frac{|C_m|^2B_0^2}{(k_\theta^2v_A^2)_{r=r_0}}|(\omega_A^2)'\omega_0|.
} From \eqref{2.44} and \eqref{power}, we see that \al{
\frac{\delta B_\parallel}{B_0}=\frac{D_A}{k_\theta^2r}\partial_r\left(r\delta\xi_r\right)+\frac{2k_\parallel\delta\xi_r}{k_\theta r}\frac{B_{0\theta}}{B_0}
} and that only the first term on the right-hand side contributes to \eqref{power} through its imaginary part. Expanding $D_A=(\omega_0^2-\omega_A^2)/v_A^2$ near the resonance, \al{
D_A\approx-\frac{\left(\omega_A^{2}\right)'}{v_A^2}\left(r-r_0\right),
} this term is, ignoring the part not giving imaginary contribution, \al{
-\frac{\omega_A^{2'}}{v_A^2}\frac{r_0^2}{m^2}C_m,
} namely the resonant absorption only depends on $C_m$ and not on $D_m$. Substituting the external solution \eqref{antennasol} in the Poynting flux and the expression of power, we obtain, only considering the contributions that lead to finite power flux, \al{
\DP{}{x}\left(x\frac{\delta\xi_r}{a}\right)\approx-\frac{\nu}{k_\parallel}\frac{\delta\hat{B}_{rh}}{B_0}\left(y'_{1\nu}(r)+\alpha y'_{2\nu}(r)\right),
} and the time and surface averaged power flux \al{\begin{aligned}
\overline{S}_r&=\frac{c}{8\pi}\text{Re}\left(\delta \mathbf{E}_m\times\delta \mathbf{B}_m^*\right)_r=-\frac{c}{8\pi}\text{Im}\left(\frac{\omega_0 B_0^2D_A}{ck_\theta^2r}\delta\xi_r^*\partial_r\left(r\delta\xi_r\right)\right)=\\
&=-\frac{B_0^2}{8\pi}\text{Im}\left[\frac{\omega_0 D_Aa^3}{m^2}\left(x\frac{\delta\xi_r}{a}\right)^*\partial_x\left(x\frac{\delta\xi_r}{a}\right)\right]=\\
&=-\frac{B_0^2\omega_0\nu a^2D_A}{8\pi m^2k_\parallel^2}\left(\frac{\delta\hat{B}_{rh}}{B_0}\right)^2\text{Im}\alpha\left[(y_{1\nu}(r)y_{2\nu}'(r)-y_{1\nu}'(r)y_{2\nu}(r)\right]=\\
&=\frac{B_0^2\omega_0 a^2}{8\pi m^2r}\left(\frac{\delta\hat{B}_{rh}}{B_0}\right)^2\text{Im}\alpha\left[(K_{|m|}(\nu d)I_{|m|}(\nu b)-K_{|m|}(\nu b)I_{|m|}(\nu d)\right],\end{aligned}
} having used in the last step the wronskian identity \al{
y_{1\nu}(r)y_{2\nu}'(r)-y_{1\nu}'(r)y_{2\nu}(r)=\frac{1}{\nu r}\left[(K_{|m|}(\nu d)I_{|m|}(\nu b)-K_{|m|}(\nu b)I_{|m|}(\nu d)\right]
} and assumed $D_A\approx-k_\parallel^2$ in "vacuum". The flux is nonzero only when $\text{Im}\alpha\neq0$: so (the inverse of) $\alpha$ is a measure of the plasma response at fixed antenna perturbation, and its imaginary part gives information on the plasma absorption rate by phase mixing. In conclusion, \al{
\overline{S}_r=-\overline{S}_{rh}\left(d/r\right),
} as expected, being, for the (modulus of the) Poynting flux calculated at the antenna, \al{
\overline{S}_{rh}:=-\frac{B_0^2\omega_0}{8\pi m^2d}\left(\frac{\delta \hat{B}_{rh}}{B_0}\right)^2\text{Im}\alpha\left[(K_{|m|}(\nu d)I_{|m|}(\nu b)-K_{|m|}(\nu b)I_{|m|}(\nu d)\right].
} Similarly, for the plasma solution \eqref{intextmatch}, we also have \al{
S_r&=-S_0D_A\text{Im}\left[\left(C_1+\alpha C_2\right)^*y_2+\left(D_1+\alpha D_2\right)^*y_1\right]\left[\left(C_1+\alpha C_2\right)y'_2+\left(D_1+\alpha D_2\right)y'_1\right]=\nonumber\\
&=-S_0D_A\left(y_1y'_2-y_2y'_1\right)\text{Im}\left[\left(C_1+\alpha C_2\right)\left(D_1+\alpha D_2\right)^*\right]=\\
&=\frac{S_0a}{Rr}\text{Im}\alpha\left(C_1D_2-D_1C_2\right)\nonumber,
} where $S_0:=\omega_0 B_0^2a^3/(8\pi m^2)$. By comparison with the power flux in vacuum, \al{
\left(C_1D_2-D_1C_2\right)\frac{a^2}{R^2}=\left(\frac{\delta\hat{B}_{rh}}{B_0}\right)^2\left[K_{|m|}(\nu d)I_{|m|}(\nu b)-K_{|m|}(\nu b)I_{|m|}(\nu d)\right],
} which is an identity, as expected. These equations essentially say that there is a power flux directed inside which goes, both in vacuum and in the plasma (up to $r_0$), as $1/r$; meanwhile, it suddenly goes to zero for $r<r_0$ and for $r>d$, as it must be for the conservation of the power flux generated by the antenna in stationary regime and for its absorption at the resonant surface. This holds in the present, MHD limit; in the next chapter, we shall see that the Poynting flux apparently vanishing at $r=r_0$ actually describes the mode conversion to the kinetic Alfvén wave.

\section{Equilibrium configuration and numerical results}\label{equilibriumprofile}

In order to solve the field equation, we assign the plasma equilibrium profiles, namely density, temperature, current, magnetic field, and pressure, and choose antenna parameters controlling the excitation of SAW.

\subsection{Equilibrium profile}\label{equilprofile}

\begin{figure}
	\centering
	\includegraphics[scale=0.3]{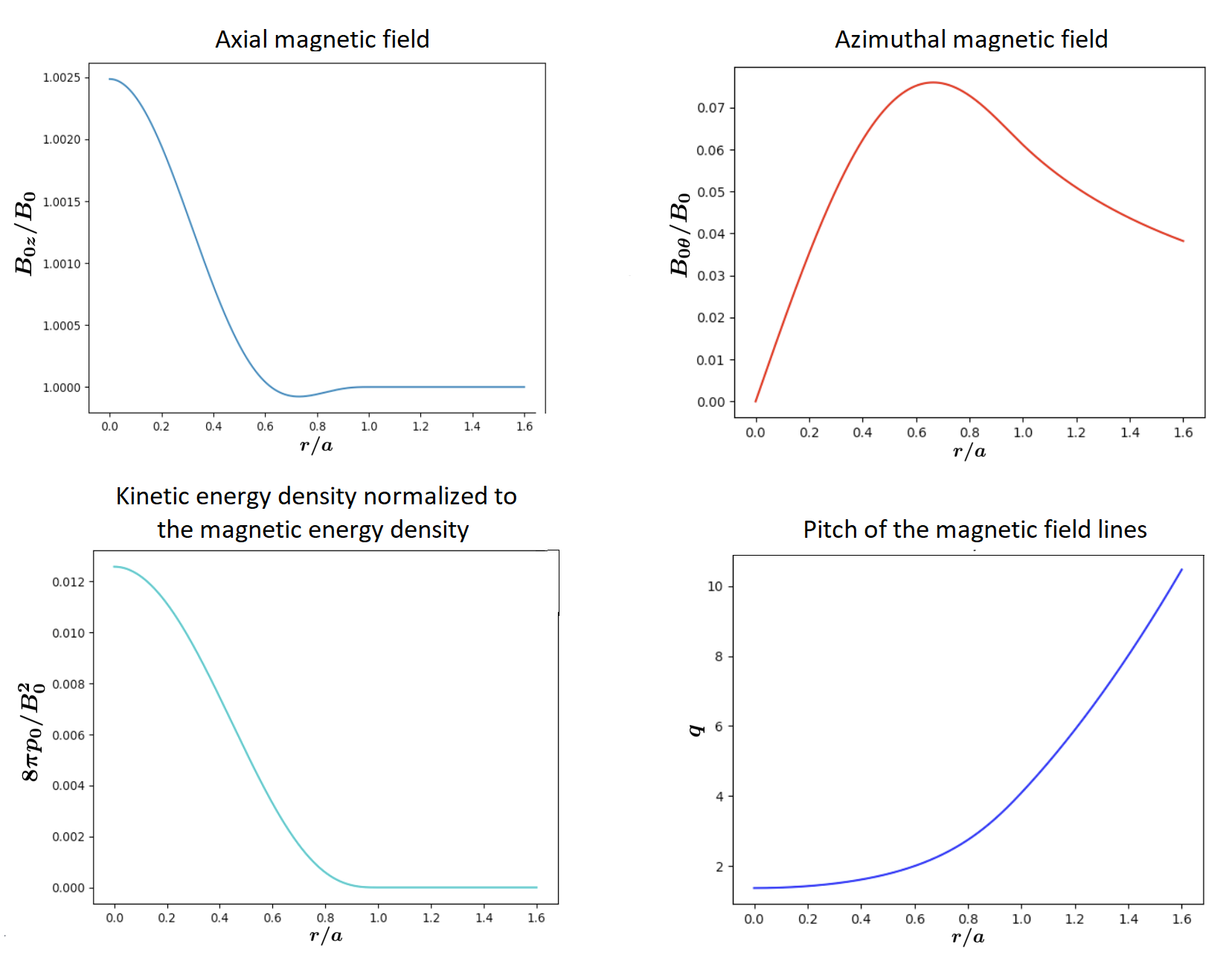}
	\includegraphics[scale=0.3]{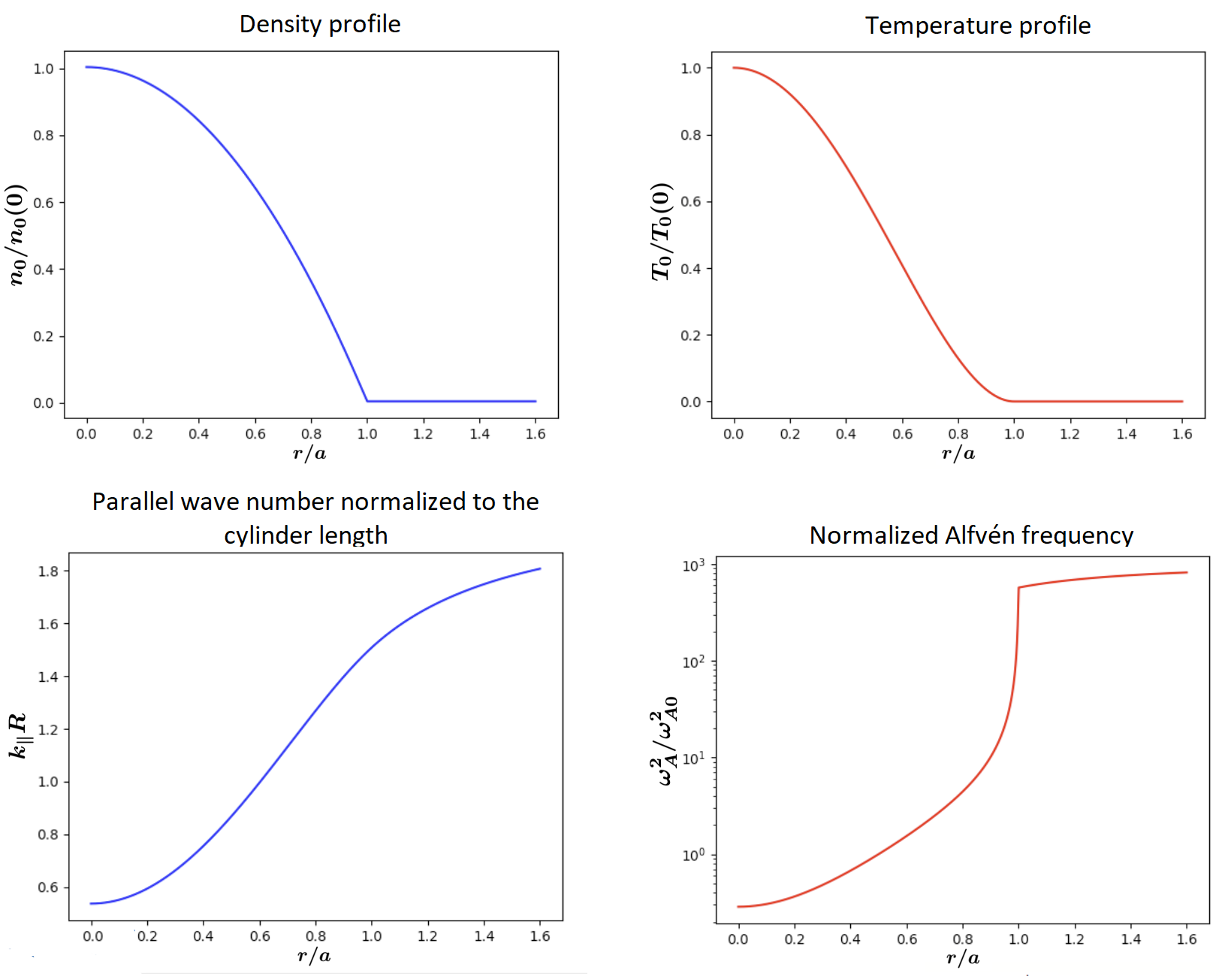}
	\caption{\emph{Fig.\ref{equil1}} Plasma equilibrium ($B_0$ is the vacuum field, while $\omega_{A0}:=B_0(0)(4\pi n_0 m_i)^{-1/2}/R$).}
	\label{equil1}
\end{figure}

We assume a mass density profile \al{
\varrho_0=m_in_0=m_in_0(0)\left[\left(1-x^2\right)^{\alpha_n}+n_\varepsilon\right];
} the last term lets a negligible number of plasma particles to exist at a distance $r>a$ from the axis, the practical reason for this (definitely realistic) choice being that, otherwise, with an infinitesimal density, the Alfvén velocity would diverge and cause difficulties with the numerical analysis. The temperature profile is given by \al{
T_{0e,i}=T_{0e,i}(0)\left(1-x^2\right)^{\alpha_T},
} with $T_{0e}(0)/T_{0i}(0)=\tau=7/3$ and $T_{0e}=T_{0i}=0$ in the vacuum region. The current density, \al{
J_{0z}=J_{0z}(0)\left(1-x^2\right)^{\alpha_J},
} goes to zero at the boundary and remains zero throughout the vacuum region. This current determines the component $B_{0\theta}$ of the magnetic field through \eqref{B0phi}, that is, \al{
B_{0\theta}=\left\{ \begin{array}{ll}\frac{2\pi aJ_{0z}(0)}{c\left(\alpha_J+1\right)x}\left[1-\left(1-x^2\right)^{\alpha_J+1}\right]&\quad\text{for}\quad x<1,\\
\frac{2\pi aJ_{0z}(0)}{c\left(\alpha_J+1\right)x}&\quad\text{for}\quad x\geq1. \end{array}\right.\label{B0thetaeq}
} Given the definition \al{
\overline{p}_0:=p_0+\frac{B_{0z}^2-B_{0z}^2(1)}{8\pi},\label{pdefinition}
} the equilibrium condition \eqref{PBequilibrium} yields \al{\begin{aligned}
4\pi\D{\overline{p}_0}{x}&=-\frac{1}{2}\D{B_{0\theta}^2}{x}-\frac{B_{0\theta}^2}{x}=\\
&=-\frac{4\pi^2a^2J_{0z}^2(0)}{c^2\left(\alpha_J+1\right)^2x^2}\left[2x\left(1-\left(1-x^2\right)^{\alpha_J+1}\right)\left(\alpha_J+1\right)\left(1-x^2\right)^{\alpha_J}\right].\end{aligned}
} The $\overline{p}_0$ field can be integrated analytically by using the confluent hypergeometric function; namely, for $x\leq1$, \al{\begin{aligned}
\overline{p}_0&=\frac{2\pi a^2J_{0z}^2(0)}{c^2\left(\alpha_J+1\right)}\Bigg[\frac{\left(1-x^2\right)^{\alpha_J+1}}{2\left(\alpha_J+1\right)}{_2}F_1\left(1,\alpha_J+1,\alpha_J+2,1-x^2\right)\\&-\frac{\left(1-x^2\right)^{2\alpha_J+2}}{4\left(\alpha_J+1\right)}{_2}F_1\left(1,2\alpha_J+2,2\alpha_J+3,1-x^2\right)\Bigg],\end{aligned}
} automatically vanishing at the border: more generally, we set $\overline{p}_0=0$ for $x\geq1$. In our case, \al{
p_0=n_0(0)T_0(0)\left(1-x^2\right)^{\alpha_n+\alpha_T},
} with $T_0(0) = T_{0i}(0) + T_{0e}(0)$, which ultimately gives $B_{0z}$ as a function of $x$ by direct solution of \eqref{pdefinition}. In particular, $\overline{p}_0-p_0>0$: as a consequence (and as it can be seen from the first plot of fig.\ref{equil1}), $B_{0z}$ is slightly larger than $B_{0z}(1)$, corresponding to a small paramagnetic effect. In our model also $B_{0z}^2\gg B_{0\theta}^2$ holds, so that $B_{0z}(1) \approx B_0$.

\subsection{Parameters}\label{parameters}
The values chosen for the several parameters are realistic but mostly indicative and don't correspond to any kind of optimization. The major radius of the cylinder is $R=2$ m, the minor radius is $a=0.5$ m, the antenna is located at $d=0.65$ m and the wall at $b=0.8$ m. The antenna frequency is $\omega_0=3.14$ MHz, and the azimuthal and axial mode numbers are $(m,n)=(2,2)$. The remaining parameters are listed below ($B_0$ is the vacuum magnetic field).
\begin{center}
\begin{multicols}{3}
\noindent
$n_0(0)=10^{14}$ cm$^{-3}$\\
$n_\varepsilon=4\cdot10^{-9}$ cm$^{-3}$\\
$\alpha_n=1$\\
$T_{0e}(0)=5$ keV\\
$T_{0i}(0)=1.5$ keV\\
$\alpha_T=2$\\
$J_{0z}(0)=0.7\cdot10^{12}$ statA\\
$\alpha_J=2$\\
$B_0=4\cdot10^4$ G
\end{multicols}
\end{center}
The equilibrium functions and parameters set the resonant layer at about $x_0=0.509$.

\subsection{Numerical solution}\label{numsol2.45simple}

\begin{figure}[h]
	\centering
	\includegraphics[scale=0.52]{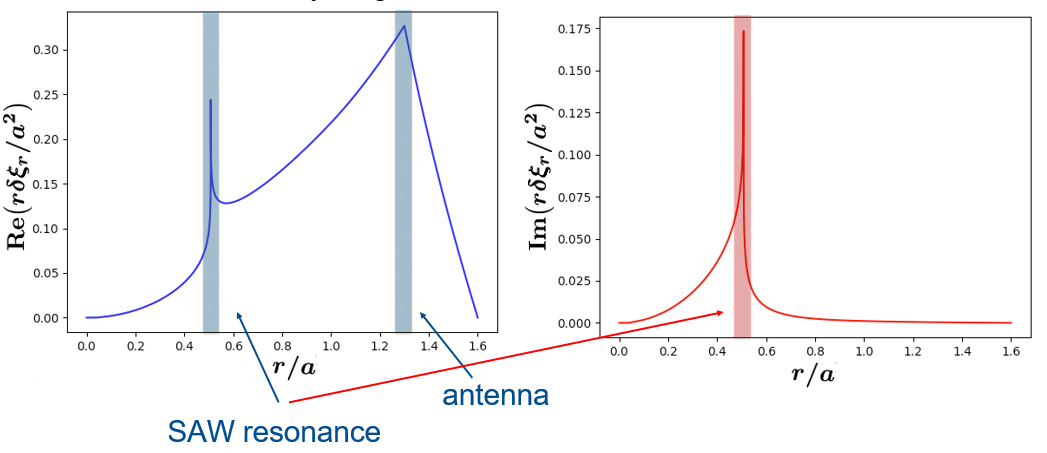}
	\caption{\emph{Fig.\ref{xi}} Radial displacement (connected to the radial variation of the magnetic field), in units of $(-\delta \hat B_{rh}/B_0)$.}
	\label{xi}
\end{figure}

\begin{figure}
	\centering
	\includegraphics[scale=0.52]{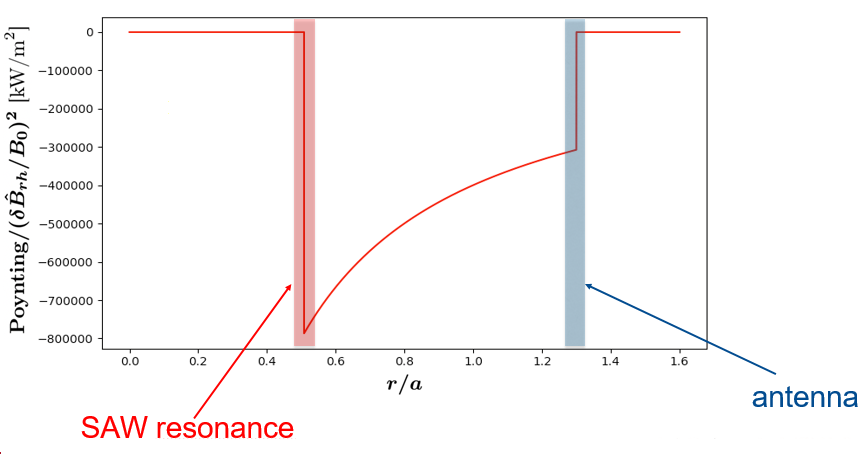}
	\caption{\emph{Fig.\ref{poy}} Poynting flux. The flux directed inside goes, both in vacuum and in the plasma (up to the SAW resonant layer), as $1/r$; it is zero in the region on the left (due to wave absorption) and in the region between the antenna and the external wall (due to wave reflection).}
	\label{poy}
\end{figure}

In the cylindrical plasma equilibrium described so far and illustrated by the plots of fig.\ref{equil1}, with the listed plasma equilibrium parameters, and boundary and regularity conditions on the magnetic axis discussed in subsection \ref{matching}, the problem has been numerically formulated and then solved, for the wave equation in the form \eqref{2.45simple}, by means of a self-designed python script which uses a standard library integrator\footnote{In particular, the solution is obtained by shooting method using the solve$\_$ivp package from the scipy.integrate library based on the 4th-oder Runge-Kutta-Fehlberg method (RK45).}. Fig.\ref{xi} shows the plot of the solution expressed as normalized (to the imposed antenna perturbation $-\delta \hat B_{rh}/B_0$, see \eqref{solmatch}) plasma radial displacement $y:=x\delta\xi_r/a$, as a function of the normalized distance $x:=r/a$ from the axis. The discontinuity is at the resonant surface, where the antenna frequency satisfies the local SAW dispersion relation and Alfvén waves develop a radial (logarithmic) singular structure; at the antenna, the discontinuity is on the first derivative (cusp) and simply corresponds to the crossing of the antenna current sheet, viz. to the Poynting flux launched by the antenna, treated as a point source, at steady state (time asymptotically); what the antenna launches towards the outside is completely reflected at the wall, so the net flux for $d<r<b$ is zero. The flux determines the finite energy absorption rate (fig.\ref{poy}). The MHD solution gives, stopping at the numerical truncation error, $\text{Re}\alpha=-0.65052655$ and $\text{Im}\alpha=-0.00075303$.

\chapter{\textsc{\textsc{kinetic alfvén waves in a cylindrical plasma:\\ linear description}}}\label{kawlinchap}
\vspace{100pt}

The existence of a continuous spectrum with oscillation frequency varying in the non-uniformity direction determines the singularity of shear Alfvén wave radial structure at the resonance surface, which is due to a spatial phase mixing process. Kinetic analysis modifies this picture at scales comparable with the ion Larmor radius (beyond the MHD resolution), and shows that the shear Alfvén wave evolves into the kinetic Alfvén wave (KAW). As a consequence, further investigations demand a gyrokinetic approach.

As a matter of fact, detailed descriptions of shear Alfvén wave resonant absorption involve processes that are not accounted for by the linear MHD description and are both of kinetic and of nonlinear kind: in this chapter, we present the liner gyrokinetic analysis of kinetic Alfvén waves in cylindrical plasmas, while descriptions based on nonlinear gyrokinetics are provided in chapter \ref{nonlinearGK}. For understanding how the radial singular structure is removed, the linear gyrokinetic approach can be adopted for our problem assuming $k_\perp^2\rho_i^2\ll1$ \cite{Hasegawa1975,Hasegawa1976}. This allows us to derive a fourth-order ordinary differential equation, which reduces to the MHD equations for $k_\perp^2\rho_i^2\rightarrow0$ and whose kinetic solutions match the MHD long-wavelength solutions outside the radial singular layer at resonance radius. This analysis will be presented first in the following. Afterwards, and as a comparison for the long-wavelength limit, both WKB and wave packet studies will be presented, valid for arbitrary values of $k_\perp^2 \rho_i^2$. In the strong damping limit, when the kinetic Alfvén wave is completely absorbed before it travels too far away from the shear Alfvén wave resonance/mode conversion layer, earlier results by Hasegawa and Chen \cite{Hasegawa1975,Hasegawa1976} and by Itoh and coworkers \cite{Itoh1982,Itoh1983} are recovered. This is the case that attracted all attention in the past, due to its possible application to wave-induced heating in fusion plasmas.

As novel result of the present work, also the weak damping limit is discussed: despite its irrelevance for plasma heating, this case is of major interest here, since the mode converted kinetic Alfvén wave can give rise to the variety of nonlinear behaviors that will be presented in the next chapter. Interestingly, the present analysis illustrates how the plasma region inside the mode conversion layer acts as a resonant cavity, where the external antenna can excite large amplitude standing waves at given frequencies.

\section{Model and KAW equation}

The model is the one introduced in section \ref{Model}, namely the periodic cylindrical plasma of $2\pi R$ length and radius $a$, axis directed along $\vec{z}$, non-uniformity in the radial direction, and equilibrium magnetic field in the screw pinch configuration $\mathbf{B}_0=(0,B_{0\theta}(r),B_{0z}(r))$. The plasma is surrounded by a conducting wall of radius $b$ and perturbed by an antenna launched wave with the model antenna (see \eqref{jz}) located at a distance $d$.

The essential results of chapter \ref{mhdchapter} are of general validity; however, for the plasma wave equation, we now have to introduce a KAW equation in place of the SAW equation \eqref{2.45simple}. The starting point is the linear gyrokinetic equation \eqref{gyroeqlin}: we have obtained this equation from \eqref{gyroeq} neglecting collisionality as well as magnetic and diamagnetic drift and nonlinear terms. Let's see in more detail, and in cylindrical geometry, if and to what extent the drift terms can actually be neglected in our linear description.

\subsection{Linear gyrokinetic equation and ordering of drift terms}\label{neglectingdrift} 
Let's start from the (ion) gyrokinetic equation with drift contributions \al{
\left(\partial_t+v_\parallel\nabla_\parallel+\mathbf{V}_d\cdot\nabla_\perp\right)\delta g=\frac{eF_M}{T}\DP{\langle\delta L_g\rangle_\alpha}{t}-\frac{c}{B_0}\mathbf{e}_\parallel\times\nabla\langle\delta L_g\rangle\cdot\nabla F_M.\label{gyrodrift}
} From the macroscopic point of view, the drift contributions drop out since, in general, the magnetic drift velocity is of order $\rho_i/R$ with respect to the thermal velocity; however, for the purposes of a more detailed analysis, we must also consider the microscopic fluctuations involved. So, let's concentrate on the third term on the left-hand side: the magnetic drift velocity, at low $\beta$, is given by \al{
\mathbf{V}_d\approx\frac{\mu B_0+v_\parallel^2}{\Omega_i}\mathbf{e}_\parallel\times\boldsymbol{\kappa},\label{vdrift}
} where the curvature $\boldsymbol{\kappa}$ can be expressed as \al{
\boldsymbol{\kappa}=-\frac{B_{0\theta}^2}{B_0^2r}\hat{\mathbf{r}}\label{curvature}
} and, keeping in mind that $B_0\approx B_{0z}$, scales as \al{
\kappa\sim\frac{r}{R^2},
} while $\Omega_i\sim v/\rho_i$, where implicitly $v=v_i$. Because of the vector product in \eqref{vdrift}, $\nabla_\perp$ only enters through the component $k_\theta$, and we conclude that the third term of \eqref{gyrodrift} is of the order \al{
\frac{\rho_ik_\theta vr}{R^2}.
} Now, the second term is of the order $v/R$, and, according to the "optimal" frequency ordering in gyrokinetics, to be investigated more in detail in the following, the same can be assumed for the first term, i.e. $\omega\sim v/R$. So, the order of the drift term with respect to both the first and the second velocity term is \al{
\frac{\rho_ik_\theta r}{R}\ll1: \label{scale1}
} in gyrokinetics, this is typically true since $r\ll R$, a condition the resonant layer $r_0$ generally satisfies, unless it gets very close to the boundary; meanwhile, the result \eqref{scale1} still holds near the boundary in the long-wavelength limit, where $|k_\theta\rho_i|\ll1$ (unless $m$ is very high). Here, it is important to note that these results rely on \eqref{curvature}, which expresses the proportionality relation between $\boldsymbol{\kappa}$ and $\hat{\mathbf{r}}$: in a cylinder the curvature is purely radial, and orthogonal to the flux surface, implying that the radial singularities of the fluctuations due to the mode conversion of SAW to KAW do not enter, ultimately allowing the drift motion to be neglected. Things are different in a toroidal fusion plasma, whose geodesic curvature is different from zero.

While we have made the estimate above based on the typical gyrokinetic ordering for plasma ions, the relative importance of magnetic drift with respect to the parallel free streaming term in \eqref{scale1} holds for electrons as well. In fact, particle mass does not enter explicitly except in the Larmor radius; thus, because of \eqref{scale1}, magnetic drift correction can be neglected a fortiori. What will be different, between the cases of ions and electrons, is the relative ordering of the free streaming term with respect to the explicit time variation. Both terms, however, are kept on the same footing in our analysis, which, therefore, allows us to analyze the effect of (Landau) damping for both electrons and ions, as reported below. 

As far as the diamagnetic term (second term on the right-hand side) is concerned, it scales as \al{
\frac{ck_\theta r}{B_0a^2},
} while the order of first term on the right-hand side, assuming a typical Maxwellian distribution for plasma particles, is \al{
\frac{e}{T}\omega\approx\frac{e}{T}\frac{v_{Ti}}{R\beta_i^{1/2}},
} having denoted by $T$ the thermal energy and estimated $\omega \sim v_A/R \sim \beta_i^{-1/2} v_{Ti}/R$ for typical SAW frequency launched as antenna spectrum. So, dividing by this quantity, we obtain the relative order of the diamagnetic contribution with respect to the first term on the right-hand side, \al{
\frac{TR\beta_i^{1/2}c}{ev_{Ti}}\frac{k_\theta r}{B_0a^2}=\frac{v_{Ti}\beta_i^{1/2}}{\Omega_i}\frac{k_\theta rR}{B_0a^2}=\rho_ik_\theta\frac{rR\beta_i^{1/2}}{a^2};\label{scale2}
} according to the realistic parameters chosen for our model, $\beta_i \sim 10^{-2}$, $r/a\sim 1/60$ and $R/a\sim 3$, so $rR/a^2\sim1/20<\frac{a^2}{R^2}\sim1/10$. Thus, similar to the analysis of the relevance of the magnetic drift term, we conclude that diamagnetic corrections can be neglected for investigating the mode converted KAW excited by the antenna unless the $|k_\theta \rho_i|\ll1$ condition is violated. In chapter \ref{nonlinearGK}, when investigating the nonlinear physics induced by KAW launched by the external antenna, we will see that this condition can indeed be violated by the nonlinearly generated spectrum, and that diamagnetic effects can have a crucially important role in the dynamics. Because diamagnetic effects are typically $\sim (R^2/a^2)\beta_i^{1/2}$ larger than magnetic drift effects, as evident by inspection from a comparison of \eqref{scale1} and \eqref{scale2}, only the former will be kept into account (see chapter \ref{nonlinearGK}).

\subsection{Vorticity equation}

From the gyrokinetic equation \eqref{gyroeqlin}, the vorticity equation \eqref{vorticity} follows, which we write in the form \al{
\nabla\cdot\left[\frac{\omega_{k}^2}{v_A^2}\frac{(1-\Gamma_k)}{b_k}\nabla_\perp\delta\phi_{k}-k_\parallel^2\nabla_\perp\delta\psi_{k}\right]+\left(\frac{1}{r}\partial_rk_\parallel^2\right)\delta\phi_{k}=0.\label{vort}
} This equation is linear in general, and has been written in such a way that the long wavelength MHD limit of \eqref{2.45simple} is readily recovered. At the same time, the WKB expression of \eqref{vorticity} is also reproduced for $b_k \approx 1$. Let us first approach \eqref{vort} it in the limit $k_\perp^2\rho_i^2\ll1$; in this limit \al{
\Gamma_k=I_0e^{-b_k}\approx1-b_k+\frac{3}{4}b_k^2
} and so the quasineutrality equation \eqref{quasi} becomes \al{
\delta\phi_{k}=\left(1-\frac{\tau(1-\Gamma_k)}{\epsilon_{s{k}}}\right)\delta\psi_{k}\approx\left(1+\frac{\tau b_k}{\epsilon_{s{k}}}\right)\delta\psi_{k},\label{deltaphideltapsi}
} transforming \eqref{vort} in \al{
\nabla\cdot\left[\frac{\omega_{k}^2}{v_A^2}\left(1-\frac{3}{4}b_k\right)-k_\parallel^2\left(1+\frac{\tau b_k}{\epsilon_{s{k}}}\right)\right]\nabla_\perp\delta\phi_{k}+\left(\frac{1}{r}\DP{}{r}k_\parallel^2\right)\delta\phi_{k}=0,\label{vortlin}
} or, also, keeping in mind that $b_k:=\rho_i^2k_\perp^2$ and further defining \al{
f_{k}(r):=\rho_i^2\left(\frac{3\omega_{k}^2}{4v_A^2(r)}+\frac{\tau k_\parallel^2(r)}{\epsilon_{s{k}}}\right),
} in \al{
\nabla\cdot\left(D_A-f_{k}k_\perp^2\right)\nabla_\perp\delta\phi_{k}+\left(\frac{1}{r}\DP{}{r}k_\parallel^2\right)\delta\phi_{k}=0.\label{vortlin2}
} Thus, a fourth-order, ordinary linear differential equation, which describes our problem including kinetic effects, is obtained for $b_k\ll1$, as well as neglecting diamagnetic and magnetic drift effects. The dielectric constant $\epsilon_{s{k}}$ is responsible for wave absorption and becomes $\epsilon_{s{k}}=1$ in the ideal no-absorption (no-damping) limit. As in chapter \ref{mhdchapter}, we can identify $\delta\phi$ with $r\delta\xi_r$. In the MHD limit $b_k\rightarrow0$ and the quasineutrality equation \eqref{deltaphideltapsi} gives again the result $\delta\phi=\delta\psi$ we already derived from the assumption $\delta E_\parallel=0$ in section \ref{kawsection}; conversely, finite $\delta E_\parallel$ is connected with finite ion Larmor radius. Also, combining \eqref{deltaEpar} and \eqref{deltaphideltapsi} one sees that $\delta E_\parallel$ and $\delta\phi$ are in quadrature unless complex behavior is brought about by the plasma dispersion function $Z$ (on which $\epsilon_{s{k}}$ depends). With $b_k=0$, \eqref{vortlin} reduces to the 2nd-order equation \al{
\nabla\cdot\left(D_A\nabla_\perp\delta\phi_{k}\right)+\frac{\partial_rk_\parallel^2}{r}\delta\phi_{k}=0,
} which exactly is the MHD master equation \eqref{2.45simple}. Taking into account that \al{
\nabla_\perp=\hat{\mathbf{r}} \frac{\partial}{\partial r}-\text{i}\hat{\boldsymbol{\theta}} \frac{m}{r},
} as well as that \al{
k_\perp^2(r)=-\frac{1}{r}\DP{}{r}\left(r\DP{}{r}\right)+\frac{m^2}{r^2},
} equation \eqref{vortlin2} explicitly reads \al{\begin{aligned}
&f_{k}\delta\phi_{k}''''+\left(\frac{2f_{k}}{r}+\partial_rf_{k}\right)\delta\phi_{k}'''+\left(D_{A{k}}+\frac{\partial_rf_{k}}{r}-\frac{2\left(m^2+1\right)f_{k}}{r^2}\right)\delta\phi_{k}''\\
&+\left(\frac{\partial_r\left(rD_{A{k}}\right)}{r}-\frac{m^2\partial_rf_{{k}}}{r^2}+\frac{\left(6m^2+1\right)f_{k}}{r^3}\right)\delta\phi_{k}'\\
&+\left(\frac{\partial_rk_\parallel^2}{r}-\frac{m^2D_{A{k}}}{r^2}+\frac{\left(m^4-5m^2\right)f_{{k}}}{r^4}\right)\delta\phi_{k}=0.
\end{aligned}\label{master}
}

\subsection{Reduced equation}

At $r=r_0$, $D_A$ vanishes, so, approaching the resonance radius, we can write \al{
\frac{\omega_{k}^2}{v_A^2}-k_\parallel^2=-a_1\left(r-r_0\right),
} where the minus sign is due to the fact that $\omega_{k}^2>k_\parallel^2v_A^2$ for $r<r_0$; also, with $a_2:=\omega_{k}^2/v_A^2\approx k_\parallel^2$, we obtain that $f_{k}b_k\approx a_2\hat{\rho}^2k_\perp^2$ and that the variation of $k_\parallel^2$ is negligible in a narrow layer around $r=r_0$. The dielectric constant in $\hat{\rho}^2$ can be approximated by $\epsilon_{sk}\approx1+\tau\left(1-\Gamma_k\right)+\text{i}\sqrt{\pi}\xi_{ke}$, where we used $|\xi_{ki}|=|\omega_{k}/(k_\parallel v_{Ti})|\gg1$ to neglect exponentially small ion Landau damping, and $|\xi_{ke}|=|\omega_k/(|k_\parallel| v_{Te})| \ll 1$ to keep into account algebraically small but finite electron Landau damping \cite{Hasegawa1975,Hasegawa1976,Itoh1982,Itoh1983}. We will see this point more in detail in the wave-packet analysis of KAW. For the time being, we simply note that the corresponding expression of $\hat \rho^2$ can be used to describe the strong (damping) absorption of KAWs, which is accounted for by its small but finite negative imaginary part. Meanwhile, in the weak-damping/long-wavelength limit, $\epsilon_{sk}$ can be further simplified to \al{
\epsilon_{sk}\approx1.\label{eskeq1}
} So, for $r-r_0\ll r_0$ and with the definition $\hat{a}:=a_1/a_2>0$, \eqref{vortlin} reduces to \al{
\DP{}{r}\left[\hat{\rho}^2\DP{^2}{r^2}-\hat{a}\left(r-r_0\right)\right]\DP{}{r}\delta\phi_{k}=0.\label{redeq}
} Noting \eqref{xirpot} with $B_0$ taken as constant, and \eqref{solmatch} to connect physical scalar fields with dimensionless quantities, we can switch from $\delta\phi$ to $y$ and, defining $z:=\partial_x y$ and introducing an integration constant $C$, we can also rewrite the above equation as the Airy equation with a non-homogeneous term \al{
\left[\hat{\rho}^2\DP{^2}{r^2}-\hat{a}\left(r-r_0\right)\right]z=-\hat{a}C.\label{redeqz}
} We can scale \eqref{redeqz} by introducing $s:=(r-r_0)/(\hat{\rho}^2/\hat{a})^{1/3}$ and $K:=-C/(\hat{\rho}^2/\hat{a})^{1/3}$ and finally obtain \al{
\left(\DP{^2}{s^2}-s\right)z=K.\label{eqz}
} The $s$ variable is adimensional, and linked with the adimensional variable $x$ through \al{
s=\left(\frac{\overline{\rho}^2}{\overline{a}}\right)^{-1/3}\left(x-x_0\right),\label{zandx}
} being $\overline{\rho}:=\hat{\rho}/a$ and $\overline{a}:=\hat{a}a$. An equation of the same kind as \eqref{eqz} is satisfied, near the resonance, by the azimuthal magnetic field: in fact, in this case the spatio-temporal evolution of KAWs is regulated by the analog of \eqref{deltab}, i.e., \al{
\left(\frac{\hat{\rho}^2}{r}\frac{\partial}{\partial r}r\frac{\partial}{\partial r}-1-\frac{1}{\omega_A^2(r)}\frac{\partial^2}{\partial t^2}\right)\delta B_\theta(r,t)=0,\label{waveeqkinetic}
} which in the steady state also gives, using Laplace transform with $s=-\text{i}\omega$, the equation \al{
\left(\frac{\hat{\rho}^2}{r}\frac{\partial}{\partial r}r\frac{\partial}{\partial r}-1+\frac{\omega^2}{\omega_A^2(r)}\right)\delta B_\theta(r)=\delta B_{\theta0}.\label{37}
} Near $r_0$, we are allowed to perform the series expansion $\omega_A^2\approx\omega_0^2+(\omega_A^2)'(r-r_0)\sim\omega_0^2-(\omega_0^2/L_A)(r-r_0)$, being $L_A$ the scale length of $\omega_A$, and, considering that $r-r_0\ll L_A$, to transform this equation into \al{ 
\left(\frac{\hat{\rho}^2}{r}\frac{\partial}{\partial r}r\frac{\partial}{\partial r}+\frac{r-r_0}{L_A}\right)\delta B_\theta(r)=\delta B_{\theta0}.
} Now, we can scale this equation by introducing $s':=-(r-r_0)/(\hat{\rho}^2L_A)^{1/3}$ and $K':=L_A\delta B_{\theta 0}/(\hat{\rho}^2L_A)^{1/3}$ so to finally obtain \cite{Hasegawa1975,Hasegawa1976,Zoncarev} the Airy equation \al{
\left(\DP{^2}{s'^2}-s'\right)\delta B_\theta=K'.
} We summarize in the following the strategy of analytic solution for equation \eqref{eqz}, also taking into account \cite{Zoncarev,Hasegawa1975,Hasegawa1976}. First of all, we know that the general solution can be expressed as a linear combination of Airy's functions $\text{Ai}(s)$ and $\text{Bi}(s)$ (which solves the associated homogeneous equation) plus a particular solution, which is to be found in the form \al{
z=c_1(s)\text{Ai}(s)+c_2(s)\text{Bi}(s);\label{zsol}
} applying the method of variation of arbitrary constants, the conditions we must impose are \al{\begin{aligned}
c_1'\text{Ai}+c_2'\text{Bi}&=0,\\
c_1'\text{Ai}'+c_2'\text{Bi}'&=K,\end{aligned}
} and, given the wronskian $W=\text{AiBi}'-\text{Ai}'\text{Bi}=1/\pi$, yield \al{\begin{aligned}
c_1'=-K\frac{\text{Bi}}{W}&=-\pi K\text{Bi},\\
c_2'= K\frac{\text{Ai}}{W}&=\pi K\text{Ai},\end{aligned}
} that is, \al{
c_1&=\pi K\left(d-\int_0^s\text{Bi}(s')\,\text{d}s'\right),\\
c_2&=\pi K\left(\int_0^s\text{Ai}(s')\,\text{d}s'-\frac{1}{3}\right).\label{c2}
} The choice of the integration constants depends on the boundary conditions one has to satisfy: in our case, \eqref{c2} has been adjusted such that $c_2\rightarrow0$ as $s\rightarrow\infty$ (keep in mind that $\int_0^\infty\text{Ai}(s)=1/3$), in other words, such to asymptotically eliminate the divergent $\propto\text{Bi}$ part in the solution \eqref{zsol}, which is thus allowed to correctly match the MHD one (for his part, $\text{Ai}$ is exponentially infinitesimal). The constant $K$ is derived from $C$ and so it is known, as a function of $\alpha$, from the external MHD solution, according to \eqref{MHDcond30}. Defining Scorer's function $\text{Gi}(s)$ as \al{
\text{Gi}(s):=\text{Bi}(s)\int_s^\infty\text{Ai}(s')\,\text{d}s'+\text{Ai}(s)\int_0^s \text{Bi}(s')\,\text{d}s',\label{Scorer}
} the general (homogeneous plus particular) solution to \eqref{eqz} is \al{
z=K\pi\left(d\text{Ai}(s)-\text{Gi}(s)\right).\label{soleqz}
} Due to \eqref{y1y2res}, the derivative of the MHD solution near the resonance goes as $-\hat{C}/x$, so it remains to verify that $z\sim-K/s$ for $s\gg1$. In the region $s>0$, the asymptotic behaviors of Airy's functions for $s\gg1$ are \al{
\text{Ai}(s)&\sim\frac{1}{2\sqrt{\pi}}s^{-1/4}e^{-\frac{2}{3}s^{\frac{3}{2}}},\\
\int_\infty^s \text{Ai}(s')\,\text{d}s'&\approx\frac{1}{2\sqrt{\pi}}s^{-3/4}e^{-\frac{2}{3}s^{\frac{3}{2}}},\\
\text{Bi}(s)&\sim\frac{1}{\sqrt{\pi}}s^{-1/4}e^{\frac{2}{3}s^\frac{3}{2}},\\
\int_0^s \text{Bi}(s')\,\text{d}s'&\approx\frac{1}{\sqrt{\pi}}s^{-\frac{3}{4}}e^{\frac{2}{3}s^{\frac{3}{2}}},
} from which, looking at the definition \eqref{Scorer}, the asymptotic behavior of Scorer's function is \al{
\text{Gi}(s)\approx\frac{1}{\pi s};\label{Gasym}
} furthermore, the above expansions also show that the asymptotic contribution of $\text{Ai}(s)$ is exponentially small with respect to the contribution of Scorer's function, so, for large $s$, \eqref{soleqz} behaves as \al{
z\approx-\frac{K}{s}=-\frac{\hat{C}}{x},
} representing the correct boundary condition matching the MHD solution. In conclusion, for $r\ll r_0$ the solutions are short-wavelength functions which grow or decay exponentially in space because the KAW is "cut-off" (not propagating) for $\omega_k^2 < k_\parallel^2 v_A^2$. Due to Landau damping, the exponentially growing mode (in space) corresponds to incoming short wavelength, while the decaying mode is outgoing short wavelength, consistently with causality. In fact, for $r>r_0$ we recover the long-wavelength solution connecting to MHD: no short-wavelength mode is entering the SAW resonance layer from the outer region since no such mode is generated by the antenna, and, at the same time, no short-wavelength mode can propagate from within the plasma region surrounded by the resonance layer. This confirms the fact that at $r=r_0$ there is mode conversion of the long wavelength fluctuation generated by the external antenna.

Having clarified how the MHD-like wave (solution) passes through the SAW resonance and undergoes mode conversion, there remains to extend this solution to the whole region $s<0$, where both $\text{Gi}$ and $\text{Ai}$ are oscillatory, up to the magnetic axis, where a regularity conditions must be applied. We shall examine the solution in this region for two limiting cases: strong and no (or weak) damping.

\subsection{Strong damping limit}\label{strongdamping}

The strong damping limit means that the propagating solution inside $r = r_0$ $(s<0)$ is entirely absorbed (decays exponentially because of Landau damping) without reaching the magnetic axis: \al{
y(s_{-\infty})=y(s\gg1)+\int_{s\gg1}^{s\ll-1} z(s')\,\text{d}s'=0.\label{strongcondition1}
} Here, $y(s\gg1)=D-\hat{C}\ln x(s)$ is the MHD solution approaching the resonance, and $z$ for $s\ll-1$ represents a wave traveling towards the magnetic axis, since the KAW is completely absorbed by the plasma by electron Landau damping before it reaches $x=0$ and is reflected back. As anticipated in the discussion above \eqref{eskeq1}, the absorption is due to the small but finite negative imaginary part of $\hat \rho^2$, which, for real $x < x_0$, results in a small but finite negative imaginary part of $s$ for $s<0$. Using the relation between $x$ and $s$ given by \eqref{zandx}, condition \eqref{strongcondition1} reads \al{
0=D+\frac{\hat{C}\ln\left(\overline{a}/\overline{\rho}^2\right)}{3}-\hat{C}\ln s+\left(\frac{\overline{a}}{\overline{\rho}^2}\right)^{-1/3}\int_{s\gg1}^0z(s')\text{d}s'+\left(\frac{\overline{a}}{\overline{\rho}^2}\right)^{-1/3}\int_0^{-\infty}z(s)\text{d}s.
} Since the asymptotic form in the region $s\ll-1$ of the Airy functions is \al{
\text{Ai}(s)\approx\frac{\sin\left(\frac{2}{3}\left(-s\right)^{\frac{3}{2}}+\frac{\pi}{4}\right)}{\sqrt{\pi}\left(-s\right)^{\frac{1}{4}}},\qquad \text{Bi}(s)\approx\frac{\cos\left(\frac{2}{3}\left(-s\right)^{\frac{3}{2}}+\frac{\pi}{4}\right)}{\sqrt{\pi}\left(-s\right)^{\frac{1}{4}}},
} and since \al{
\int_{-\infty}^{+\infty}\text{Ai}(s)\text{d}s=1\qquad\text{and}\qquad\int_0^{-\infty}\text{Bi}(s)\text{d}s=0,
} we see that $\text{Gi}(s)\approx\text{Bi}(s)$ for $s\rightarrow-\infty$: this suggests $d=-\text{i}$ in \eqref{soleqz} as the right choice to have an asymptotic $z$ of the type \al{
z\approx-\frac{k\sqrt{\pi}}{\left(-s\right)^{1/4}}e^{\text{i}\left[\frac{2}{3}\left(-s\right)^{3/2}+\frac{\pi}{4}\right]},
} which represents a wave traveling towards the magnetic axis. From this expression, it is also clear that, with $s$ characterized by a small but finite negative imaginary part for $s<0$ as anticipated above, the value of $z$ will be exponentially decaying while traveling inward. Using the definition of $K$, we have \al{
0&=D+\hat{C}\left(\frac{\ln\left(\overline{a}/\overline{\rho}^2\right)}{3}-\ln s+\pi\int_{s\gg1}^0\left(d\text{Ai}(s)-\text{Gi}(s')\right)\text{d}s'\right)\nonumber\\&+\pi\hat{C}\int_0^{-\infty}(d\text{Ai}(s)-\text{Gi}(s))\text{d}s,
} or, for convenience, \al{
0&=D+\frac{\hat{C}\ln\left(\overline{a}/\overline{\rho}^2\right)}{3}+\pi\hat{C}\Bigg[\text{i}\int_{-\infty}^{+\infty}\text{Ai}(s)\text{d}s+\int_{-\infty}^0\text{Gi}(s)\text{d}s\nonumber\\&+\lim_{s\rightarrow+\infty}\left(\int_0^s\text{Gi}(s')\text{d}s'-\frac{1}{\pi}\ln s\right) \Bigg]. \label{sd1}
} The first integral is 1, while the second integral can be rewritten and numerically computed as \al{
\int_{-\infty}^0\text{Gi}(s)\text{d}s=\int_{-\infty}^0\text{Bi}(s)\text{d}s-\int_{-\infty}^0\text{Hi}(s)\text{d}s\approx-1.36554-1.12654.\label{Ginumest}
} Finally, the last term can be numerically estimated as $\approx0.239$ (note that the two functions in the argument are both divergent when taken individually). As a consequence, \eqref{sd1} reads \al{
0&=D+\frac{\hat{C}\ln\left(\overline{a}/\overline{\rho}^2\right)}{3}+\pi\hat{C}\left(\text{i}+\Lambda\right),\label{alpharelation}
} where $\Lambda\approx-2.25308$. This is a relation among constants ultimately giving a value for $\alpha$, since both $\hat{C}$ and $D$ have been defined as known linear combinations of $\alpha$ once the solution for $r> r_0$ is scaled by $(- \delta \hat B_{rh}/B_0)$ and boundary conditions are applied on the conducting wall. In conclusion, the case of strong damping or total absorption implies two conditions: only inward propagating wave for $r< r_0$ and boundary conditions at the conducting wall at $r = b$. These completely determine the 2nd-order problem, giving the two constants $d$ and $\alpha$ a value. In section \ref{kinlinnumsol}, we will show that the value of $\alpha$ determined in this way is very close to that determined within the ideal MHD model, consistent with the original findings of \cite{Hasegawa1976}, demonstrating that the plasma impedance is essentially the same in the MHD and kinetic analysis. This is a consequence of the fact that, in the strong damping case, the inward propagating mode converted KAW is fully absorbed by electron Landau damping after propagating over a few wavelengths; that is, locally.

\subsection{No-damping limit}

A no- (or weak-)damping limit, as opposite to the strong damping limit studied so far, is obtained for $\epsilon_{s{k}}\approx 1$, which is the consequence of $\xi_{ke}\ll1$ (algebraically small electron-Landau damping) and $\xi_{ki}\gg1$ (exponentially small ion-Landau damping). In this case, the wave can propagate further inward and the local expansion for analyzing the KAW propagation near the SAW resonance, adopted in \eqref{redeqz} and \eqref{eqz}, is no longer valid. More generally, the wave is allowed to reach the magnetic axis and to be possibly reflected. Nonetheless, the general solution to \eqref{eqz} is still needed because it gives the correct boundary conditions at the resonance layer $s=0$ and matches the external MHD antenna solution. Denoting with $s_\infty$ a $s\gg1$ where MHD holds, the solution around the resonance, namely the solution to \eqref{eqz}, is, in general, \al{
y(s)&=y(s_\infty)+\left(\frac{\overline{a}}{\overline{\rho}^2}\right)^{-1/3}\int_{s_\infty}^s z(s')\,\text{d}s',
} or, more explicitly, \al{\begin{aligned}
y(s)&=D+\frac{\hat{C}\ln(\overline{a}/\overline{\rho}^2)}{3}+\pi\hat{C}\Bigg[d\int_{+\infty}^s\text{Ai}(s')\text{d}s'+\int_s^0\text{Gi}(s')\text{d}s'\\&+\lim_{s_\infty\rightarrow+\infty}\left(\int_0^{s_\infty}\text{Gi}(s')\text{d}s'-\frac{1}{\pi}\ln s_\infty\right) \Bigg]=\\&=D+\frac{\hat{C}\ln(\overline{a}/\overline{\rho}^2)}{3}+\pi\hat{C}\Bigg[d\int_{+\infty}^s\text{Ai}(s')\text{d}s'+\int_s^0\text{Gi}(s')\text{d}s'+0.239\Bigg],\end{aligned}\label{reduced}
} from which \al{
y(0)=D+\frac{\hat{C}\ln(\overline{a}/\overline{\rho}^2)}{3}+\pi\hat{C}\left(0.239-\frac{d}{3}\right),\label{bond1}\\
y'(0)=z(0)=\pi\hat{C}\left(\frac{\overline{a}}{\overline{\rho}^2}\right)^{1/3}\left(d\text{Ai}(0)-\text{Gi}(0)\right),\label{bond2}
} and so for the second and third derivative, \al{
y''(0)=\pi\hat{C}\left(\frac{\overline{a}}{\overline{\rho}^2}\right)^{2/3}\left(d\text{Ai}'(0)-\text{Gi}'(0)\right),\label{bond3}\\
y'''(0)=\pi\hat{C}\left(\frac{\overline{a}}{\overline{\rho}^2}\right)\left(d\text{Ai}''(0)-\text{Gi}''(0)\right)=\pi\hat{C}\left(\frac{\overline{a}}{\overline{\rho}^2}\right).\label{bond4}
} Thus, we have the four boundary conditions needed by the fourth-order equation \eqref{master}, whose solution can be obtained numerically between the magnetic axis and the resonant layer; these conditions are functions of $d$ and $\alpha$, which are given by the two axial regularity conditions \al{
y(r=0^+)=y'(r=0^+)=0\label{regularity}
} on the solution to \eqref{master}.

Actually, no-damping is an ideal limit case of weak damping, for which, in general, nonlinearities are expected to play an important role, due to the nature of the interesting phenomena connected to weak absorption, so the general (nonlinear) gyrokinetic formulation is needed (chapter \ref{nonlinearGK}).

\subsection{Weak damping: wave-packet analysis}\label{wavepacketanalysis}
For our purposes, the most interesting case is weak damping, for which, in the previous subsection, we discussed the solution of the vorticity equation reduced to a 4th-order ordinary differential equation in the long-wavelength limit. Let us now return to \eqref{vort} without the assumption $b_k\ll 1$, and with the dielectric constant \eqref{dielectricconstant} approximated as \al{
\epsilon_{sk}\approx1+\tau\left(1-\Gamma_k\right)+\text{i}\sqrt{\pi}\xi_{ek},
} where we have used $\xi_{ik} Z(\xi_{ik}) \approx -1$ for $|\xi_{ik}| \gg 1$ and, for the plasma dispersion function $Z(\xi_{ek})$ introduced by \eqref{plasmadispersionf}, we have adopted the power series expansion for $|\xi_{ek}|\ll 1$ \cite{Chenint} \al{
Z(\xi_{ek})=\text{i}\sqrt{\pi}e^{-\xi_{ek}^2}-2\xi_{ek}\left(1-\frac{2}{3}\xi_{ek}^2+...\right)\approx\text{i}\sqrt{\pi}.
} By direct substitution into \eqref{deltaphideltapsi}, the relation between the perturbations $\delta\psi_{k}$ and $\delta\phi_{k}$ is \al{
\delta\psi_{k}\approx\left[1+\tau\left(1-\Gamma_k\right)\left(1-\text{i}\sqrt{\pi}\xi_{ek}\right)\right]\delta\phi_{k}.
} So, \eqref{deltaphideltapsi0} implies (also see \cite{Hasegawa1976}) the dispersion relation \al{
\frac{\omega_{{k}}^2}{k_\parallel^2v_A^2}=\frac{b_k}{1-\Gamma_k}+b_k\tau\left(1-\text{i}\sqrt{\pi}\xi_{ek}\right),\label{dispersionrelwp}
} and, in itself, reads \al{
\left(D_R^0+\text{i}D_A^1\right)\delta\phi_{k}=0,
} being \begin{gather}
D_R^0:=\frac{\omega_{{k}}^2}{k_\parallel^2v_A^2}-\frac{b_k}{1-\Gamma_k}-b_k\tau,\\
D_A^1:=b_k\tau\sqrt{\pi}\xi_{ek};\label{DAsch}\end{gather}
meanwhile, \eqref{vort} can be cast as \al{
\left(D_R+\text{i}D_A^1\right)\delta\phi_{k}=0,\label{vortD}
} with \al{
D_R:=D_R^0+D_R^1,\quad D_R^1:=-\frac{\rho_i^2}{r\left(1-\Gamma_k\right)}\frac{\partial_rk_\parallel^2}{k_\parallel^2}.
} At the leading order, $D_R=0$ and \al{
k_\perp^2\rho_i^2=k_\perp(\omega_{k},x),\label{vortDleading}
} whereas, at the next order, and for a wave packet \al{
\delta\phi_{k}\approx e^{-\text{i}\omega_{{k}} t+\text{i}\int_{r_0}^rk_r\text{d}r'}A(r,t),
} \eqref{vortD} yields \cite{Chen2016,Zonca2014,Zonca2015b,Zonca2021} \al{
\DP{}{t}\left(\DP{D_R}{\omega_{{k}}}A^2\right)-\DP{}{r}\left(\DP{D_R}{k_r}A^2\right)+2D_A^1A^2+\text{i}A\DP{^2D_R}{k_r^2}\DP{^2}{r^2}A=0.\label{Schrodingerlike}
} This equation can be solved as a fixed boundary eigenvalue problem, that is for $\partial_t=0$: the corresponding solution or eigenvalue provides the radial mode envelope $A(r)$. In the next chapter, this equation is extended to non-linear interactions.

Before providing the numerical solution of \eqref{Schrodingerlike}, we can find the approximate dispersion relation generated by \eqref{Schrodingerlike} by WKB analysis. Neglecting the second-order derivative in $r$, only important near turning points, the solution is \al{
A\approx\frac{\text{constant}}{\sqrt{\partial_{k_r}D_R}}.\label{A}
} Furthermore, near magnetic axis, \al{
k_\perp^2=k_{\perp0}^2-gr,
} with $g$ a constant that can be determined by solving \eqref{vortDleading} numerically. As a consequence, the equation for $\delta\phi_{k}$ as $r\rightarrow0$ is \al{
\left[\frac{1}{r}\DP{}{r}\left(r\DP{}{r}\right)-\frac{m^2}{r^2}\right]\delta\phi_{k}=-k_{\perp0}^2\delta\phi_{k}.
} Imposing regularity (no divergence) at $r=0$, this equation has the solution \al{
\delta\phi_{k}=A_0J_{|m|}(k_{\perp0}r)\label{phiA0}
} which, for $k_{\perp0}r\gg1$, gives \al{
\delta\phi_{k}\sim\sqrt{\frac{2}{\pi k_{\perp0}r}}A_0\cos\left(k_{\perp0}r-\frac{|m|\pi}{2}-\frac{\pi}{4}\right)\label{deltaphibessel}
} and \al{
\DP{}{x}\delta\phi_{k}\sim-\sqrt{\frac{2k_{\perp0}a}{\pi x}}A_0\sin\left(k_{\perp0}r-\frac{|m|\pi}{2}-\frac{\pi}{4}\right).\label{wpdeltaphi}
} At the resonance layer $x=x_0$, this solution must match \eqref{reduced} and its derivative $z$. Neglecting $D_R^1$ for simplicity, we note that \begin{gather}
\DP{D_R}{k_r}\approx-2k_ra^2\overline{\rho}^2=-2k_ra^2\left(\frac{3}{4}+\tau\left(1-\text{i}\sqrt{\pi}\xi_{ek}\right)\right)\frac{\rho_i^2}{a^2},\\
k_\perp^2a^2\overline{\rho}^2\approx k_r^2a^2\overline{\rho}^2\approx-\overline{a}\left(x-x_0\right),
\end{gather}
from which \al{
\DP{D_R}{k_r}=-2a\left(\overline{a}\overline{\rho}^4\right)^{1/3}\left(-s\right)^{1/2};
} hence, the WKB solution matching the mode converted KAW at resonance is given by \al{\begin{aligned}
\D{y}{x}=\frac{\sqrt{2\pi\left(\overline{a}\overline{\rho}^4\right)^{1/3}}K}{\sqrt{-\partial_{k_r}D_R/a}}\Bigg[&\frac{d}{2\text{i}}\left(e^{\frac{\text{i}\pi}{4}-\text{i}\int_{r_0}^rk_r\text{d}r'}-e^{-\frac{\text{i}\pi}{4}+\text{i}\int_{r_0}^rk_r\text{d}r'}\right)\\&-\frac{1}{2}\left(e^{\frac{\text{i}\pi}{4}-\text{i}\int_{r_0}^rk_r\text{d}r'}+e^{-\frac{\text{i}\pi}{4}+\text{i}\int_{r_0}^rk_r\text{d}r'}\right)\Bigg],\end{aligned}\label{DyDxmatch}
} where, in the linear limit, the constant $K$ controls the (arbitrary) amplitude of the solution, which, in turn, is controlled by the amplitude of the antenna solution. Matching between this equation and \eqref{wpdeltaphi} yields \al{\begin{aligned}
\lambda\left(d-\text{i}\right)&=e^{\text{i}\left(\frac{|m|\pi}{2}-\Phi\right)},\\
\lambda\left(d+\text{i}\right)&=e^{-\text{i}\left(\frac{|m|\pi}{2}-\Phi\right)},\end{aligned}\label{lambdas}
} where \al{\begin{gathered}
\lambda:=\frac{\pi K\left(\overline{a}\overline{\rho}^4\right)^{1/6}(\overline{r}_0/a)^{1/2}}{A_0\sqrt{-k_{\perp0}\partial_{k_r}D_R}}\vert_{x\rightarrow\overline{r}_0/a},\\
\Phi:=\int_{\overline{r}_0}^{r_0}k_r\,\text{d}r+k_{\overline{r}_0}\overline{r}_0=\Phi_R+\text{i}\Phi_I,
\end{gathered}\label{philambda}
} being $\overline{r}_0\ll r_0$ and denoting the maximum value for $k_r$. In fact, \al{
k_r^2=\frac{b_k}{\rho_i^2}-\frac{m^2}{r^2};
} thus, $k_r$ increases for decreasing $r$, due to the behavior of $b_k(r)$, and, after reaching its maximum near the axis, it sharply decreases becoming very large, and complex, approaching $r\rightarrow0$. This embeds the expected Bessel function behavior for $r\rightarrow0$, represented by \eqref{deltaphibessel}. To avoid needless complications with WKB representation, it is reasonable to interpret \eqref{deltaphibessel} with the local $k_r$ near $\overline{r}_0$ where the matching between \eqref{deltaphibessel} and \eqref{DyDxmatch} is made: this naturally yields the expression for $\Phi$ as in \eqref{philambda}. Also, note that $\lambda\rightarrow0$ and $\Phi_I>0$ as a consequence of finite dissipation (due to electron Landau damping). Because of \eqref{lambdas} and \eqref{philambda}, and, in particular, of the fact that the second of \eqref{lambdas} is exponentially small, the strong damping limit is recovered for $d=-\text{i}$, as expected \cite{Hasegawa1976}. Equations \eqref{lambdas} and \eqref{philambda} give \al{\begin{gathered}
d=-\text{i}\frac{e^{-2\text{i}\Phi_R}+(-1)^{|m|}e^{-2\Phi_I}}{e^{-2\text{i}\Phi_R}-(-1)^{|m|}e^{-2\Phi_I}},\\
\lambda=\frac{\text{i}^{|m|+1}e^{\text{i}\phi_R+\Phi_I}}{2}\left(e^{-2\text{i}\Phi_R}-(-1)^{|m|}e^{-2\Phi_I}\right)\end{gathered}
} and yield the strong damping limit for $d=-\text{i}$ and $\lambda\gg1$ ($|A_0|\ll|K|$, fluctuation on axis exponentially small).

In summary, solving \eqref{Schrodingerlike} by WKB yields $A_0$ (fluctuation strength on axis) by means of $\lambda$, and $d$, that is the mix of in- and out- going KAW inside the resonance radius. The same can be solved more exactly by means of \eqref{Schrodingerlike} itself. The wave-packet KAW structure, similarly, can be compared with the WKB envelope, given by \eqref{A}, \eqref{phiA0}, \eqref{wpdeltaphi} and \eqref{DyDxmatch} in the different regions.

The value of $d$ being determined, one can calculate $\alpha$ from the same equation as in the strong damping case. To see this more clearly, let us consider again \eqref{reduced} and note that, for $s$ large and positive, \al{
y\approx -\hat{C}\ln s+\frac{\hat{C}}{3}\ln\left(\frac{\overline{a}}{\overline{\rho}^2}\right)+D
} must match \al{
y=y_{-\infty}+\pi\hat{C}\left(d-\int_{-\infty}^s\text{Gi}(s')\,\text{d}s'\right),
} where $y_{-\infty}$ is the value of the solution propagating from the magnetic axis toward the resonance. Typically, such value is negligible, while the corresponding derivatives are not: thus, it can be neglected in the WKB calculation of $\alpha$, which reflects only the value of $d$. The matching with outer region, that is with the ideal MHD solution, is the same as \eqref{alpharelation}, except for $d$, namely \al{
0&=D+\frac{\hat{C}\ln\left(\overline{a}/\overline{\rho}^2\right)}{3}-\pi\hat{C}\left(d+\Lambda\right),
} and for the fact that now $\Lambda\approx-1.431$, due to the different numerical value of \eqref{Ginumest}. The choice of where $y_{-\infty}$ actually goes to zero gives an uncertainty that may shift in frequency the position of "resonant cavity modes": the reasonable solution is to control the $s_{-\infty}$ value where $y_{-\infty}\rightarrow0$ to physically match where the solution actually goes to 0. Considering the limit of negligible Landau damping, $\overline{\rho}^2$ is real and $\phi_I=0$, and, as an example, $|m|=2$ and $d=\text{ctg}\Phi_R$, we get \al{
\alpha=\frac{\pi\hat{C}_1\left[\text{ctg}\Phi_R-\Lambda-(1/3\pi)\ln\left(\overline{a}/\overline{\rho}^2\right)\right]-D_1}{D_2-\pi\hat{C}_2\left[\text{ctg}\Phi_R-\Lambda-(1/3\pi)\ln\left(\overline{a}/\overline{\rho}^2\right)\right]}.\label{alphawp}
} We recall that $\alpha$ is the ratio of the solution to the homogeneous equation to the particular forced solution (say, plasma solution to antenna solution), namely a kind of field/per\-tur\-ba\-tion ratio, and expresses the Poynting flux the antenna launches from outside into the inner region. The imaginary part of $\alpha$ is related to absorption (it is proportional to Poynting flux) and so vanishes in the no-damping limit. The real part of $\alpha$ going to infinity means a resonance, occurring at natural frequencies where the plasma core responds to the mode converted KAW as a cavity resonator. Weak damping means, thus, that the wave can propagate back and forth inside the SAW resonance sufficiently many times to set up a standing wave, in contrast to the strong damping case, where the mode converted KAW is absorbed locally. For this reason, the radial mode structure is expected to be nearly independent on the actual absorption in the weak damping limit, while the fluctuation amplitude will increase for decreasing absorption (increasing gain) as in a typical resonant cavity. A frequency shift $\Delta\omega_k$ in the antenna frequency brings a relative modification in $C_1$, $C_2$, $D_1$, and $D_2$ of the order $\Delta\omega_k/\omega_k$ and similarly in $\Delta x_0/x_0$. Note that the dominant shift in \eqref{alphawp} is due to $\text{ctg}\phi_R$ since \al{
\Phi_R=\text{Re}\int_0^rk_r\,\text{d}r.
} From \eqref{dispersionrelwp} it is clear that $\Delta k_r/k_r\sim\Delta\omega_k/\omega_k$ and, then, \al{
\Delta\Phi_R=k_rr_0\frac{\Delta\omega_k}{\omega_k},\qquad \frac{\Delta\omega_k}{\omega_k}\sim\frac{\Delta \Phi_R}{k_rr_0}\sim\frac{\pi}{k_rr_0},\label{resonancedistance}
} where $\Delta\omega_k$ expresses the typical distance between two subsequent resonances in $\alpha$.

\subsection{Weak damping: solution of the Schr\"odinger envelope equation}
Let's rewrite \eqref{Schrodingerlike} as \al{
\omega_0\DP{D_R}{\omega_0}\DP{A}{\tau}=\frac{1}{a}\DP{D_R}{k_r}\DP{A}{x}-D_IA-\frac{\text{i}}{2a^2}\DP{^2D_R}{k_r^2}\DP{^2A}{x^2},\label{Schrodingerlike2}
} where we have introduced the time normalization $\tau=\omega_0t$ and $\omega_0$ represents the reference antenna frequency, which we consider as a monochromatic spectrum. When comparing this equation with \eqref{Schrodingerlike}, we need to consider that the anti-Hermitian response is \al{
D_A:=D_I+\frac{1}{2a}\DP{^2D_R}{x\partial k_r},
} instead of the simpler $D_A=D_I$ of \eqref{DAsch} at the leading order. Equation \eqref{Schrodingerlike2} can be made even simpler noting that it is written for the slowly varying envelope behavior only, having extracted $\sim e^{\text{i}\int k_r\text{d}r'}$. To better examine the connection with the antenna solution, let us introduce the normalized solution $z:=\text{d}y/\text{d}x$, with \al{
A:=ze^{-\text{i}\int_{x_0}^xak_r\text{d}x'},
} so that \al{\begin{gathered}
A'=\left(z'-\text{i}k_raz\right)e^{-\text{i}\int_{x_0}^xak_r\text{d}x'},\\
A''=\left(z''-2\text{i}z'k_ra-\text{i}k_r'az-k_r^2a^2z\right)e^{-\text{i}\int_{x_0}^xak_r\text{d}x'},\end{gathered}\label{gath1}
} where the superscripts $'$ and $''$ mean, respectively, the first and second derivative with respect to $x$. In order to exploit the $b_k$ dependence of $D_R$, we can write \al{\begin{gathered}
\frac{1}{a}\DP{D_R}{k_r}=\frac{2k_r}{a}\DP{D_R}{\left(k_r^2\right)},\\
\frac{1}{2a^2}\DP{^2D_R}{k_r^2}=\frac{1}{a^2}\DP{D^2}{k_r^2}+\frac{2k_r^2}{a^2}\DP{^2D_R}{\left(k_r^2\right)^2}.\end{gathered}\label{gath2}
} Substituting \eqref{gath1} and \eqref{gath2} back into \eqref{Schrodingerlike2}, we obtain \al{\begin{aligned}
\omega_0\DP{D_R}{\omega_0}\DP{A}{\tau}e^{\text{i}\int_{x_0}^xak_r\text{d}x'}&=-D_Iz-\text{i}\left(\frac{\partial D_R}{a^2\partial\left(k_r^2\right)}+\frac{2k_r^2}{a^2}\DP{^2D_R}{\left(k_r^2\right)^2}\right)\left(z''+k_r^2a^2z-\text{i}k_r'az\right)\\&+\frac{4\text{i}k_r^2}{a^2}\DP{^2D_R}{\left(k_r^2\right)^2}\left(\text{i}z'+k_raz\right)k_ra;\raisetag{2.6\normalbaselineskip}\end{aligned}\label{Schrodingerlike3}
} if $z$ is a WKB solution, namely a solution $\sim e^{\text{i}\int_{x_0}^xak_r\text{d}x'}$, this equation is identically satisfied as it should be for $\partial_\tau A\sim\partial_\tau z=0$. Now note that, for the problem of interest, \al{
\left|\frac{\partial D_R}{a^2\partial\left(k_r^2\right)}\right|\gg\left|\frac{k_r^2}{a^2}\DP{^2D_R}{\left(k_r^2\right)^2}\right|.
} Thus, equation \eqref{Schrodingerlike2}, via \eqref{Schrodingerlike3}, can be simplified to \al{
\omega_0\DP{D_R}{\omega_0}\DP{z}{\tau}=-D_Iz-\text{i}\DP{D_R}{b_{k_0}}\frac{\rho_i^2}{a^2}\left(z''+z\frac{b_{k_0}}{\rho_i^2}a^2-\frac{m^2}{x^2}z\right),\label{Schrodingerlike4}
} with $b_{k_0}(\omega_0,x)$ satisfying the WKB dispersion relation. However, note that this equation is still missing the antenna drive. Integrating \eqref{vort} in radius near the SAW resonance, where radial singular structure dominates, it is readily shown that \eqref{Schrodingerlike4} is modified into \al{
\omega_0\DP{D_R}{\omega_0}\DP{z}{\tau}=-D_Iz-\text{i}\DP{D_R}{b_{k_0}}\frac{\rho_i^2}{a^2}\left(z''+z\frac{b_{k_0}}{\rho_i^2}a^2-\frac{m^2}{x^2}z-\frac{\overline{a}}{\overline{\rho}^2}\right),\label{Schrodingerlike5}
} having further assumed that the solution $z$ is normalized to $\hat{C}$. In fact, the last equation reduces to \eqref{redeqz} considering \al{
b_{k_0}\frac{a^2}{\rho_i^2}-\frac{m^2}{x^2}+\text{i}\frac{D_Ia^2}{\rho_i^2}\left(\DP{D_R}{b_{k_0}}\right)^{-1}\approx-\frac{\overline{a}\left(x-x_0\right)}{\overline{\rho}^2},
} as $x\rightarrow x_0$.

\section{Numerical solution}\label{kinlinnumsol}
In the cylindrical plasma equilibrium described in subsection \ref{equilprofile}, and with the plasma equilibrium parameters listed in subsection \ref{parameters}, the problem has been solved both in the case of strong damping and in the case of weak damping. The ion Larmor frequency is nearly two orders of magnitude higher than the antenna frequency, so the use of the gyrokinetic approach is justified. In the following subsections, we discuss the numerical solution for the KAW wave equation in its general form and in the various limits discussed theoretically in the previous section.

The numerical method remains the same as in chapter \ref{mhdchapter}\footnote{That is, the solution is obtained by shooting method using the solve$\_$ivp package from the scipy.integrate library based on the 4th-order Runge-Kutta-Fehlberg method (RK45).}; what changes in this chapter are the various forms of the governing wave equations. The first model is \eqref{master}, where the long-wavelength limit is adopted and the leading order correction in finite Larmor radius to the ideal MHD model is used. This model still contains all equilibrium non-uniformity and geometry effects. We speculate that this model must reduce to that of \eqref{redeqz}, proposed by Hasegawa and Chen \cite{Hasegawa1975,Hasegawa1976}, in the local limit that is justified for strong absorption. For weak absorption, the wave packet can travel far away (in wavelength units) and $b_k=k_\perp^2 \rho_i^2$ may become of order unity. In order to account for this important change, we cast the wave equation in the form of the wave packet intensity equation \eqref{Schrodingerlike}, which we solve analytically by WKB method as well as numerically, after casting it in the form of \eqref{Schrodingerlike5}. In particular, \eqref{Schrodingerlike5} reduces to \eqref{redeqz} near the mode conversion layer. Thus, all various wave equations used in Chapter \ref{mhdchapter} are consistent for strong damping. Meanwhile, for weak damping, they account for cylindrical geometry and equilibrium nonuniformity, see \eqref{master}, but still in the long-wavelength limit, or also address the finite-wavelength wave-packet response, see \eqref{Schrodingerlike} and \eqref{Schrodingerlike5}, which is necessary when KAWs propagate away from the mode conversion layer.

\subsection{Strong damping}

\begin{figure}[!h]
	\centering
	\includegraphics[scale=0.23]{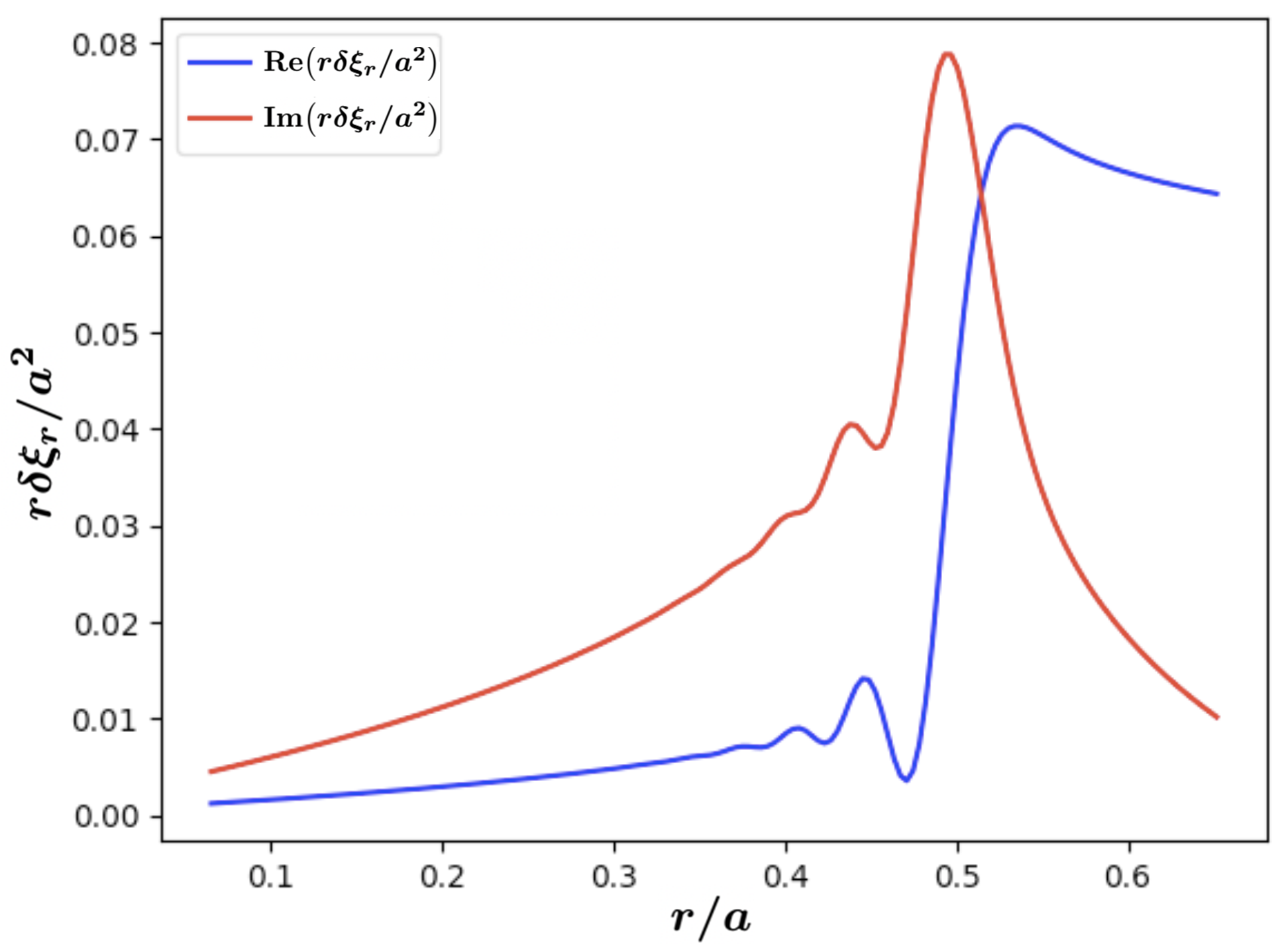}
	\caption{\emph{Fig.\ref{kaw}} Radial displacement, normalized to $(-\delta \hat B_{rh}/B_0)$, according to the linear gyrokinetic solution for strong damping. The normalized distance $r/a$ goes from the axis to just beyond the SAW resonance.}
	\label{kaw}
\end{figure}
\begin{figure}
    \centering
	\includegraphics[scale=0.7]{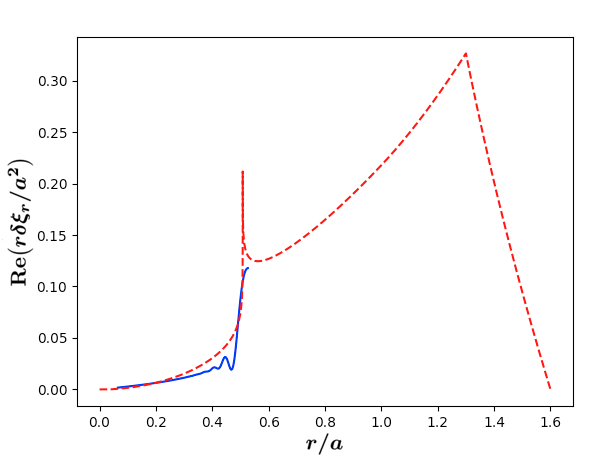}
	\caption{\emph{Fig.\ref{match1}} Matching between the linear gyrokinetic solution for strong damping (solid blue line) and the rescaled MHD solution (dashed red line) computed with the proper kinetic value of $\alpha$.}
	\label{match1}
	\includegraphics[scale=0.7]{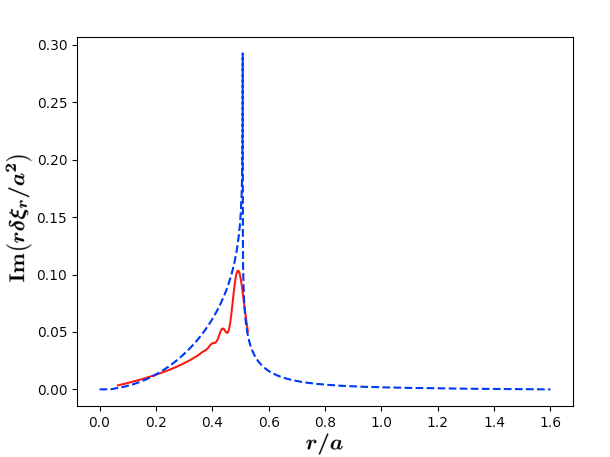}
	\caption{\emph{Fig.\ref{match2}} Matching between the linear gyrokinetic solution for strong damping (solid red line) and the rescaled MHD solution (dashed blue line) computed with the proper kinetic value of $\alpha$.}
	\label{match2}
\end{figure}

In this case, the problem is essentially solved analytically as explained in subsection \ref{strongdamping}, the numerical part being devoted to the computation of the integrals over the Airy functions and to the plotting of the solution, expressed as normalized plasma radial displacement $y:=x\delta\xi_r/a$ as a function of the normalized distance $x:=r/a$ from the magnetic axis (fig.\ref{kaw}). Results show that the Poynting flux isn't absorbed at a given singular point, rather, it is converted in a kinetic Alfvén wave and absorbed by thermal electrons (through Landau damping) in a spatial interval whose width depends on the wavelength and on the damping coefficient: in practice, within few oscillations, in the radial direction, of the radial electric field. The linear gyrokinetic solution removes the singularity and slightly reduces the amplitude, giving a correspondingly slightly larger absorption with respect to the MHD estimate, with, from \eqref{alpharelation}, $\text{Re}\alpha=-0.65033704$ and $\text{Im}\alpha=-0.00127338$ (stopping at the numerical truncation error). The internal KAW solution in terms of Airy functions matches by construction the external MHD solution (fig.\ref{match1} and fig.\ref{match2}), namely the MHD solution with the appropriate kinetic $\alpha$. Clearly, the plotted kinetic solution is valid around the resonance and would not be accurate for the long-wavelength branch.

\subsection{Weak damping}

\begin{figure}[!h]
	\centering
	\includegraphics[scale=0.3]{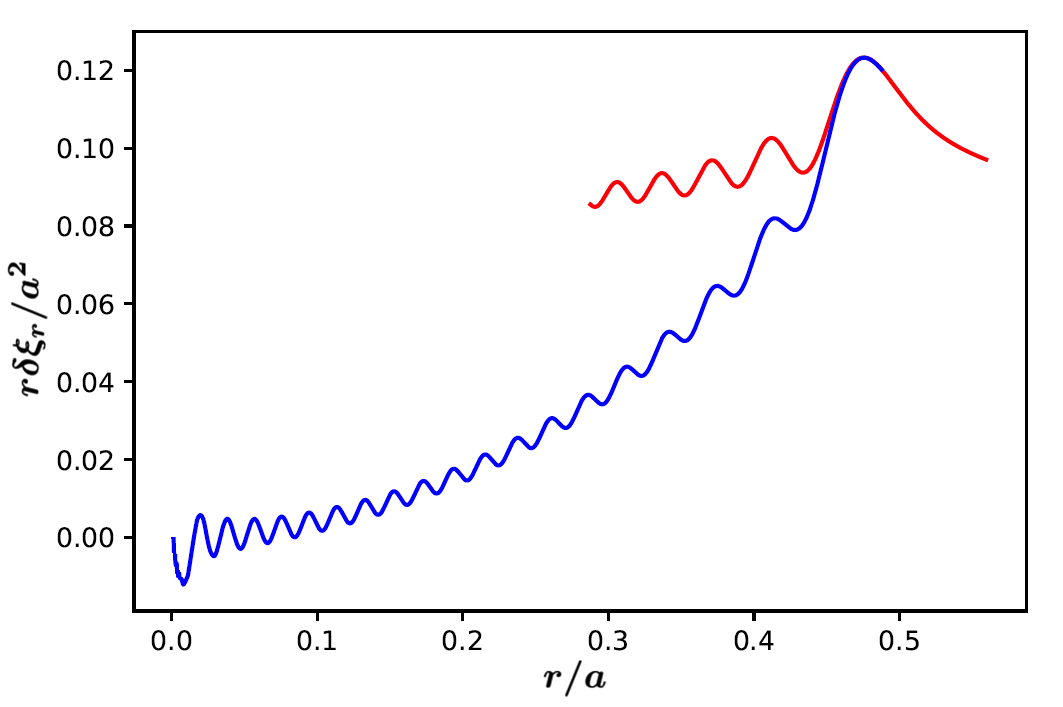}
	\caption{\emph{Fig.\ref{ynum1}} Radial displacement, normalized to $(- \delta \hat B_{rh}/B_0)$, for the linear gyrokinetic case in the no-damping limit (blue line). The red line represents the local solution, equation \eqref{reduced}, which clearly deviates from the complete solution in the long-wavelength limit away from the SAW resonance layer. For this plot, $B_0=5\cdot10^5$ \emph{G} and $J_{0z}(0)=10^{12}$ \emph{statA}. The normalized distance $r/a$ goes from the axis to just beyond the SAW resonance.}
	\label{ynum1}
\end{figure}

\begin{figure}[!h]
	\centering
	\includegraphics[scale=0.81]{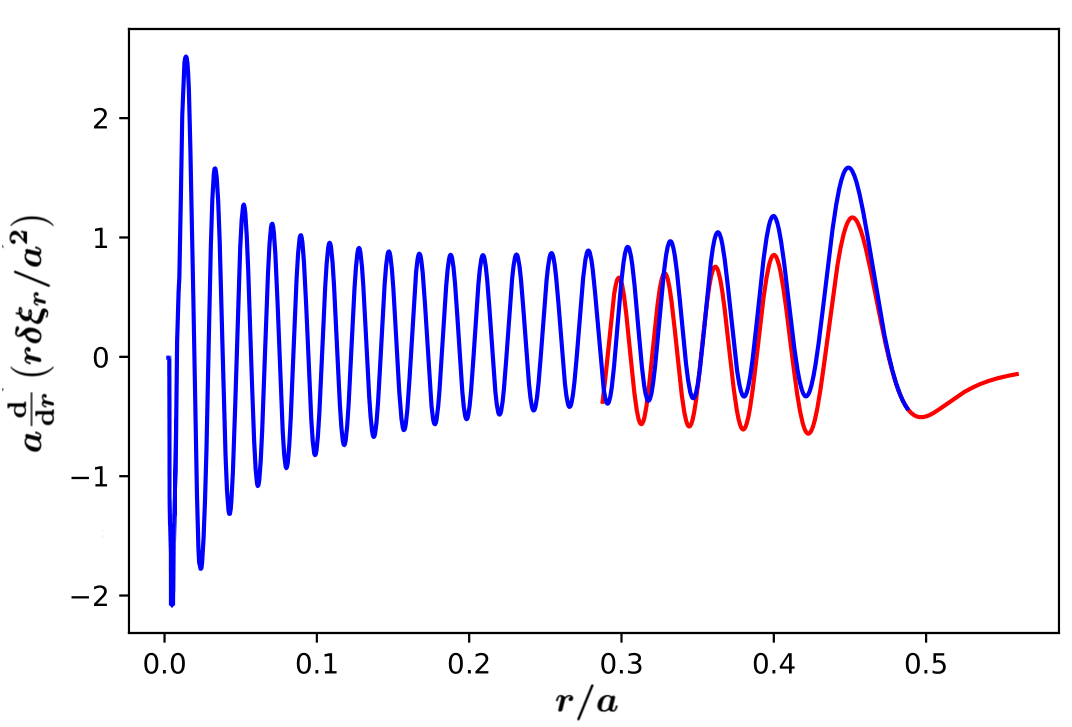}
	\caption{\emph{Fig.\ref{ynum2}} First derivative of the radial displacement, normalized to $(- \delta \hat B_{rh}/B_0)$, for the linear gyrokinetic case in the no-damping limit (blue line). The red line represents the local solution, namely the solution $z$ of the inhomogenous Airy equation. For this plot, $B_0=5\cdot10^5$ \emph{G} and $J_{0z}(0)=10^{12}$ \emph{statA}. The normalized distance $r/a$ goes from the axis to just beyond the SAW resonance.}
	\label{ynum2}
\end{figure}

In the weak/no damping case, we start with the numerical solution of the wave equation \eqref{master}, which is valid in the long-wavelength limit. We solve this equation between the axis and the resonant layer as a boundary eigenvalue problem. Switching from $\delta\phi_{k}$ to the dimensionless field variable $y$ (and dropping the subscript ${k}$ for brevity), the fourth-order equation \eqref{master} is written as the equivalent system of four first-order equations \al{\begin{aligned}
\begin{cases}
&y'=y^\text{\tiny I},\qquad \left(y^\text{\tiny I}\right)'=y^\text{\tiny II},\qquad \left(y^\text{\tiny II}\right)'=y^\text{\tiny III},\vspace{0.3cm}\\
&f\left(y^\text{\tiny III}\right)'=-\left(\frac{2f}{r}+\partial_rf\right)y^\text{\tiny III}-\left(D_{A}+\frac{\partial_rf}{r}-\frac{2\left(m^2+1\right)f}{r^2}\right)y^\text{\tiny II}\\
&-\left(\frac{\partial_r\left(rD_{A}\right)}{r}-\frac{m^2\partial_rf}{r^2}+\frac{\left(6m^2+1\right)f}{r^3}\right)y^\text{\tiny I}-\left(\frac{\partial_rk_\parallel^2}{r}-\frac{m^2D_{A}}{r^2}+\frac{\left(m^4-5m^2\right)f}{r^4}\right)y.
\end{cases}\end{aligned}\label{master4}
} The boundary conditions are the matching conditions \eqref{bond1}, \eqref{bond2}, \eqref{bond3}, \eqref{bond4} at the resonant layer, where the integration starts, and the regularity conditions \eqref{regularity} close to the magnetic axis. In order to make the dependence on $\alpha$ and $d$ explicit, based on \eqref{bond1}, we write $y$ as the following superposition of two independent solutions of the homogeneous problem, $y_1$ and $y_2$, plus a particular non-homogeneous solution $y_3$: \al{
y=Dy_1+\pi d\hat{C}y_2-\hat{C}y_3.\label{ynumerical}
} Note that the further independent solution of the homogeneous problem is eliminated by construction, imposing the matching condition with the external antenna solution at the SAW resonance layer. These solutions evidently satisfy the initial conditions \al{
y_1(0)=1,\qquad y^\text{\tiny I}_1(0)=0,\qquad y^\text{\tiny II}_1(0)=0,\qquad y^\text{\tiny III}_1(0)=0,
} \al{
y_2(0)=-\frac{1}{3},\quad y^\text{\tiny I}_2(0)=\left(\frac{\overline{a}}{\overline{\rho}^2}\right)^{1/3}\text{Ai}(0),\quad y^\text{\tiny II}_2(0)=\left(\frac{\overline{a}}{\overline{\rho}^2}\right)^{2/3}\text{Ai}'(0),\quad y^\text{\tiny III}_2(0)=0,
} \al{
\begin{gathered}
y_3(0)=-\frac{1}{3}\ln\left(\frac{\overline{a}}{\overline{\rho}^2}\right),\qquad y^\text{\tiny I}_3(0)=\pi\left(\frac{\overline{a}}{\overline{\rho}^2}\right)^{1/3}\text{Gi}(0),\\ y^\text{\tiny II}_3(0)=\pi\left(\frac{\overline{a}}{\overline{\rho}^2}\right)^{2/3}\text{Gi}'(0),\qquad y^\text{\tiny III}_3(0)=-\frac{\overline{a}}{\overline{\rho}^2}.
\end{gathered}
} With these conditions, we do three distinct integrations of \eqref{master4} to obtain $y_1$, $y_2$, and $y_3$; then, we compute the parameters $\alpha$ and $d$ applying \eqref{regularity}, which can be suitably rewritten as \al{\begin{cases}
k_1y_1(r=0^+)+k_2y_2(r=0^+)-y_3(r=0^+)=0,\\
k_1y^\text{\tiny I}_1(r=0^+)+k_2y^\text{\tiny I}_2(r=0^+)-y^\text{\tiny I}_3(r=0^+)=0,\end{cases}
} where $k_1:=D/\hat{C}$ and $k_2:=\pi d$. Having determined $k_1$ and $k_2$, where $\alpha$ and $d$ can be reconstructed from, the general solution \eqref{ynumerical} is finally obtained and displayed in figs.\ref{ynum1}-\ref{ynum2}. Meanwhile, the analytic solution \eqref{reduced} for the mode converted KAW, expressed as linear superposition of Airy functions integrals, is also smoothly connected with the external MHD solution by construction, but is valid only in proximity of the mode conversion layer. This is made visible in fig.\ref{ynum1} and fig.\ref{ynum2} by the comparison of blue and red curves. 

\begin{figure}[ht]
	\centering
	\includegraphics[scale=0.55]{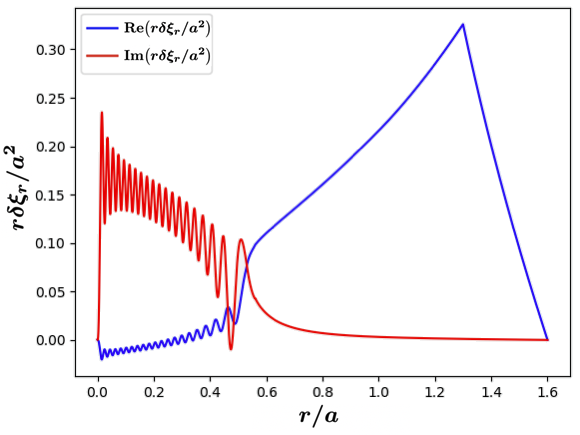}
	\caption{\emph{Fig.\ref{Structure1}} Radial displacement, normalized to $(- \delta \hat B_{rh}/B_0)$, from the solution of equation \eqref{Schrodingerlike5} and reduced damping (1\% of the nominal electron Landau damping) with respect to the case of \emph{fig.\ref{kaw}}. The mode conversion from MHD-antenna solution to KAW is evident.}
	\label{Structure1}
\end{figure}

\begin{figure} [!h]
	\centering
	\includegraphics[scale=1]{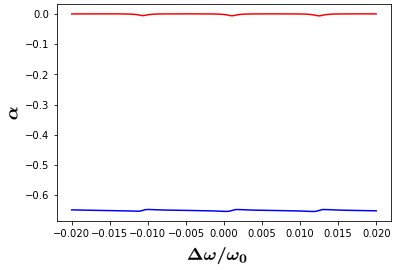}
	\caption{\emph{Fig.\ref{wkb1}} \emph{Re}$\alpha$ (blue line) and \emph{Im}$\alpha$ (red line) from the WKB solution.}
	\label{wkb1}
	\vspace{2cm}
	\includegraphics[scale=0.73]{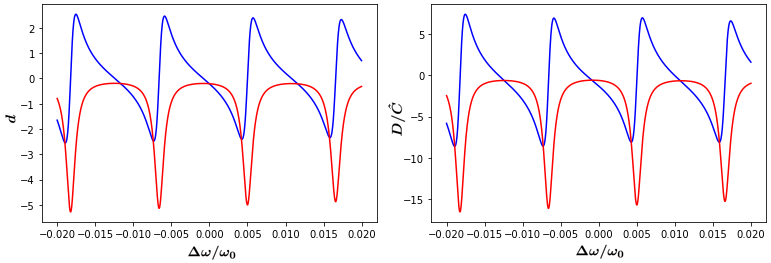}
	\caption{\emph{Fig.\ref{wkb2}} Real (blue lines) and imaginary (red lines) parts of $d$ and $D/\hat{C}$ from the WKB solution.}
	\label{wkb2}
\end{figure}

\begin{figure}
	\centering
	\includegraphics[scale=0.46]{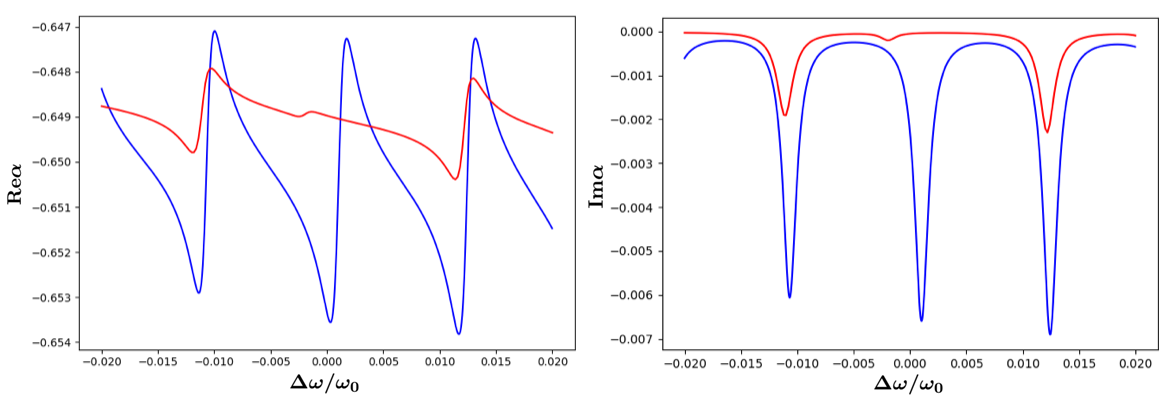}
	\caption{\emph{Fig.\ref{Spectrum1}} Comparison between the WKB $\alpha$ (blue line) and the wave-packet $\alpha$ (red line) for the nominal value of electron Landau damping, as functions of the antenna frequency in a neighborhood of the reference frequency $\omega_0$.}
	\label{Spectrum1}
	\includegraphics[scale=0.46]{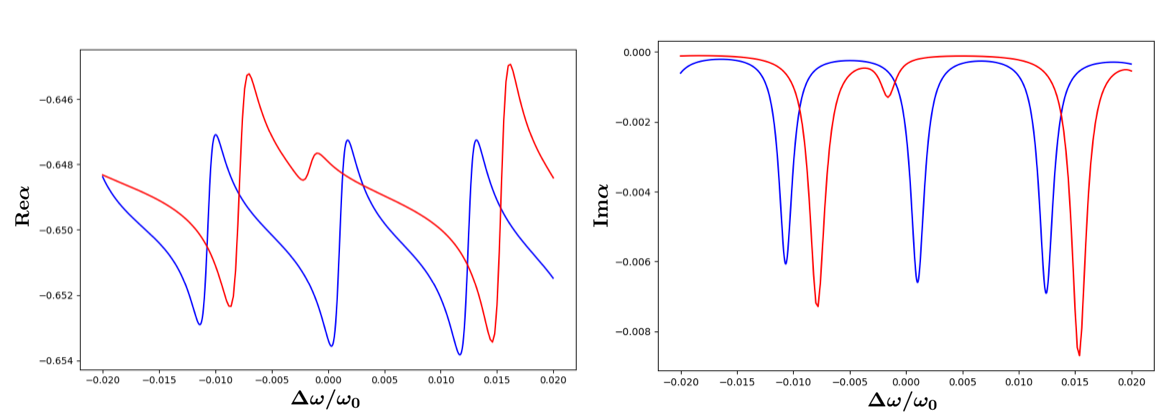}
	\caption{\emph{Fig.\ref{Spectrum2}} Comparison between the WKB $\alpha$ (blue line) and the wave-packet $\alpha$ (red line), for reduced (by a factor $10^2$) electron Landau damping, as functions of the antenna frequency in a neighborhood of the reference frequency $\omega_0$.}
	\label{Spectrum2}
	\vspace{1cm}
	\includegraphics[scale=1]{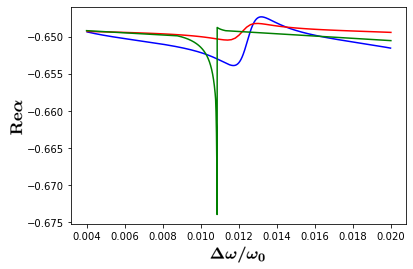}
	\caption{\emph{Fig.\ref{Radice}} Dispersion relation about a "resonant cavity" mode: wave-packet solution (red), WKB solution (blue) and no-damping $b_k\ll1$ solution (green).}
	\label{Radice}
\end{figure}

We have solved the problem also for the general weak damping case with $b_k\sim1$ using the wave-packet formalism, see \eqref{Schrodingerlike} and/or \eqref{Schrodingerlike5}. Results of numerical solutions of \eqref{Schrodingerlike5} are shown in fig.\ref{Structure1} for 1\% of the nominal electron Landau damping that causes the strong local absorption of the mode converted KAW in fig.\ref{kaw}. At the resonant layer there is, similarly to the strong damping case, a mode conversion from MHD-antenna solution to KAW, but now the KAW is only weakly absorbed and propagates inside, leading to "resonant cavity" modes that are of crucial importance for the discussion of nonlinear phenomena, presented in chapter \ref{nonlinearGK}. Also in this case, the external MHD solution is exactly matched by construction with the gyrokinetic solution when the kinetic absorption coefficients are used, consistently with \cite{Hasegawa1976}, predicting plasma impedance remains unchanged at leading order. The results of numerical solutions of \eqref{Schrodingerlike5} are compared with those obtained in the WKB approximation, see \eqref{DyDxmatch}. In particular, we have solved the WKB dispersion relation (see section \ref{wavepacketanalysis}), plotting $\alpha$, $d$ and $D/ \hat{C}$ vs. the normalized frequency shift externally imposed at the antenna, $\Delta \omega/\omega_0$ (see figs. \ref{wkb1}-\ref{wkb2}) using nominal electron Landau damping first. Recall that the (negative) imaginary part of $\alpha$ (red line in fig.\ref{wkb1}) gives us information on wave absorption (Poynting flux); fig.\ref{wkb2}, in particular, demonstrates, thus, the existence of a discrete spectrum of closely spaced antenna frequencies that cause a particularly strong plasma response for fixed applied antenna perturbation amplitude. The closely spaced spectrum is consistent with the estimated frequency separation of \eqref{resonancedistance}. The KAW spectra, namely $\alpha$ in its real and imaginary parts, and of $d$ and $D/ \hat{C}$, have been plotted as functions of the normalized frequency shift $\Delta \omega/\omega_0$ both for the WKB solution and for the wave-packet solution (fig.\ref{Spectrum1}). Wave-packet solution and WKB approximation are consistent except for the slightly different amplitude of the peaks, which suggests that WKB approximation more easily breaks down near magnetic axis for every second radial eigenstate. The peaks correspond to "resonant cavity" excitations, which are very mild for the nominal value of electron Landau damping. In order to illustrate how a very small antenna perturbation may generate a finite wave amplitude response inside the plasma, we have repeated the solution of \eqref{Schrodingerlike5} for 1\% of the nominal electron Landau damping, resulting in the KAW radial mode structure of fig.\ref{Structure1}, to be compared with that of fig.\ref{kaw}. So, fig.\ref{Spectrum2} is the same as fig.\ref{Spectrum1} but specialized for this reduced damping case. By direct comparison, it is clear that excitation of "resonant cavity" modes in the KAW spectrum is more pronounced in this case and that this is the explanation of the much stronger plasma response obtained for selected frequencies. In order to better appreciate the robustness of this "resonant cavity" phenomenon, in fig.\ref{Radice}, there's a zoom about the third natural frequency of fig.\ref{Spectrum1}, also including the limiting case of no-damping. The small shift of the green peak is due to the more stringent assumptions in the no-damping case, in particular to the fact that we assumed the long-wavelength limit for reducing the KAW equation to the 4th-order ordinary differential equation form \eqref{master} or \eqref{master4}.

\begin{figure}
	\centering
	\includegraphics[scale=0.6]{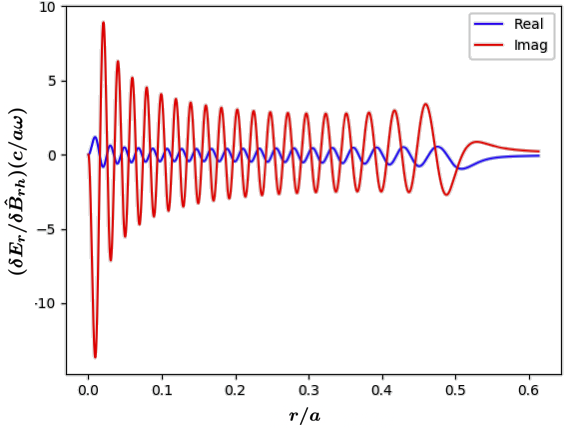}
	\caption{\emph{Fig.\ref{Structure2}} Normalized KAW radial electric field $\delta E_r$ (for reduced electron Landau damping).}
	\label{Structure2}
\end{figure}

\begin{figure}
	\centering
	\includegraphics[scale=0.61]{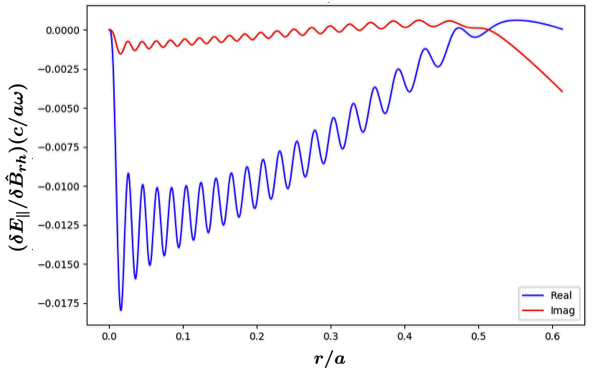}
	\caption{\emph{Fig.\ref{Structure3}} Normalized KAW parallel electric field $\delta E_\parallel$ (for reduced electron Landau damping).}
	\label{Structure3}
\end{figure}

\begin{figure}
	\centering
	\includegraphics[scale=0.8]{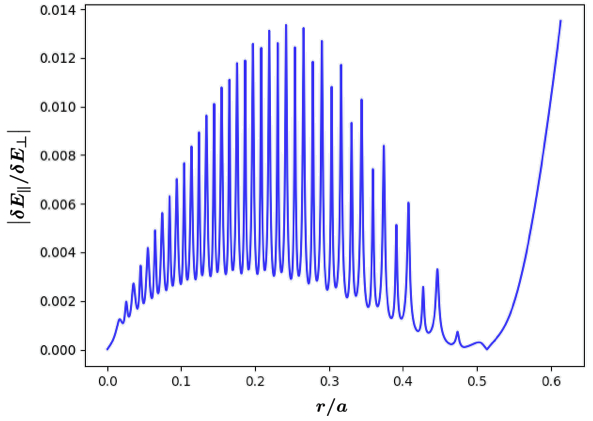}
	\caption{\emph{Fig.\ref{Structure4}} Modulus of KAW electric field ratio $\delta E_\parallel/\delta E_\perp$ (for reduced electron Landau damping).}
	\label{Structure4}
\end{figure}

The last plots (figs.\ref{Structure2}-\ref{Structure4}) represent the normalized components of the electric field, given by the relations \al{
\frac{c}{a\omega}\frac{\delta E_r}{\delta\hat{B}_{rh}}=-\frac{1}{m}\DP{y}{x},\qquad \frac{c}{a\omega}\frac{\delta E_\parallel}{\delta\hat{B}_{rh}}=\frac{\text{i}a\tau k_\parallel\left(1-\Gamma_k\right)}{m}y,
} and their ratio.

\vspace{1cm}

\section{Summary}
In this chapter, we have revisited the ideal MHD analysis of chapter 3 using gyrokinetic theory, highlighting the microscale phenomena that underlie the resonant absorption of the SAW at the radial location where the imposed antenna frequency matches the continuously varying SAW frequency spectrum; consistent with earlier analyses \cite{Hasegawa1975,Hasegawa1976}, the resonant absorption consists in the mode conversion to the short wavelength kinetic Alfvén wave, which is typically absorbed within a short distance from the resonant layer and within a few spatial oscillation period of the mode converted wave (strong absorption or strong damping). Despite this case was and has been of interest for existing studies so far due to its direct relevance in plasma heating \cite{Hasegawa1975,Hasegawa1976,Itoh1982,Itoh1983}, here we have rather focused on the opposite weak absorption case, where, even with a modest amplitude SAW launched by the antenna, large KAWs can be excited inside the resonant (mode conversion) layer and yield a number of interesting nonlinear behaviors. As new novel results of chapter 4, it has been demonstrated that the plasma region inside the mode conversion layer behaves as a "resonant cavity"; that is, when the antenna frequency matches a given discrete spectrum, the plasma response is stronger, exhibiting the clear behavior of a resonantly driven and weakly damped oscillator. This is a result of relevance for chapter 5, in which we shall investigate nonlinear behaviors induced by large amplitude kinetic Alfvén waves, fine-tuned in phase and amplitude by the external antenna.

\chapter{\textsc{\textsc{kinetic alfvén waves in a cylindrical plasma:\\ nonlinear description}}}\label{nonlinearGK}
\vspace{100pt}

Within the MHD description, the shear Alfvén wave launched by the antenna develops radial singular structures at the radial position where the antenna frequency matches the local shear Alfvén continuous spectrum frequency. Correspondingly, the incoming Poynting flux vanishes at the resonant surface, suggesting that the long wavelength fluctuation is completely absorbed at that very position. Within the linear gyrokinetic analysis, we have shown how the power flux actually transfers to the kinetic Alfvén wave by the mode conversion process. The new kinetic wave is damped and, for typical parameters, absorbed by thermal plasma electrons via Landau damping. This process was extensively studied in the literature from the discovery of kinetic Alfvén waves \cite{Hasegawa1975,Hasegawa1976} to the many application studies looking at plasma heating by Alfvén waves, e.g., in \cite{Itoh1982,Itoh1983}. The scope of this thesis work is to investigate the exactly opposite limit, where the mode converted kinetic Alfvén wave is weakly absorbed. While it was shown in chapter \ref{kawlinchap} that this case is of little interest for Alfvén wave heating, we also noted that it is characterized by the plasma behavior as a "resonant cavity" inside the mode conversion layer. Thus, with a relatively small perturbation amplitude imposed at the antenna, a significantly large kinetic Alfvén wave can be excited with consequences on the nonlinear plasma behaviors. In the following, we adopt the nonlinear gyrokinetic description and show that the power that is mode converted to kinetic Alfvén waves is only partly absorbed by electrons through Landau damping, while the dominant fraction forms a nonlinear convective cell which is characterized by $k_\parallel \equiv 0$. Our study of the generation of convective cells by modulational instability, summarized in this chapter, generalizes a previous work in uniform magnetized plasmas \cite{Zonca2015}, and a local analysis of magnetostatic convective cells in non-uniform plasmas \cite{Kaw1983}. This problem is of interest both for the physics of planetary magnetospheres and for the production of longitudinal electric fields in laboratory plasmas by external excitation of high-frequency waves (particle acceleration). The present theoretical analyses, yielding both analytical as well as numerical new original results, can be the basis for the proposal of a small table-top experiment as well as the test-bed for novel numerical simulation codes, such as TRIMEG \cite{Lu2019,Lu2021}, STRUPHY \cite{Holderied2021} and GeFi \cite{Lin2005}, addressing the same fundamental nonlinear interactions. 

Since the modulational instability that underlies the formation of convective cells has similarities with the parametric decay instability, we start introducing the qualitative features of this universal process. The concept of convective cells is then introduced, explaining their role in cross-field transport in magnetized plasmas as a reason that has attracted significant attention in plasma research. We also explain the excitation of convective cells by modulational instability and the special role played by kinetic Alfvén waves in this process. After this first general preamble, we carry out a systematic analysis of the nonlinear interplay of kinetic Alfvén waves and convective cells in cylindrical plasmas. In the more general, nonlinear regime, fluctuations are allowed to be of finite amplitude with, however, the constraint that the corresponding nonlinear frequencies, $\omega_{nl}$, be much less than the cyclotron frequency. In other words, consistent with the linear gyrokinetic orderings, \al{
\left|\omega_{nl}\right|\sim\left|\delta\mathbf{u}_\perp\cdot\nabla_\perp\right|\sim\left|\omega\right|\sim\mathcal{O}(\varepsilon)\left|\Omega_i\right|,
} with $\delta\mathbf{u}_\perp$ representing the fluctuation-induced particle (guiding-center) jiggling velocity, for example $\delta\mathbf{u}_\perp\approx v_\parallel\delta\mathbf{B}_\perp/B_0$ or $\delta\mathbf{u}_\perp\approx (c/B_0^2)(\delta \mathbf{E}_\perp \times \mathbf{B}_0)$ \cite{Zoncarev}, and with $\varepsilon= \rho_i/L \sim |\omega/\Omega_i|$ the small parameter introduced in subsection \ref{gyromodsub}.

As a result of our nonlinear gyrokinetic analysis, we demonstrate that optimal conditions for convective cell generation in cylindrical plasmas are those of a "$\theta$-pinch" equilibrium, that is, with an azimuthal magnetic field only and no magnetic curvature. Earlier results of convective cell excitation by kinetic Alfvén waves in uniform plasmas \cite{Zonca2015} are recovered in the proper limiting case. Meanwhile, the new original results of this thesis work are presented. In particular, the effect of plasma nonuniformity is shown to importantly affect the convective cell stability both qualitatively and quantitatively via the diamagnetic response \cite{Chen2022,Kaw1983}. In fact, the convective cell growth is significantly stronger (typically up to an order of magnitude) than in uniform plasmas, and the unstable parameter region is significantly broader. Another element of crucial importance is the subtle interplay between nonlinearity and plasma nonuniformity, which ultimately impacts the plasma self-organization that can be controlled by fine-tuning the amplitude of the antenna driven mode converted kinetic Alfvén wave.

\section{Parametric decay instabilities}
The parametric decay instability (PDI) is a fundamental nonlinear process \cite{Jackson1967,Liu1976} involving three nonlinear coupled waves/oscillators: a large-amplitude pump ("mother") wave decays into two daughter waves once its amplitude exceeds a nonlinear threshold. Since the pump wave can be either spontaneously or externally excited, PDI is an important channel for wave energy transfer along with its associated consequences on plasma heating, acceleration and transports. Consider, as a prototype, a system of two coupled driven harmonic oscillators of equations \al{\begin{aligned}
\D{^2x_1}{t^2}+\omega_1^2x_1&=c_1x_2E_0,\\
\D{^2x_2}{t^2}+\omega_2^2x_2&=c_2x_1E_0,\end{aligned}\label{PDIcouple}
} where $\mathbf{k}_{1,2}$ and $\omega_{1,2}$ are taken to be the resonant wave vectors and frequencies, $c_{1,2}$ are the constants representing the strength of the coupling and $E_0=\overline{E}_0\cos(\mathbf{k}_0\cdot\mathbf{x}-\omega_0t)$ is the amplitude of the driver wave, namely the pump. Assuming $x_1=\overline{x}_1\cos(\mathbf{k}_1\cdot\mathbf{x}-\omega t)$ and $x_2=\overline{x}_2\cos(\mathbf{k}_2\cdot\mathbf{x}-\omega't)$, the second of \eqref{PDIcouple} can be written as \al{\begin{aligned}
\left(\omega_2^2-\omega'^2\right)\overline{x}_2\cos\left(\mathbf{k}'\cdot\mathbf{x}-\omega't\right)&=c_2\overline{E}_0\overline{x}_1\cos\left(\mathbf{k}_0\cdot\mathbf{x}-\omega_0t\right)\cos\left(\mathbf{k}\cdot\mathbf{x}-\omega t\right)=\\&=\frac{c_2\overline{E}_0\overline{x}_1}{2}\big\{\cos\left[\left(\mathbf{k}_0+\mathbf{k}\right)\cdot\mathbf{x}-\left(\omega_0+\omega\right)t\right]\\&+\cos\left[\left(\mathbf{k}_0-\mathbf{k}\right)\cdot\mathbf{x}-\left(\omega_0-\omega\right)t\right]\big\};\end{aligned}\label{pdicouple2}
} so, in particular, the driving terms on the right can excite oscillators $x_2$ with frequencies $\omega'=\omega_0\pm\omega$, and oscillators $x_1$ and $x_2$ constitute a couple of decay waves. Equation \eqref{pdicouple2} suggests that nonlinear driving terms are still effective assuming a finite frequency shift; so $\omega'$ does not need to be exactly $\omega_2$, but only approximately equal to $\omega_2$. Furthermore, $\omega'$ can be complex because there can be damping or growth, so the oscillator $x_2$ can respond to a range of frequencies. In conclusion, the frequency and wave-vector matching conditions are \al{
\omega_0\approx\omega_1\pm\omega_2,\\
\mathbf{k}_0\approx\mathbf{k}_1\pm\mathbf{k}_2.
} A parametric decay instability may occur when an incident electromagnetic wave of high frequency and large phase velocity excites two waves with $\mathbf{k}_1\approx-\mathbf{k}_2$. Keep in mind that even a small amount of damping (either collisional or collisionless) will prevent parametric instability unless the pump wave is strong enough. To calculate the threshold amplitude for wave excitation, damping rates $\gamma_1$ and $\gamma_2$ in the first and second of \eqref{PDIcouple} are introduced by means of damping terms $2\gamma_{1,2}\dot{x}_{1,2}$, ultimately obtaining \al{
\left(\omega_1^2-\omega^2-2\text{i}\omega\gamma_1\right)x_1(\omega)=c_1x_2E_0,\\
\left[\omega_2^2-\left(\omega-\omega_0\right)^2-2\text{i}\left(\omega-\omega_0\right)\gamma_2\right]x_2(\omega-\omega_0)=c_2x_1E_0.
} The PDI can be either resonant, if both decay waves are marginally stable or weakly damped normal modes, or non-resonant, if one of the decay waves is a heavily damped quasi mode. The conditions $\omega\approx\omega_1$ and $\omega_0-\omega\approx\omega_2$ give, from the above equations, an instability threshold as a function of the damping coefficients, that is, expressing $x_1$ and $x_2$ with their peak amplitudes, \al{
{\left(c_1c_2\overline{E}_0^2\right)}_{min}=16\omega_1\omega_2\gamma_1\gamma_2;
} thus, the threshold goes to zero for vanishing damping.

Let now the three interacting waves be the pump KAW $\boldsymbol{\Omega}_0=(\omega_0,\mathbf{k}_0)$, the low-frequency daughter ion sound wave (ISW) $\boldsymbol{\Omega}_s=(\omega_s,\mathbf{k}_s)$ and the daughter KAW $\boldsymbol{\Omega}_-=(\omega_-,\mathbf{k}_-)$ with $\omega_-=\omega_s-\omega_0$ and $\mathbf{k}_-=\mathbf{k}_s-\mathbf{k}_0$, consistent with frequency and wave-vector matching conditions. Let the small but finite pump wave amplitude be denoted as $\Phi_0=e\delta\phi_0/T_e$. As $\Omega_s$ could be a quasi mode, we then need to retain terms of order $|\Phi_0|^2$ in view of a proper account of non-resonant PDI. The KAW PDI dispersion relation is \cite{Zoncarev,Chen2011} \al{
\epsilon_{sk}\left(\epsilon_{Ak-}+\chi_{A-}^{(2)}\right)=C_k\left|\Phi_0\right|^2,\label{PDIdisp}
} with the ISW dielectric constant linearly approximated as \al{
\epsilon_{sk}=1+\tau+\tau\Gamma_s\xi_sZ(\xi_s),
} while $\epsilon_{Ak-}$ is the linear dielectric constant of the KAW decay wave, \al{
\epsilon_{Ak-}=\frac{1-\Gamma_-}{b_-}-\sigma_-\left(\frac{k_\parallel^2v_A^2}{\omega^2}\right)_-,
} where $\Gamma_-$, $b_-$ and $\sigma_-$ are the usual functions introduced in section \ref{kawsection} and as yet expressed with the generic subscript $k$; furthermore, $\chi_{A_{-}}^{(2)}$ is a function accounting for nonlinear ion Compton scattering \cite{Sagdeev1969,Chen2011,Zoncarev} and $C_k$ is the nonlinear coupling coefficient between the two daughter waves via the pump mode \cite{Chen2011,Zoncarev}. It is beyond the illustrative scope of PDI given in this section to provide detailed derivation of the $\chi_{A -}^{(2)}$ and $C_k$ terms. So, they will be omitted here. In the case of resonant decay, for a frequency $\omega_s=\omega_{sr}+\text{i}\gamma$, and denoted $\gamma_{dA_{-}}$ and $\gamma_{ds}$ the linear damping rates of the KAW and ISW daughter waves, respectively, relation \eqref{PDIdisp} reduces to \al{
\left(\gamma+\gamma_{dA_{-}}\right)\left(\gamma+\gamma_{ds}\right)=C_k\left|\Phi_0\right|^2\left[-\DP{\epsilon_{skr}}{\omega_{sr}}\DP{\epsilon_{Ak_{-}r}}{\omega_{A_{-}r}}\right]^{-1}.
} We see that the parametric decay instability growth rates increase with $|C_k|\Phi_0|^2|$, which scales with $\left|k_\perp\rho_i\right|^4\left|\delta\mathbf{B}_{\perp0}/B_0\right|^2$ for $\left|k_\perp\rho_i\right|^2\ll1$ and $\left|\delta\mathbf{B}_{\perp0}/B_0\right|^2/\left|k_\perp\rho_i\right|$ for $\left|k_\perp\rho_i\right|^2\gg1$ \cite{Zoncarev,Chen2011}. The decay instabilities are, thus, strongest when $\left|k_\perp\rho_i\right|\sim1$, and it clearly demonstrates the necessity of keeping finite ion Larmor radius kinetic effects in dealing with the decay instabilities of KAWs. 

Finally, it is illuminating to compare the decay instabilities of KAWs versus those of SAWs in the MHD regime. Essentially, by employing the ideal MHD fluid theory the PDI dispersion relation takes the form similar to the KAW PDI dispersion relation \eqref{PDIdisp}, with KAW terms replaced by corresponding SAW terms. The more fundamental change lies in the nonlinear coupling term, that is, $C_k$ replaced by $C_I$, with \cite{Zoncarev} \al{
C_I\propto\cos^2\theta_0,\qquad C_k\propto\sin^2\theta_0,\label{CICK}
} where $\theta_0$ is the angle between $\mathbf{k}_{\perp0}$ and $\mathbf{k}_{\perp-}$; in particular, \al{
\frac{C_k}{C_I}\sim\mathcal{O}\left(\left|\Omega_i/\omega_0\right|^2\right)\left|k_\perp\rho_i\right|^4,\label{CkCi}
} for $\left|k_\perp\rho_i\right|^2\ll1$, and \al{
\frac{C_k}{C_I}\sim\mathcal{O}\left(\left|\Omega_i/\omega_0\right|^2\right),
} for $\left|k_\perp\rho_i\right|^2\sim1$. The formula \eqref{CkCi} indicates that, for $1>\left|k_\perp\rho_i\right|^2 >\left|\omega_0/\Omega_i\right|$, nonlinear couplings via kinetic effects dominate. Noting that $\left|\omega_0/\Omega_i\right|\sim10^{-3}$ in typical laboratory plasmas, the validity regime of MHD fluid theory for the SAW nonlinear physics is rather limited. Furthermore, at the $\left|k_\perp\rho_i\right|\sim1$ regime where KAW nonlinear effects maximize, we have $\left|C_k\right|/\left|C_I\right|\sim\left|\Omega_i/\omega_0\right|^2$ for typical parameters. In summary, in addition to the significantly enhanced PDI growth rates, there is, perhaps, more significant qualitative difference between KAW and SAW PDI in terms of the wave vector of the scattered daughter wave with respect to that of the pump wave. Indeed, since $C_I\propto\cos^2\theta_0$ according to \eqref{CICK}, the SAW scattering maximizes around $\theta_0=0,\pi$, i.e., when $\mathbf{k}_{\perp-}$ is parallel or anti-parallel to $\mathbf{k}_{\perp0}$. In contrast, the behavior of $C_k$ implies that the KAW scattering maximizes around $\theta_0=\pm\pi/2$, i.e., $\mathbf{k}_{\perp0}$ and $\mathbf{k}_{\perp-}$ are orthogonal. This difference affects the nonlinear evolution of KAW turbulence and the charged particle transports induced by the KAW decay processes \cite{Zoncarev}. In fact, assuming the pump wave is a mode converted KAW and $\mathbf{k}_{\perp 0}$ is predominantly radial, the daughter KAW will also have $\mathbf{k}_{\perp -}$ predominantly radial in the ideal MHD analysis, with little effect on plasma transport. Meanwhile, $\mathbf{k}_{\perp -}$ of the daughter KAW will be predominantly azimuthal in the kinetic regime, with important consequences on plasma transport.

\section{Convective cells}
When, along with the confining magnetic field, an electric field is present, the motion of particles is the sum of the usual circular Larmor gyration plus a drift of the guiding center, resulting in an additional transverse component of velocity \cite{Chen2011}, \al{
\mathbf{v}_E=\frac{c\,\mathbf{E}\times\mathbf{B}}{B^2}.
} For long wavelengths, electron and ions have similar responses; however, on shorter scales of the order of the ion Larmor radius, electron and ion responses become unbalanced and this, in turn, generates local charge separation and therefore strong radial electric fields. The ion flow forms two-dimensional (in the plane perpendicular to the ambient magnetic field) drift patterns named \textit{convective cells} \cite{Taylor1971,Okuda1973,Cheng1977,Chu1978,Lin1978,Shukla1981,Shukla1984,Diamond2005}, namely perturbations which are constant along the field line but change in the perpendicular direction; as a consequence, these perturbations play an important role in the vortex dynamics and in the diffusion and transport of plasma across the confining magnetic field. A classification scheme for convective cells (CCs) can be made between electrostatic convective cells (ESCCs), where $\delta E=\delta E_\perp$ \cite{Okuda1973,Taylor1971}, and magnetostatic convective cells (MSCCs), with $\delta B=\delta B_\perp$ \cite{Chu1978}. Two-dimensionality means, for convective cells, $k_\parallel=0$, with typically $\text{Re}(\omega)\approx0$. Convective-cell motion is therefore similar to that of two-dimensional vortexes in an incompressible fluid, both motions being divergence free. In the presence of collisions, the cells are damped because of ion viscosity, leading to a normal mode with purely imaginary frequency. Their slow motion can lead to anomalously rapid plasma transport across the magnetic field even in thermal equilibrium, mostly when they are connected to instability mechanisms such as a turbulence in an inhomogeneous plasma: initially given linearly unstable drift-waves, namely the waves or collective excitations driven by a pressure gradient destabilized by differences between ion and electron motion, can nonlinearly interact and excite them. Another nonlinear route leading to the formation of convective cells, which is of central importance for the present work, is through Alfvén waves and the modulational instability mechanism discussed in the following.

It is worthwhile recalling, here, the connection of the modulational instability with the parametric decay instability, which was discussed in Sec. 5.1 and is the first example that was historically studied, where the importance of wave scattering in plasmas was recognized and explained \cite{Itoh1983,Lin2005}. This connection between modulational instability and parametric deday instability will be further elaborated in the following.

\subsection{Modulational instability}
In laboratory fusion plasmas, nonlinear excitations of convective cells usually occur via mode-mode couplings of ambient drift-wave and/or Alfvén-wave instabilities. Since convective cells have no parallel wavenumber, their nonlinear interactions involve couplings between co-propagating Alfvén waves with the same $k_\parallel$, which vanishes for SAWs in the ideal MHD limit. Thus, it has long been recognized that only non-ideal MHD fluctuations, such as KAWs, can nonlinearly excite convective cells. Furthermore, since convective cells are excited at $\omega=0$, it is also recognized that their nonlinear excitation takes the form of a \textit{modulational instability}. In general, a modulational instability is a deviation from the wave periodic behavior further reinforced by nonlinearity: a weakly modulated continuous wave in a nonlinear medium grows and leads to spectral sidebands and possibly to the breaking of the periodic fluctuation into modulated pulses. In space, it transforms weakly modulated plane waves into spatially periodic patterns. So, modulational instability is formally quite similar to a PDI, where the carrier wave is the pump, and the two sidebands behave as analogs of the daughter waves. Previous theoretical studies about convective cells excited by KAWs via modulational instability suffer from being restricted to the two-fluid or drift-kinetic descriptions (ignoring finite Larmor radius effects), and/or the limiting assumption that ESCCs and MSCCs are decoupled. These hypotheses have been adopted in order to simplify the theoretical analysis but lead to erroneous conclusions on the nonlinear excitation mechanisms. Notably, spontaneous excitation of CCs can only occur when one keeps the crucial finite ion Larmor radius (FILR) effects, $|k_\perp\rho_i|\sim1$; and in this regime one finds that ESCCs and MSCCs are, in general, intrinsically coupled \cite{Zonca2015}. As new original element with respect to previous works \cite{Kaw1983,Zonca2015}, we will adopt nonuniform cylindrical plasma geometry. Meanwhile, we will employ the nonlinear gyrokinetic equations and demonstrate that both the FILR as well as the finite coupling between ESCCs and MSCCs play qualitatively crucial roles in the dynamics of the modulational excitations of convective cells.

\section{Nonlinear gyrokinetic equations}\label{nonlineqsec}
In the cylindrical plasma equilibrium of interest here, which was introduced in chapter \ref{mhdchapter}, the vorticity equation can be written as (see also \cite{Zonca2015}) \al{\begin{gathered}
v_A^2\rho_i^2\nabla\cdot\left[k_\parallel^2\nabla_\perp\delta\psi_{k}+\frac{1-\Gamma_k}{b_kv_A^2}\DP{^2}{t^2}\nabla_\perp\delta\phi_k\right]=-\frac{c}{B_0}\Lambda_{k'}^{k''}\DP{}{t}\bigg[\rho_i^2v_A^2\bigg(k_\parallel''\frac{\delta\psi_{k''}}{\omega_{k''}}k_\parallel'k_\perp'^2\frac{\delta\psi_{k'}}{\omega_{k'}}\\
-k_\parallel'\frac{\delta\psi_{k'}}{\omega_{k'}}k_\parallel''k_\perp''^2\frac{\delta\psi_{k''}}{\omega_{k''}}\bigg)+\left(\delta\phi_{k''}\Gamma_{k'}\delta\phi_{k'}-\delta\phi_{k'}\Gamma_{k''}\delta\phi_{k''}\right)\bigg];\end{gathered}\label{kawnonlinear}
} here $\mathbf{k}=\mathbf{k}'+\mathbf{k}''$, where $\mathbf{k}'$ and $\mathbf{k}''$ are two
wave vectors involved in the three-wave interaction with $\mathbf{k}$ (similar to parametric decay instability), and $\omega_{{k}}=\omega_{{k}'}+\omega_{{k}''}$ from the frequency matching condition. Furthermore, $\Lambda_{k'}^{k''}=\left(\mathbf{k}'\times\mathbf{k}''\right)\cdot\hat{\mathbf{z}}$, while $\Gamma_k:=I_0(b_k)e^{-b_k}$ with $I_0$ the modified Bessel function. The KAW generated by the external antenna via mode conversion interacts with low-amplitude convective cells and generates sidebands. In particular, as we show in the following, an externally excited KAW pump of sufficiently high amplitude can spontaneously generate a convective cell and the two coupling sidebands which are due to the beat between convective cell and pump mode. As already mentioned in the previous section, we know \cite{Zonca2015} that spontaneous excitation of CCs can only occur when one keeps the crucial finite ion Larmor radius (FILR) effects, i.e., in the $|k_\perp\rho_i|\sim1$ regime, and that ESCCs and MSCCs are intrinsically coupled. Also note that \eqref{kawnonlinear} readily reduces to the already discussed linear limit. In WKB form, \eqref{kawnonlinear} can be cast as \al{\begin{gathered}
\omega_k\left(\frac{k_\parallel^2v_A^2}{\omega_k^2}\right)\left(1-\Gamma_k\right)\left[\frac{\omega_k^2\delta\phi_k}{k_\parallel^2v_A^2}-\frac{b_k\delta\psi_k}{1-\Gamma_k}\right]=-\text{i}\frac{c}{B_0}\Lambda_{k'}^{k''}\bigg[\left(\Gamma_{k''}-\Gamma_{k'}\right)\delta\phi_{k'}\delta\phi_{k''}\\
-\frac{k_\parallel'k_\parallel''v_A^2}{\omega_{k'}\omega_{k''}}\left(b_{k'}-b_{k''}\right)\delta\psi_{k'}\delta\psi_{k''}\bigg].\end{gathered}\label{kawnonlinearwkb}
} The mode decomposition is made by writing explicitly the expressions for $\delta\phi$: \al{
\delta\phi_0:=\frac{T_i}{2e}\left[\delta\hat{\phi}_0e^{\text{i}\left(\frac{n}{R}z-m\theta-\omega_0t\right)}\right]+c.c.\qquad&\text{(KAW pump)},\label{kawpump}\\
\delta\phi_\pm:=\frac{T_i}{2e}\left[\delta\hat{\phi}_\pm e^{\text{i}\left(\pm\frac{n}{R}z-(m_z\pm m)\theta-(\omega_z\pm\omega_0)t\right)}\right]+c.c.\qquad&\text{(KAW sidebands)},\label{kawsidebands}\\
\delta\phi_z:=\frac{T_i}{2e}\left[\delta\hat{\phi}_z e^{-\text{i}\left(m_z\theta+\omega_zt\right)}\right]+c.c.\qquad&\text{(CC)}\label{cc};
} so the subscript $0$ is for the pump, the subscripts $\pm$ are for the two sidebands and the subscript $z$ is for the convective cell. The expressions for $\delta\psi$ follow consistently. Note that the normalization used here is fully consistent with \cite{Chen1999} and slightly different from \cite{Zonca2015}. The fields $\delta\hat{\phi}_0$, $\delta\hat{\phi}_\pm$, $\delta\hat{\phi}_z$ are dimensionless: the corresponding magnetic stream functions are \al{
\delta\hat{\psi}_0=\frac{\omega_0}{k_\parallel c}\delta\hat{A}_{\parallel 0},\qquad \delta\hat{\psi}_\pm=\pm\frac{\omega_z\pm\omega_0}{k_\parallel c}\delta\hat{A}_{\parallel\pm},\qquad \delta\hat{\psi}_z=\frac{\omega_0}{k_\parallel c}\delta\hat{A}_{\parallel z},\label{streamfunctions}
} where $k_\parallel = k_{\parallel 0}$ and $\psi_{0,\pm,z}$ are the components of the effective induced parallel potential (they are
proportional to the vector potential). Direct substitution of \eqref{kawpump}, \eqref{kawsidebands}, \eqref{cc}, and \eqref{streamfunctions} into \eqref{kawnonlinearwkb} gives the following coupled nonlinear vorticity equations for KAW pump, sidebands and CC: \al{\begin{gathered}
\omega_0\frac{k_\parallel^2v_A^2}{\omega_0^2}\left(1-\Gamma_0\right)\left[\frac{\omega_0^2}{k_\parallel^2v_A^2}\delta\hat{\phi}_0-\frac{b_0}{1-\Gamma_0}\delta\hat{\psi}_0\right]=-\frac{\text{i}}{2}\omega_{ci}\rho_i^2\bigg[\Lambda_{k_+}^{k_z^*}\bigg(\left(\Gamma_z-\Gamma_+\right)\delta\hat{\phi}_+\delta\hat{\phi}_z^*\\
-\frac{k_\parallel^2v_A^2\left(b_+-b_z\right)}{\omega_0\left(\omega_0+\omega_z\right)}\delta\hat{\psi}_+\delta\hat{\psi}_z^*\bigg)+\Lambda_{k_-^*}^{k_z}\bigg(\left(\Gamma_z-\Gamma_-\right)\delta\hat{\phi}_-^*\delta\hat{\phi}_z-\frac{k_\parallel^2v_A^2\left(b_--b_z\right)}{\omega_0\left(\omega_0-\omega_z^*\right)}\delta\hat{\psi}_-^*\delta\hat{\psi}_z\bigg)\bigg],\end{gathered}\label{coupledpump}
} \al{\begin{gathered}
\left(\omega_z\pm\omega_0\right)\frac{k_\parallel^2v_A^2}{\left(\omega_z\pm\omega_0\right)^2}\left(1-\Gamma_\pm\right)\left[\frac{\left(\omega_z\pm\omega_0\right)^2}{k_\parallel^2v_A^2}\delta\hat{\phi}_\pm-\frac{b_\pm}{1-\Gamma_\pm}\delta\hat{\psi}_\pm\right]=\\=-\frac{\text{i}}{2}\omega_{ci}\rho_i^2\Bigg[\left(\begin{array}{c}
\Lambda_{k_0}^{k_z} \\ \Lambda_{k_0^*}^{k_z}\end{array}
\right)\left(\Gamma_z-\Gamma_0\right)\delta\hat{\phi}_z\left(\begin{array}{c}
\delta\hat{\phi}_0 \\ \delta\hat{\phi}_0^*\end{array}
\right)-\left(\begin{array}{c}
\Lambda_{k_0}^{k_z} \\ \Lambda_{k_0^*}^{k_z}\end{array}
\right)\frac{k_\parallel^2v_A^2\left(b_0-b_z\right)}{\omega_0^2}\delta\hat{\psi}_z\left(\begin{array}{c}
\delta\hat{\psi}_0 \\ \delta\hat{\psi}_0^*\end{array}
\right)\Bigg],\raisetag{3.5\normalbaselineskip}\end{gathered}\label{coupledsidebands}
} \al{\begin{gathered}
\omega_z\left(1-\Gamma_z\right)\delta\hat{\phi}_z=-\frac{\text{i}}{2}\omega_{ci}\rho_i^2\bigg[\Lambda_{k_0}^{k_-}\bigg(\left(\Gamma_--\Gamma_0\right)\delta\hat{\phi}_0\delta\hat{\phi}_--\frac{k_\parallel^2v_A^2\left(b_0-b_-\right)}{\omega_0\left(\omega_0-\omega_z\right)}\delta\hat{\psi}_0\delta\hat{\psi}_-\bigg)\\
+\Lambda_{k_0^*}^{k_+}\bigg(\left(\Gamma_+-\Gamma_0\right)\delta\hat{\phi}_0^*\delta\hat{\phi}_+-\frac{k_\parallel^2v_A^2\left(b_0-b_+\right)}{\omega_0\left(\omega_0+\omega_z\right)}\delta\hat{\psi}_0^*\delta\hat{\psi}_+\bigg)\bigg].\end{gathered}\label{coupledCC}
} Note that, in the ideal MHD limit, the right-hand sides of these equations go to 0 since the differences involving $\Gamma_k$ functions and the terms with $b_k$ vanish. On the other hand, the measure of nonlinear behavior is given by the coupling coefficients $\Lambda_{k'}^{k''}$, which are maximized when the pump wavenumber is orthogonal to the sidebands wavenumbers. These equations must be closed by the quasineutrality conditions \al{
\left(1+\frac{1}{\tau}\right)\delta\phi_k=\frac{T_i}{ne}\langle J_k\delta g_{ki}-\delta g_{ke}\rangle,\label{deltagnl}
} where $\delta g_k$ satisfies the Frieman-Chen nonlinear gyrokinetic equation for electrons and ions \cite{Frieman1982,Zonca2015} which, in uniform plasma, reads \al{
\text{i}\left(k_\parallel v_\parallel-\omega_k\right)\delta g_k-\frac{c}{B_0}\Lambda_{k'}^{k''}\left[\langle\delta L_g\rangle_{k'}\delta g_{k''}-\langle\delta L_g\rangle_{k''}\delta g_{k'}\right]=-\frac{\text{i}e\omega_k}{T}F_M\langle\delta L_g\rangle_k.
} Solving \eqref{deltagnl} for $k_\parallel\neq0$ we obtain \cite{Zonca2015}, at the lowest (linear) order, \al{
\delta g_{ki}^{(1)}\approx\frac{e F_{Mi}J_k}{T_i}\delta L_k,\qquad \delta g_{ke}^{(1)}\approx-\frac{e F_{Me}}{T_e}\delta\psi_k,\label{lowestorder}
} with $\delta L_k=\delta\phi_k-(v_\parallel/c)\delta A_\parallel$. Substitution back into \eqref{deltagnl} yields \al{
\sigma_k\delta\phi_k-\delta\psi_k=\frac{T_e}{ne}\langle J_k\delta g_{ki}^{(2)}-\delta g_{ke}^{(2)}\rangle,\label{deltagnl2}
} with $\sigma_k$ as defined in \ref{kawsection}; however, including Landau damping gives $\sigma_k$ a negative complex imaginary part, $\sigma_k=1+\tau(1-\Gamma_k)(1-\text{i}\sqrt{\pi}\xi_e)$. The anti-Hermitian response is predominantly due to electron-Landau damping. At the next order, \al{
\delta g_{ki}^{(2)}\approx0,\qquad k_\parallel\delta g_{ke}^{(2)}\approx-\text{i}\frac{c}{B_0}\Lambda_{k'}^{k''}\left(\delta g_{k'e}\frac{\delta A_{\parallel k''}}{c}-\delta g_{k''e}\frac{\delta A_{\parallel k'}}{c}\right).\label{nextorder}
} Noting (keep in mind \eqref{streamfunctions}) \al{
\delta g_{ze}^{(1)}\approx-\frac{e}{T_e}F_{Me}\left(\delta\phi_z-\frac{v_\parallel}{c}\delta A_{\parallel z}\right)\label{deltagze}
} and substituting back into \eqref{deltagnl2}, we obtain \al{\begin{aligned}
\sigma_0\delta\hat{\phi}_0-\delta\hat{\psi}_0&=-\frac{\text{i}}{2}\omega_{ci}\rho_i^2\bigg[\Lambda_{k_+}^{k_z^*}\left(\frac{\delta\hat{\psi}_z^*}{\omega_0}-\frac{\delta\hat{\phi}_z^*}{\omega_0+\omega_z}\right)\delta\hat{\psi}_+\\&+\Lambda_{k_-^*}^{k_z}\left(\frac{\delta\hat{\psi}_z}{\omega_0}-\frac{\delta\hat{\phi}_z}{\omega_0-\omega_z^*}\right)\delta\hat{\psi}_-^*\bigg]\end{aligned}\label{electron1}
} and \al{
\hat{\sigma}_\pm\delta\hat{\phi}_\pm-\delta\hat{\psi}_\pm=\mp\frac{\text{i}\omega_{ci}\rho_i^2}{2\omega_0}\left(\begin{array}{c}\Lambda_{k_0}^{k_z} \\ \Lambda_{k_0^*}^{k_z}\end{array}\right)\left(\delta\hat{\psi}_z-\delta\hat{\phi}_z\right)\left(\begin{array}{c}\delta\hat{\psi}_0 \\ \delta\hat{\psi}_0^*\end{array}\right).\label{electron2}
} In order to close the system, the equation for $\delta\hat{\psi}_z$ must be determined independently of \eqref{deltagnl2}, which is degenerate with \eqref{coupledCC} for the CC. For wavelength longer than the collisionless skin depth, the equation for $\delta\hat{\psi}_z$ is obtained imposing the vanishing of $\delta J_{\parallel z}$ for electrons: \al{
\frac{k_\perp^2c^2}{\omega_{pe}^2}\ll1,\qquad \delta J_{\parallel ze}=-e\int v_\parallel\delta f_{ze}\,\text{d}\mathbf{v}\approx0.\label{kperpcomega}
} For $k_\parallel=0$, the nonlinear gyrokinetic equation becomes \cite{Zonca2015} \al{
\DP{\delta f_{ze}}{t}=\frac{ev_\parallel F_{Me}}{cT_e}\left[\DP{\delta A_{\parallel z}}{t}+\frac{c}{B_0}\Lambda_{k'}^{k''}\left(\delta A_{\parallel k'}\delta\psi_{k''}-\delta A_{\parallel k''}\delta\psi_{k'}\right)\right],\label{deltafze}
} from which \eqref{deltagze} readily follows. By substitution of \eqref{deltafze} into \eqref{kperpcomega} we finally obtain \al{
\DP{\delta\hat{\psi}_z}{t}=-\frac{\omega_{ci}\rho_i^2}{2}\left[\Lambda_{k_0}^{k_-}\left(\delta\hat{\psi}_0\delta\hat{\psi}_--\frac{\delta\hat{\psi}_-\delta\hat{\psi}_0}{1-\omega_z/\omega_0}\right)+\Lambda_{k_0^*}^{k_+}\left(\delta\hat{\psi}_0^*\delta\hat{\psi}_+-\frac{\delta\hat{\psi}_+\delta\hat{\psi}_0^*}{1+\omega_z/\omega_0}\right)\right],
} that is, \al{
\delta\hat{\psi}_z=\text{i}\frac{\omega_{ci}\rho_i^2}{2\omega_0}\left[\Lambda_{k_0}^{k_-}\frac{\delta\hat{\psi}_-\delta\hat{\psi}_0}{1-\omega_z/\omega_0}-\Lambda_{k_0^*}^{k_+}\frac{\delta\hat{\psi}_+\delta\hat{\psi}_0^*}{1+\omega_z/\omega_0}\right].\label{electron3}
} The nonlinear gyrokinetic vorticity equations \eqref{coupledpump}, \eqref{coupledsidebands} and \eqref{coupledCC} and the electron parallel force balance/quasineutrality equations \eqref{electron1}, \eqref{electron2} and \eqref{electron3} provide a complete set of nonlinear equations that describe the formation of CCs by KAW decay in uniform plasmas. They extend the results of \cite{Zonca2015} since they include, with \eqref{coupledpump} and \eqref{electron1}, the feedback of CC and sideband onto the KAW pump. These equations will be further extended to nonuniform plasmas in the next section. So we have six equations for 6 unknowns, to be resolved in time-dependent or time-independent way: for assigned $\delta\psi_0$ and $\delta\phi_0$, that is for an assigned pump, we reduce to 4 algebraic equations. Parametric decay instability can be investigated by \eqref{coupledsidebands} and \eqref{coupledCC} coupled with \eqref{electron2} and \eqref{electron3}, assuming a fixed amplitude KAW pump. After determining the nonlinear threshold, the further nonlinear evolution of the system can be studied by exploring the whole set of equations.

Varying the NL coupling coefficients $\Lambda_{k'}^{k''}$, as a function of $\mathbf{k}_0$ and $\mathbf{k}_z$ (with $\mathbf{k}_\pm=\mathbf{k}_z\pm \mathbf{k}_0$), we can study the local growth rate of modulational instability as a function of wavenumber and amplitude. Since for a CC $k_\parallel=0$, with $k_\parallel$ given by \eqref{kcyl}, we see that, in general screw pinch cylindrical plasma equilibria, CC existence implies a strong constraint: $n=m=0$. Our interest, meanwhile, is to maximize nonlinear coupling. Thus, it seems reasonable to consider a $\theta$-pinch equilibrium as a special case of the general screw pinch equilibrium with vanishing azimuthal magnetic field and $q \rightarrow \infty$. In this way, the CC existence imposes $n=0$ and we are allowed to freely choose $m_z$ to maximize nonlinear coupling. Meanwhile, due to the intrinsic stability of the screw pinch configuration, we can properly choose electron and ion temperature profiles to minimize Landau damping. In other words, this special case of cylindrical plasma equilibrium is particularly suitable for our studies intended at investigating CC generation by high amplitude mode converted KAW excited by an external antenna.

\subsection{Nonuniform plasma in $\theta$-pinch configuration}

Considering that $|k_\parallel v_{Ti}|\ll\omega\ll|k_\parallel v_{Te}|$, drifts effects are possibly changing ion gyrokinetic response in the short-wavelength limit. However, from equilibrium condition \eqref{radmomeq} in a $\theta$-pinch magnetic field configuration, that is, \al{
\nabla_\perp\left(\frac{B_0^2}{8\pi}+p_0\right)=0,
} one readily shows that $v_d\sim(\beta/2)v_*$ for thermal ions. Thus, the characteristic ion magnetic drift velocity, $v_d$, is typically much smaller that the corresponding ion magnetic drift, $v_*$. Therefore, we may include plasma nonuniformity only via ions and electrons diamagnetic responses. In particular, for the sake of simplicity, let us consider the effect of density gradient only, while ion and temperature profiles are taken as constants; see also \cite{Chen2022}. The former equation can be solved, with $k_\parallel\neq0$, at the lowest linear order given by \al{
\delta g_{ki}^{(1)}\approx\frac{e F_{Mi}J_k}{T_i}\left(1-\frac{\omega_{*i}}{\omega}\right)_k\delta L_k,\qquad \delta g_{ke}^{(1)}\approx-\frac{e F_{Me}}{T_e}\left(1-\frac{\omega_{*e}}{\omega}\right)_k\delta\psi_k,\label{lowestorder2}
} with the diamagnetic frequencies defined as \al{
\omega_{*i}&:=\frac{cT_i}{eB_0}k_\theta\frac{\partial_r n_0}{n_0},\label{diamfreq}\\
\omega_{*e}&:=-\frac{cT_e}{eB_0}k_\theta\frac{\partial_r n_0}{n_0}=-\tau\omega_{*i}.
} This modifies the quasineutrality condition (see \eqref{deltagnl2}) of $k_\parallel\neq0$ modes as \al{
\left[1+\tau-\tau\Gamma_k\left(1-\frac{\omega_{*i}}{\omega}\right)_k\right]\delta\phi_k-\delta\psi_k\left(1-\frac{\omega_{*e}}{\omega}\right)_k=\frac{T_e}{ne}\left\langle J_k\delta g_{ki}^{(2)}-\delta g_{ke}^{(2)}\right\rangle;
} for electrons, this equation follows the solution of \eqref{nextorder}, and yields \al{
\frac{2e}{T_i}\left\langle-\frac{T_e}{ne}\delta g_{ke}^{(2)}\right\rangle_\pm=\pm\frac{\text{i}\omega_{ci}\rho_i^2}{2\omega_0}\left(\begin{array}{c}\Lambda_{k_0}^{k_z}\delta\hat{\psi}_0 \\ \Lambda_{k_0^*}^{k_z}\delta\hat{\psi}_0^*\end{array}\right)\left[\delta\hat{\phi}_z\left(1-\frac{\omega_{*e}}{\omega}\right)_z-\delta\hat{\psi}_z\left(1-\frac{\omega_{*e}}{\omega}\right)_0\right].
} Meanwhile, $\delta g_{ki}^{(2)}$ contribution does not vanish (it vanishes for $\omega_{*i}=0$) and becomes \al{
\frac{2e}{T_i}\left\langle\frac{T_e}{ne}\delta g_{ki}^{(2)}\right\rangle_\pm=\frac{\text{i}\omega_{ci}\rho_i^2\tau}{2\left(\omega_z\pm\omega_0\right)}\left(\begin{array}{c}\Lambda_{k_0}^{k_z}F_{1+}\delta\hat{\phi}_0 \\ \Lambda_{k_0^*}^{k_z}F_{1-}\delta\hat{\phi}_0^*\end{array}\right)\delta\hat{\phi}_z\left[\left(\frac{\omega_{*i}}{\omega}\right)_0-\left(\frac{\omega_{*i}}{\omega}\right)_z\right],
} where $F_{1+}=\langle J_zJ_{k0}J_{k+}F_{Mi}/n_0\rangle$, $F_{1-}=\left\langle J_zJ_{k0^*}J_{k-}F_{Mi}/n_0\right\rangle$ and where terms proportional to $\delta A_\parallel$ have been neglected as they are $\mathcal{O}(\beta_i)$. Equation \eqref{electron2} is modified into the following sidebands quasineutrality condition \al{\begin{aligned}
\hat{\sigma}_\pm\delta\hat{\phi}_\pm-\delta\hat{\psi}_\pm&=\pm\frac{\text{i}\omega_{ci}\rho_i^2}{2\omega_0\left(1-\frac{\omega_{*e}}{\omega}\right)_\pm}\left(\begin{array}{c}\Lambda_{k_0}^{k_z}\delta\hat{\psi}_0 \\ \Lambda_{k_0^*}^{k_z}\delta\hat{\psi}_0^*\end{array}\right)\left[\delta\hat{\phi}_z\left(1-\frac{\omega_{*e}}{\omega}\right)_z-\delta\hat{\psi}_z\left(1-\frac{\omega_{*e}}{\omega}\right)_0\right]\\
&\pm\frac{\text{i}\omega_{ci}\rho_i^2}{2\left(\omega_0\pm\omega_z\right)\left(1-\frac{\omega_{*e}}{\omega}\right)_\pm}\left(\begin{array}{c}\Lambda_{k_0}^{k_z}F_{1+}\delta\hat{\phi}_0 \\ \Lambda_{k_0^*}^{k_z}F_{1-}\delta\hat{\phi}_0^*\end{array}\right)\tau\delta\hat{\phi}_z\left[\left(\frac{\omega_{*i}}{\omega}\right)_0-\left(\frac{\omega_{*i}}{\omega}\right)_z\right],\raisetag{3\normalbaselineskip}\end{aligned}\label{quasineutsid}
} with $\hat{\sigma}_\pm=\left[1+\tau-\tau\Gamma_\pm(1-\omega_{*i}/\omega)_\pm\right]/(1-\omega_{*e}/\omega)_\pm$, as shown by \cite{Chen2022}. A similar procedure can be adopted for computing the parallel force balance for electrons, keeping in mind, in massless approximation for electrons, \eqref{nextorder}, \al{
\delta g_{ze}^{(2)}\approx\frac{\text{i}cv_\parallel}{\omega_zB_0c}\Lambda_{k'}^{k''}\frac{eF_{Me}}{T_e}\left[\left(1-\frac{\omega_{*e}}{\omega}\right)_{k''}\delta\psi_{k''}\delta A_{\parallel k'}-\left(1-\frac{\omega_{*e}}{\omega}\right)_{k'}\delta\psi_{k'}\delta A_{\parallel k''}\right].\label{nextorder2}
} Following the same procedure delineated after \eqref{nextorder}, and looking for a solution to $\delta J_{\parallel ze}=0$, one can verify that \eqref{quasineutsid} and \eqref{nextorder2} yield to a simplification of electron diamagnetic effects and that \eqref{electron3} still applies, \al{
\delta\hat{\psi}_z=\text{i}\frac{\omega_{ci}\rho_i^2}{2\omega_0}\left[\Lambda_{k_0}^{k_-}\frac{\delta\hat{\psi}_-\delta\hat{\psi}_0}{1-\omega_z/\omega_0}-\Lambda_{k_0^*}^{k_+}\frac{\delta\hat{\psi}_+\delta\hat{\psi}_0^*}{1+\omega_z/\omega_0}\right].\label{electron3new}
} So, closing equations by means of vorticity equations, we get \al{\begin{gathered}
\omega_k\left[\left(1-\frac{\omega_{*i}}{\omega}\right)_k\left(1-\Gamma_k\right)\delta\hat{\phi}_k-\frac{b_kk_\parallel^2v_A^2}{\omega_k^2}\delta\hat{\psi}_k\right]=\\
=-\frac{\text{i}\omega_{ci}\rho_i^2}{2}\Lambda_{k'}^{k''}\bigg\{\Big[\left(1-\frac{\omega_{*i}}{\omega}\right)_{k''}\Gamma_{k''}-\left(1-\frac{\omega_{*i}}{\omega}\right)_{k'}\Gamma_{k'}+F_1\left(\frac{\omega_{*i}}{\omega}\right)_{k''}\\
-F_1\left(\frac{\omega_{*i}}{\omega}\right)_{k'}\Big]\delta\hat{\phi}_{k'}\delta\hat{\phi}_{k''}
-\left(\frac{k_\parallel v_A}{\omega}\right)_{k'}\left(\frac{k_\parallel v_A}{\omega}\right)_{k''}\left(b_{k'}-b_{k''}\right)\delta\hat{\psi}_{k'}\delta\hat{\psi}_{k''}\bigg\},
\end{gathered}} which can be specialized to both KAW sideband, \al{\begin{gathered}
\left(\omega_z\pm\omega_0\right)\left[\left(1-\frac{\omega_{*i}}{\omega}\right)_\pm\left(1-\Gamma_\pm\right)\delta\hat{\phi}_\pm-\frac{b_\pm k_\parallel^2v_A^2}{\left(\omega_z\pm\omega_0\right)^2}\delta\hat{\psi}_\pm\right]=\\
=-\frac{\text{i}\omega_{ci}\rho_i^2}{2}\bigg\{\Big[\left(1-\frac{\omega_{*i}}{\omega}\right)_z\Gamma_z-\left(1-\frac{\omega_{*i}}{\omega}\right)_0\Gamma_0+F_1\left(\frac{\omega_{*i}}{\omega}\right)_z\\
-F_1\left(\frac{\omega_{*i}}{\omega}\right)_0\Big]\left(\begin{array}{c}\Lambda_{k_0}^{k_z}\delta\hat{\phi}_0 \\ \Lambda_{k_0^*}^{k_z}\delta\hat{\phi}_0^*\end{array}\right)\delta\hat{\phi}_z-\frac{k_\parallel^2v_A^2}{\omega_0^2}\left(b_0-b_z\right) \left(\begin{array}{c}\Lambda_{k_0}^{k_z}\delta\hat{\psi}_0 \\ \Lambda_{k_0^*}^{k_z}\delta\hat{\psi}_0^*\end{array}\right)\delta\hat{\psi}_z\bigg\},\label{newsid}
\end{gathered}} and convective cells, \al{\begin{gathered}
\omega_z\left(1-\frac{\omega_{*i}}{\omega}\right)_z\left(1-\Gamma_z\right)\delta\hat{\phi}_z=\\
=-\frac{\text{i}\omega_{ci}\rho_i^2}{2}\bigg\{\Lambda_{k_0}^{k_-}\Big[\left(1-\frac{\omega_{*i}}{\omega}\right)_-\Gamma_--\left(1-\frac{\omega_{*i}}{\omega}\right)_0\Gamma_0+F_1\left(\frac{\omega_{*i}}{\omega}\right)_-\\
-F_1\left(\frac{\omega_{*i}}{\omega}\right)_0\Big]\delta\hat{\phi}_0\delta\hat{\phi}_--\frac{k_\parallel^2v_A^2}{\omega_0\left(\omega_0-\omega_z\right)}\left(b_0-b_-\right)\Lambda_{k_0}^{k_-}\delta\hat{\psi}_0\delta\hat{\psi}_-\\
+\Lambda_{k_0^*}^{k_+}\left[\left(1-\frac{\omega_{*i}}{\omega}\right)_+\Gamma_+-\left(1-\frac{\omega_{*i}}{\omega}\right)_0\Gamma_0+F_1\left(\frac{\omega_{*i}}{\omega}\right)_+-F_1\left(\frac{\omega_{*i}}{\omega}\right)_0\right]\delta\hat{\phi}_0^*\delta\hat{\phi}_+\\
-\frac{k_\parallel^2v_A^2}{\omega_0\left(\omega_0+\omega_z\right)}\left(b_0-b_+\right)\Lambda_{k_0^*}^{k_+}\delta\hat{\psi}_0^*\delta\hat{\psi}_+\bigg\}.\label{newcon}
\end{gathered}} In summary, \eqref{quasineutsid}, \eqref{electron3new}, \eqref{newsid}, and \eqref{newcon} are the complete set of equations to solve for computing CC modulational instability growth. They include diamagnetic effects (symmetry breaking), reducing to the previous ones for $\omega_{*i,e}\rightarrow0$, and are consistent with the analysis in \cite{Chen2022}. These theoretical analyses are an original novel contribution of the present thesis work together with the corresponding numerical solutions that will be presented in the next sections.

\section{Solution of the modulational instability}

For a more focused study, let us introduce the following simplifications that are aimed at maximizing the non-linear interactions, given that the pump KAW is the mode converted wave generated by the external antenna drive with prescribed frequency $\omega_0$ as well as (low) poloidal and azimuthal mode numbers $(n,m)$: \al{\begin{gathered}
\mathbf{k}_{\perp0}\approx \mathbf{k}_{0r}\hat{\mathbf{r}},\qquad \mathbf{k}_{\parallel0}=\mathbf{k}_\parallel\hat{\mathbf{z}}=\frac{n}{R}\hat{\mathbf{z}},\\
\mathbf{k}_{\perp z}\approx-\frac{m_z}{r}\hat{\boldsymbol{\theta}},\\
\mathbf{k}_{\perp\pm}\approx-\frac{m_z}{r}\hat{\boldsymbol{\theta}}\pm k_{0r}\hat{\mathbf{r}},\qquad \mathbf{k}_{\parallel\pm}=\pm\frac{n}{R}\hat{\mathbf{z}},\end{gathered}\label{simplifications}
} in addition to the general request $k_{\parallel z}=0$. Note that the underlying assumption here is that the pump KAW has $\mathbf{k}_\perp$ predominantly in the radial direction (as we expect, since we construct the pump/mother KAW by mode conversion at the SAW resonance) and that its behavior is predominantly oscillatory with a modulation in amplitude occurring on a scale longer than $\rho_i$. Meanwhile, the CC is characterized by a macroscale radial structure with a predominant short poloidal scale that is connected with $|m_z|\gg |m|$. This feature clearly maximizes the nonlinear coupling coefficient $\Lambda_{k_0}^{k_z}$. It will be shown in the following that this condition is also consistent with the maximization of the CC excitation rate. Thanks to these assumptions, we have \al{
b_\pm\approx b_0+b_z,\qquad b_z\approx\frac{m_z^2\rho_i^2}{r^2},\qquad b_0\approx k_{0r}^2\rho_i^2,\label{postsimpli1}
} and so \al{
\Gamma_+\approx\Gamma_-,\qquad F_{1+}\approx F_{1-}=\int_0^\infty e^{-t}J_0\left(\sqrt{2b_zt}\right)J_0\left(\sqrt{2b_0t}\right)J_0\left(\sqrt{2b_+t}\right)\,\text{d}t.\label{postsimpli2}
} Finally, in order to more easily manipulate equations, let us introduce the symbolic notations $R^\pm_1$ and $R^\pm_2$ to denote, respectively, the right-hand side of \eqref{quasineutsid} and of \eqref{newsid} for the $\pm$ sign. So, from \eqref{quasineutsid} and \eqref{newsid} we obtain \al{
\left[\left(1-\frac{\omega_{*i}}{\omega}\right)_\pm\left(1-\Gamma_\pm\right)-b_\pm\frac{k_\parallel^2v_A^2}{\left(\omega_z\pm\omega_0\right)^2}\hat{\sigma}_\pm\right]\delta\hat{\phi}_\pm=\frac{R^\pm_2}{\omega_z\pm\omega_0}-\frac{b_\pm k_\parallel^2v_A^2}{\left(\omega_z\pm\omega_0\right)^2}R^\pm_1,\label{obtain1}
} \al{\begin{aligned}
\left[\left(1-\frac{\omega_{*i}}{\omega}\right)_\pm\left(1-\Gamma_\pm\right)-b_\pm\frac{k_\parallel^2v_A^2}{\left(\omega_z\pm\omega_0\right)^2}\hat{\sigma}_\pm\right]\delta\hat{\psi}_\pm&=-R^\pm_1\left(1-\frac{\omega_{*i}}{\omega}\right)_\pm\left(1-\Gamma_\pm\right)\\
&+\frac{\hat{\sigma}_\pm R_2^\pm}{\omega_z\pm\omega_0},\label{obtain2}\end{aligned}
} where, considering \eqref{simplifications}, and noting that $\Lambda_{k_0^*}^{k_z}=-\Lambda_{k_0}^{k_z}$, \al{\begin{aligned}
R_1^\pm&=\frac{\text{i}\omega_{ci}\rho_i^2}{2\omega_0}\frac{\Lambda_{k_0}^{k_z}\sigma_0}{\left(1-\frac{\omega_{*e}}{\omega}\right)_\pm}\Bigg\{\Bigg[1+\left(\frac{\omega_{*i}}{\omega}\right)_z\tau-\left(\frac{\omega_{*i}}{\omega}\right)_z\frac{\tau F_1}{\sigma_0\left(1\pm\frac{\omega_z}{\omega_0}\right)}\Bigg]\left(\delta\hat{\phi}_z-\delta\hat{\psi}_z\right)\\
&+\Bigg[\left(\frac{\omega_{*i}}{\omega}\right)_z\tau-\left(\frac{\omega_{*i}}{\omega}\right)_z\frac{\tau F_1}{\sigma_0\left(1\pm\frac{\omega_z}{\omega_0}\right)}\Bigg]\delta\hat{\psi}_z\Bigg\}\left(\begin{array}{c}\delta\hat{\phi}_0 \\ \delta\hat{\phi}_0^*\end{array}\right),\end{aligned}
} and \al{
\begin{aligned}
\pm R_2^\pm&=\frac{\text{i}\omega_{ci}\rho_i^2}{2}\Lambda_{k_0}^{k_z}\Bigg\{\Bigg[\Gamma_0-\Gamma_z-\left(\frac{\omega_{*i}}{\omega}\right)_z\left(F_1-\Gamma_z\right)-\frac{1-\Gamma_0}{b_0}\left(b_z-b_0\right)\Bigg]\delta\hat{\psi}_z\\
&+\Bigg[\Gamma_0-\Gamma_z-\left(\frac{\omega_{*i}}{\omega}\right)_z\left(F_1-\Gamma_z\right)\Bigg]\left(\delta\hat{\phi}_z-\delta\hat{\psi}_z\right)\Bigg\}\left(\begin{array}{c}\delta\hat{\phi}_0 \\ \delta\hat{\phi}_0^*\end{array}\right).\end{aligned}
} Introducing the KAW dielectric constants for sidebands \al{
\epsilon_{A\pm}:=\left(1-\frac{\omega_{*i}}{\omega}\right)_\pm\frac{1-\Gamma_\pm}{b_\pm}-\frac{k_\parallel^2v_A^2}{\left(\omega_z\pm\omega_0\right)^2}\hat{\sigma}_\pm
} and collecting terms, equations \eqref{obtain1} and \eqref{obtain2} can be cast as \al{\begin{aligned}
\left(\omega_0\pm\omega_z\right)b_\pm\hat{\epsilon}_{A\pm}\delta\hat{\phi}_\pm&=\frac{\text{i}\omega_{ci}\rho_i^2}{2}\Lambda_{k_0}^{k_z}\left[c_\phi^\pm\left(\delta\hat{\phi}_z-\delta\hat{\psi}_z\right)+d_\phi^\pm\delta\hat{\psi}_z\right]\left(\begin{array}{c}\delta\hat{\phi}_0 \\ \delta\hat{\phi}_0^*\end{array}\right),\\
\left(\omega_0\pm\omega_z\right)b_\pm\hat{\epsilon}_{A\pm}\delta\hat{\psi}_\pm&=\frac{\text{i}\omega_{ci}\rho_i^2}{2}\Lambda_{k_0}^{k_z}\left[c_\psi^\pm\left(\delta\hat{\phi}_z-\delta\hat{\psi}_z\right)+d_\psi^\pm\delta\hat{\psi}_z\right]\left(\begin{array}{c}\delta\hat{\phi}_0 \\ \delta\hat{\phi}_0^*\end{array}\right),\end{aligned}\label{twoeq}
} where \al{\begin{aligned}
c_\phi^\pm&:=\Gamma_0-\Gamma_z-\left(\frac{\omega_{*i}}{\omega}\right)_z\left(F_1-\Gamma_z\right)\\
&-\frac{b_\pm\left(1-\Gamma_0\right)}{b_0\left(1-\frac{\omega_{*e}}{\omega}\right)_\pm\left(1\pm\frac{\omega_{z}}{\omega_0}\right)}\left[1+\left(\frac{\omega_{*i}}{\omega}\right)_z\tau\left(1-\frac{F_1}{\sigma_0\left(1\pm\frac{\omega_z}{\omega_0}\right)}\right)\right],\end{aligned}\label{cphi}
} \al{\begin{aligned}
d_\phi^\pm&:=\Gamma_0-\Gamma_z-\left(\frac{\omega_{*i}}{\omega}\right)_z\left(F_1-\Gamma_z\right)-\frac{\left(b_z-b_0\right)\left(1-\Gamma_0\right)}{b_0}\\
&-\frac{b_\pm\left(1-\Gamma_0\right)}{b_0\left(1-\frac{\omega_{*e}}{\omega}\right)_\pm\left(1\pm\frac{\omega_{z}}{\omega_0}\right)}\left[\left(\frac{\omega_{*i}}{\omega}\right)_z\tau\left(1-\frac{F_1}{\sigma_0\left(1\pm\frac{\omega_z}{\omega_0}\right)}\right)\right],\end{aligned}\label{dphi}
} \al{\begin{aligned}
c_\psi^\pm&:=\hat{\sigma}_\pm\left[\Gamma_0-\Gamma_z-\left(\frac{\omega_{*i}}{\omega}\right)_z\left(F_1-\Gamma_z\right)\right]\\
&-\frac{\left(1-\Gamma_\pm\right)\left(1-\frac{\omega_{*i}}{\omega}\right)_\pm\left(1\pm\frac{\omega_{z}}{\omega_0}\right)}{\left(1-\frac{\omega_{*e}}{\omega}\right)_\pm}\sigma_0\left[1+\left(\frac{\omega_{*i}}{\omega}\right)_z\tau\left(1-\frac{F_1}{\sigma_0\left(1\pm\frac{\omega_z}{\omega_0}\right)}\right)\right],\end{aligned}\label{cpsi}
} \al{\begin{aligned}
d_\psi^\pm&:=\hat{\sigma}_\pm\left[\Gamma_0-\Gamma_z-\left(\frac{\omega_{*i}}{\omega}\right)_z\left(F_1-\Gamma_z\right)-\frac{\left(b_z-b_0\right)\left(1-\Gamma_0\right)}{b_0}\right]\\
&-\frac{\left(1-\Gamma_\pm\right)\left(1-\frac{\omega_{*i}}{\omega}\right)_\pm\left(1\pm\frac{\omega_{z}}{\omega_0}\right)}{\left(1-\frac{\omega_{*e}}{\omega}\right)_\pm}\sigma_0\left[\left(\frac{\omega_{*i}}{\omega}\right)_z\tau\left(1-\frac{F_1}{\sigma_0\left(1\pm\frac{\omega_z}{\omega_0}\right)}\right)\right].\end{aligned}\label{dpsi}
} With \eqref{twoeq}, \eqref{cphi}, \eqref{dphi}, \eqref{cpsi}, \eqref{dpsi} we can compute the modulational instability dispersion relation. First note that \al{
\hat{\sigma}_\pm=\frac{\omega_z\left(1+\tau-\tau\Gamma_\pm\right)\pm\omega_0\left(1+\tau-\tau\Gamma_\pm\right)+\tau\Gamma_\pm\omega_{*i}}{\omega_z\pm\omega_0+\tau\omega_{*i}},\label{sigmaplusminus}
} $\omega_{*i}$ being the CC diamagnetic frequency \eqref{diamfreq}. Meanwhile, using the notation $\sigma_\pm$ for $\hat{\sigma}_\pm$ with no $\omega_{*i}$, we can write \al{\begin{aligned}
b_\pm\hat{\epsilon}_{A\pm}&=\frac{\left(1-\Gamma_\pm\right)\left(1\pm\frac{\omega_z}{\omega_0}\right)\left(1\pm\frac{\omega_z}{\omega_0}\pm\tau\frac{\omega_{*i}}{\omega_0}\right)\left(1\pm\frac{\omega_z}{\omega_0}\mp\frac{\omega_{*i}}{\omega_0}\right)}{\left(1\pm\frac{\omega_z}{\omega_0}\right)^2\left(1\pm\frac{\omega_z}{\omega_0}\pm\tau\frac{\omega_{*i}}{\omega_0}\right)}\\
&-\frac{\frac{1-\Gamma_0}{b_0\sigma_0}\sigma_\pm b_\pm\left(1\pm\frac{\omega_z}{\omega_0}\pm\frac{\tau\Gamma_\pm}{\sigma_\pm}\frac{\omega_{*i}}{\omega_0}\right)}{\left(1\pm\frac{\omega_z}{\omega_0}\right)^2\left(1\pm\frac{\omega_z}{\omega_0}\pm\tau\frac{\omega_{*i}}{\omega_0}\right)},\end{aligned}
} from which we obtain \al{\begin{gathered}
b_\pm\hat{\epsilon}_{A\pm}\left(1\pm\frac{\omega_z}{\omega_0}\right)^2\left(1\pm\frac{\omega_z}{\omega_0}\pm\frac{\tau\omega_{*i}}{\omega_0}\right)=2\left(1-\Gamma_\pm\right)\left(1\pm\frac{\omega_z}{\omega_0}\right)\left(\frac{\omega_z^2}{2\omega_0^2}-\frac{\tau\omega_{*i}^2}{2\omega_0^2}\pm\frac{\omega_z}{\omega_0}-\frac{\Delta_\pm}{\omega_0}\right)\\
\pm\frac{\omega_{*i}}{\omega_0}\left(1-\Gamma_\pm\right)\left(\left(\tau-1\right)\left(1\pm\frac{\omega_z}{\omega_0}\right)^2-\frac{\tau\Gamma_\pm b_\pm\left(1-\Gamma_0\right)}{\sigma_0b_0\left(1-\Gamma_\pm\right)}\right)\Bigg],\end{gathered}\label{obtain3}\raisetag{2.6\normalbaselineskip}
} with \al{
\frac{\Delta_\pm}{\omega_0}:=\frac{\left(1-\Gamma_0\right)b_\pm\sigma_\pm-\left(1-\Gamma_\pm\right)b_0\sigma_0}{2b_0\sigma_0\left(1-\Gamma_\pm\right)}\label{obtain4}
} being the leading-order frequency mismatch between the sideband frequency and the expected normal mode frequency. In fact, \eqref{obtain3} and \eqref{obtain4} express that KAW sidebands are not normal mode of the system and, thus, $\hat \epsilon_{A\pm} \neq 0$.

In order to simplify notation, let us adopt from now on the subscript + for denoting both upper and lower KAW sidebands. Collecting terms and noting that $\Delta_\pm=\Delta_+$, $b_\pm=b_+$, $\Gamma_\pm=\Gamma_+$ and $\sigma_\pm=\sigma_+$, we can write \al{
b_\pm\hat{\epsilon}_{A\pm}\left(1\pm\frac{\omega_z}{\omega_0}\right)\left(1\pm\frac{\omega_z}{\omega_0}\pm\frac{\tau\omega_{*i}}{\omega_0}\right)=\pm\tilde{\epsilon}_{A1}-\tilde{\epsilon}_{A2},
} where \al{\begin{gathered}
\tilde{\epsilon}_{A1}=\left(1-\Gamma_+\right)\left\{2\frac{\omega_z}{\omega_0}+\frac{\omega_{*i}}{\omega_0}\left[\tau-1-\left(\frac{2\Delta_+}{\omega_0}+1\right)\frac{\tau\Gamma_+}{\sigma_+\left(1-\frac{\omega_z^2}{\omega_0^2}\right)}\right]\right\},\\
\tilde{\epsilon}_{A2}=\left(1-\Gamma_+\right)\left\{2\frac{\Delta_+}{\omega_0}-\frac{\omega_z^2}{\omega_0^2}+\tau\frac{\omega_{*i}^2}{\omega_0^2}-\frac{\omega_z\omega_{*i}}{\omega_0^2}\left[\tau-1+\left(\frac{2\Delta_+}{\omega_0}+1\right)\frac{\tau\Gamma_+}{\sigma_+\left(1-\frac{\omega_z^2}{\omega_0^2}\right)}\right]\right\}.
\end{gathered}\raisetag{4.2\normalbaselineskip}
}
Meanwhile, equations \eqref{twoeq} become \al{\begin{aligned}
\left(\pm\tilde{\epsilon}_{A1}-\tilde{\epsilon}_{A2}\right)\delta\hat{\phi}_\pm&=\frac{\text{i}\omega_{ci}\rho_i^2
}{2\omega_0}\Lambda_{k_0}^{k_z}\left(\begin{array}{c}\delta\hat{\phi}_0 \\ \delta\hat{\phi}_0^*\end{array}\right)\left[\left(c_{\phi_2}\pm c_{\phi_1}\right)\left(\delta\hat{\phi}_z-\delta\hat{\psi}_z\right)+\left(d_{\phi_2}\pm d_{\phi_1}\right)\delta\hat{\psi}_z\right],\\
\left(\pm\tilde{\epsilon}_{A1}-\tilde{\epsilon}_{A2}\right)\delta\hat{\psi}_\pm&=\frac{\text{i}\omega_{ci}\rho_i^2
}{2\omega_0}\Lambda_{k_0}^{k_z}\left(\begin{array}{c}\delta\hat{\phi}_0 \\ \delta\hat{\phi}_0^*\end{array}\right)\left[\left(c_{\psi_2}\pm c_{\psi_1}\right)\left(\delta\hat{\phi}_z-\delta\hat{\psi}_z\right)+\left(d_{\psi_2}\pm d_{\psi_1}\right)\delta\hat{\psi}_z\right],\end{aligned}\label{epsa1minusepsa2}\raisetag{2.9\normalbaselineskip}
} where, from \eqref{cphi}, \eqref{dphi}, \eqref{cpsi}, \eqref{dpsi}, we have \al{
c_{\phi_1}=\left(\frac{\omega_z}{\omega_0}+\frac{\tau\omega_{*i}}{\omega_0}\right)\left[\Gamma_0-\Gamma_z-\frac{\omega_{*i}}{\omega_z}\left(F_1-\Gamma_z\right)\right]-\frac{\tau\omega_{*i}b_+\left(1-\Gamma_0\right)F_1}{\omega_0b_0\sigma_0\left(1-\frac{\omega_z^2}{\omega_0^2}\right)},
} \al{
c_{\phi_2}=\Gamma_0-\Gamma_z-\frac{\omega_{*i}}{\omega_z}\left(F_1-\Gamma_z\right)-\frac{b_+\left(1-\Gamma_0\right)}{b_0}\left[1+\frac{\tau\omega_{*i}}{\omega_z}-\frac{\tau\omega_{*i}F_1}{\omega_z\sigma_0\left(1-\frac{\omega_z^2}{\omega_0^2}\right)}\right],
} \al{\begin{aligned}
d_{\phi_1}&=\left(\frac{\omega_z}{\omega_0}+\frac{\tau\omega_{*i}}{\omega_0}\right)\left[\Gamma_0-\Gamma_z-\frac{\omega_{*i}}{\omega_z}\left(F_1-\Gamma_z\right)-\frac{\left(1-\Gamma_0\right)\left(b_z-b_0\right)}{b_0}\right]\\
&-\frac{\tau\omega_{*i}b_+\left(1-\Gamma_0\right)F_1}{\omega_0b_0\sigma_0\left(1-\frac{\omega_z^2}{\omega_0^2}\right)},\end{aligned}
} \al{\begin{aligned}
d_{\phi_2}&=\Gamma_0-\Gamma_z-\frac{\omega_{*i}}{\omega_z}\left(F_1-\Gamma_z\right)\\
&-\frac{1-\Gamma_0}{b_0}\left\{b_z-b_0+b_+\left[\frac{\tau\omega_{*i}}{\omega_z}\left(1-\frac{F_1}{\sigma_0\left(1-\frac{\omega_z^2}{\omega_0^2}\right)}\right)\right]\right\},\end{aligned}
} and \al{\begin{aligned}
c_{\psi_1}&=\sigma_+\left(\frac{\omega_z}{\omega_0}+\frac{\tau\omega_{*i}\Gamma_+}{\omega_0\sigma_+}\right)\left[\Gamma_0-\Gamma_z-\left(\frac{\omega_{*i}}{\omega}\right)_z\left(F_1-\Gamma_z\right)\right]\\
&-\sigma_0\left(1-\Gamma_+\right)\left[\left(\frac{2\omega_z-\omega_{*i}}{\omega_0}\right)\left(1+\frac{\tau\omega_{*i}}{\omega_z}\right)-\frac{\tau\omega_{*i}\left(\omega_z-\omega_{*i}\right)F_1}{\omega_0\omega_z\sigma_0}\right],\end{aligned}
} \al{\begin{aligned}
c_{\psi_2}&=\sigma_+\left[\Gamma_0-\Gamma_z-\left(\frac{\omega_{*i}}{\omega}\right)_z\left(F_1-\Gamma_z\right)\right]\\
&-\sigma_0\left(1-\Gamma_+\right)\left[\left(\frac{\omega_z\left(\omega_z-\omega_{*i}\right)}{\omega_0^2}+1\right)\left(1+\frac{\tau\omega_{*i}}{\omega_z}\right)-\frac{\tau\omega_{*i}F_1}{\omega_z\sigma_0}\right],\end{aligned}
} \al{\begin{aligned}
d_{\psi_1}&=\sigma_+\left(\frac{\omega_z}{\omega_0}+\frac{\tau\omega_{*i}\Gamma_+}{\omega_0\sigma_+}\right)\left[\Gamma_0-\Gamma_z-\left(\frac{\omega_{*i}}{\omega}\right)_z\left(F_1-\Gamma_z\right)-\frac{\left(1-\Gamma_0\right)\left(b_z-b_0\right)}{b_0}\right]\\
&-\sigma_0\left(1-\Gamma_+\right)\left[\frac{\tau\omega_{*i}}{\omega_0}\left(2-\frac{F_1}{\sigma_0}\right)-\frac{\tau\omega_{*i}^2}{\omega_0\omega_z}\left(1-\frac{F_1}{\sigma_0}\right)\right],\end{aligned}
} \al{\begin{aligned}
d_{\psi_2}&=\sigma_+\left[\Gamma_0-\Gamma_z-\left(\frac{\omega_{*i}}{\omega}\right)_z\left(F_1-\Gamma_z\right)-\frac{\left(1-\Gamma_0\right)\left(b_z-b_0\right)}{b_0}\right]\\
&-\sigma_0\left(1-\Gamma_+\right)\frac{\tau\omega_{*i}}{\omega_z}\left[1+\frac{\omega_z\left(\omega_z-\omega_{*i}\right)}{\omega_0^2}-\frac{F_1}{\sigma_0}\right].\end{aligned}
} Considering now \eqref{electron3new}, we write \al{\begin{aligned}
\delta\hat{\psi}_z&=\text{i}\frac{\omega_{ci}\rho_i^2}{2\omega_0}\Lambda_{k_0}^{k_z}\sigma_0\left(\frac{\delta\hat{\phi}_0\delta\hat{\psi}_-}{1-\frac{\omega_z}{\omega_0}}+\frac{\delta\hat{\phi}_0^*\delta\hat{\psi}_+}{1+\frac{\omega_z}{\omega_0}}\right)=\\
&=\text{i}\frac{\omega_{ci}\rho_i^2}{2\omega_0}\Lambda_{k_0}^{k_z}\sigma_0\frac{\delta\hat{\phi}_0\delta\hat{\psi}_-+\delta\hat{\phi}_0^*\delta\hat{\psi}_++\frac{\omega_z}{\omega_0}\left(\delta\hat{\phi}_0\delta\hat{\psi}_-+\delta\hat{\phi}_0^*\delta\hat{\psi}_+\right)}{1-\frac{\omega_z^2}{\omega_0^2}},\end{aligned}
} then, using the second of the \eqref{epsa1minusepsa2}, \al{\begin{aligned}
\left(1-\frac{\omega_z^2}{\omega_0^2}\right)\left(\tilde{\epsilon}_{A2}^2-\tilde{\epsilon}_{A1}^2\right)\delta\hat{\psi}_z=-2\left|\frac{\omega_{ci}\rho_i^2}{2\omega_0}\Lambda_{k_0}^{k_z}\delta\hat{\phi}_0\right|^2\sigma_0\Bigg\{\Big[\frac{\omega_z}{\omega_0}\left(\tilde{\epsilon}_{A1}c_{\psi_2}+\tilde{\epsilon}_{A2}c_{\psi_1}\right)-\tilde{\epsilon}_{A1}c_{\psi_1}\\
-\tilde{\epsilon}_{A2}c_{\psi_2}\Big]\left(\delta\hat{\phi}_z-\delta\hat{\psi}_z\right)+\left[\frac{\omega_z}{\omega_0}\left(\tilde{\epsilon}_{A1}d_{\psi_2}+\tilde{\epsilon}_{A2}d_{\psi_1}\right)-\tilde{\epsilon}_{A1}d_{\psi_1}-\tilde{\epsilon}_{A2}d_{\psi_2}\right]\delta\hat{\psi}_z\Bigg\}.\end{aligned}\label{final1}
\raisetag{3.4\normalbaselineskip}
} This equation reduces to those in \cite{Zonca2015} and \cite{Zoncarev} in the uniform plasma limit. The same procedure can be adopted for the CC vorticity equation, which, from \eqref{newcon}, can be rewritten as \al{
\frac{\omega_z}{\omega_0}\left(1-\frac{\omega_{*i}}{\omega_z}\right)\left(1-\Gamma_z\right)\delta\hat{\phi}_z&=-\text{i}\frac{\omega_{ci}\rho_i^2}{2\omega_0}\Lambda_{k_0}^{k_z}\Bigg\{\Bigg[\Gamma_+-\Gamma_0-\frac{\left(F_1-\Gamma_+\right)\omega_{*i}\omega_z}{\omega_0^2\left(1-\frac{\omega_z^2}{\omega_0^2}\right)}\nonumber\\
&+\frac{b_z\left(1-\Gamma_0\right)}{b_0\left(1-\frac{\omega_z^2}{\omega_0^2}\right)}\Bigg]\left(\delta\hat{\phi}_0\delta\hat{\phi}_--\delta\hat{\phi}_0^*\delta\hat{\phi}_+\right)\\
&-\left[\frac{\left(F_1-\Gamma_+\right)\omega_{*i}}{\omega_0\left(1-\frac{\omega_z^2}{\omega_0^2}\right)}-\frac{b_z\left(1-\Gamma_0\right)\omega_z}{b_0\omega_0\left(1-\frac{\omega_z^2}{\omega_0^2}\right)}\right]\left(\delta\hat{\phi}_0\delta\hat{\phi}_-+\delta\hat{\phi}_0^*\delta\hat{\phi}_+\right)\Bigg\}.\nonumber
} Again, using the first of the \eqref{epsa1minusepsa2}, this equation becomes

\al{\begin{gathered}
\frac{\omega_z}{\omega_0}\left(1-\frac{\omega_{*i}}{\omega_z}\right)\left(\tilde{\epsilon}_{A2}^2-\tilde{\epsilon}_{A1}^2\right)\left(1-\Gamma_z\right)\delta\hat{\phi}_z=\left|\frac{\omega_{ci}\rho_i^2}{2\omega_0}\Lambda_{k_0}^{k_z}\delta\hat{\phi}_0\right|^2\Bigg\{\Bigg[2\Bigg(\Gamma_+-\Gamma_0\\
-\frac{\left(F_1-\Gamma_+\right)\omega_{*i}\omega_z}{\omega_0^2-\omega_z^2}\Bigg)\left(\tilde{\epsilon}_{A1}c_{\phi_2}+\tilde{\epsilon}_{A2}c_{\phi_1}\right)+\frac{2\left(F_1-\Gamma_+\right)\omega_{*i}\omega_0}{\omega_0^2-\omega_z^2}\left(\tilde{\epsilon}_{A1}c_{\phi_1}+\tilde{\epsilon}_{A2}c_{\phi_2}\right)\\
+\frac{2b_z\left(1-\Gamma_0\right)}{b_0\left(1-\frac{\omega_z^2}{\omega_0^2}\right)}\left(\tilde{\epsilon}_{A1}c_{\psi_2}+\tilde{\epsilon}_{A2}c_{\psi_1}-\frac{\omega_z}{\omega_0}\left(\tilde{\epsilon}_{A1}c_{\psi_1}+\tilde{\epsilon}_{A2}c_{\psi_2}\right)\right)\Bigg]\left(\delta\hat{\phi}_z-\delta\hat{\psi}_z\right)\\
+\Bigg[2\left(\Gamma_+-\Gamma_0-\frac{\left(F_1-\Gamma_+\right)\omega_{*i}\omega_z}{\omega_0^2-\omega_z^2}\right)\left(\tilde{\epsilon}_{A1}d_{\phi_2}+\tilde{\epsilon}_{A2}d_{\phi_1}\right)\\
+\frac{2\left(F_1-\Gamma_+\right)\omega_{*i}\omega_0}{\omega_0^2-\omega_z^2}\left(\tilde{\epsilon}_{A1}d_{\phi_1}+\tilde{\epsilon}_{A2}d_{\phi_2}\right)\\
+\frac{2b_z\left(1-\Gamma_0\right)}{b_0\left(1-\frac{\omega_z^2}{\omega_0^2}\right)}\left(\tilde{\epsilon}_{A1}d_{\psi_2}+\tilde{\epsilon}_{A2}d_{\psi_1}-\frac{\omega_z}{\omega_0}\left(\tilde{\epsilon}_{A1}d_{\psi_1}+\tilde{\epsilon}_{A2}d_{\psi_2}\right)\right)\Bigg]\delta\hat{\psi}_z\Bigg\}.\end{gathered}\label{final2}
} Also this equation reduces to those in \cite{Zonca2015} and \cite{Zoncarev} in the uniform plasma limit. Equations \eqref{final1} and \eqref{final2} are two coupled equations that, for a given power spectrum and $\mathbf{k}_{\perp z}$ give, as a solution, $\omega_z$ and the ratio $\delta\hat{\psi}_z/\delta\hat{\phi}_z$. That is, they fully solve the modulational instability problem and CC polarization, which is generally mixed. In fact, it is not possible to postulate a priori a magnetostatic and/or electrostatic CC, and the polarization has to be computed self-consistently (the polarization issue is discussed in \cite{Zonca2015} and \cite{Zoncarev}). Given this, the present analysis has some common features with \cite{Kaw1983}, in particular the consideration of the important role of plasma nonuniformity, which will be further discussed in the next section. Compared with \cite{Kaw1983}, the present work is not limited to magnetostatic CCs and is extended to cylindrical plasmas. The standard form of CC equations, by analogy with the analyses of \cite{Zonca2015,Zoncarev}, can be cast as \al{\begin{aligned}
\frac{\tilde{\epsilon}_{A2}^2-\tilde{\epsilon}_{A1}^2}{4\left(1-\Gamma_+\right)^2\left(1+\frac{\Delta_+}{\omega_0}\right)}\delta\hat{\phi}_z&=-\alpha_\phi\left(\delta\hat{\phi}_z-\delta\hat{\psi}_z\right)+\beta_\phi\delta\hat{\psi}_z,\\
\frac{\tilde{\epsilon}_{A2}^2-\tilde{\epsilon}_{A1}^2}{4\left(1-\Gamma_+\right)^2\left(1+\frac{\Delta_+}{\omega_0}\right)}\delta\hat{\psi}_z&=-\alpha_\psi\left(\delta\hat{\phi}_z-\delta\hat{\psi}_z\right)+\beta_\psi\delta\hat{\psi}_z,\end{aligned}\label{finalcouple}
} with $\alpha_{\phi,\psi}$ and $\beta_{\phi,\psi}$ defined by inspection comparing \eqref{finalcouple} with \eqref{final1} and \eqref{final2}. In the uniform plasma limit, equations \eqref{finalcouple} represent the counterpart of equation (16) in \cite{Zonca2015} and equation (120) in \cite{Zoncarev}. In nonuniform plasmas, meanwhile, the symmetry breaking is evident by transforming $\omega_{*i} \mapsto - \omega_{*i}$. This clearly impacts the modulational instability process, which is no longer characterized by $\omega_z = 0$ at the critical threshold of pump KAW amplitude for CC excitation \cite{Zonca2015,Zoncarev}. In fact, marginally unstable CCs typically rotates in the electron diamagnetic direction, opposite to $\omega_{*i}$, consistent with \cite{Kaw1983}, and this fact is expected to influence the CC effect on plasma transport. However, the symmetry breaking effect connected with finite $\omega_{*i}$ does not violate the overall cylindrical symmetry property of the system, which is reflected by the invariance of the CC dispersion relation under the transformation $\omega_{*i} \mapsto - \omega_{*i}$ and $\omega_z \mapsto - \omega_z$.

\section{Equilibrium configuration and numerical solution}\label{NLequilibriumprofile}

The plasma equilibrium profiles and parameters introduced in section \ref{equilibriumprofile} define a generic screw pinch configuration, with the axial component of the magnetic field dominant with respect to the azimuthal one. Meanwhile, in section \ref{nonlineqsec} we have demonstrated that, in order to maximize nonlinear coupling and the generation of CCs by modulational instability, the preferred cylindrical plasma equilibrium is a $\theta$-pinch, which can be seen as a limiting case of a screw pinch with vanishing azimuthal magnetic field and/or $q \rightarrow \infty$. This situation can be achieved by lowering the value of $J_{0z}$ by several orders of magnitude so that, according to \eqref{B0thetaeq}, the azimuthal component becomes zero in any practical respect. We also adjust the value of $B_{0z}(0)$ and $n_0(0)$ for further convenience. The novel equilibrium considered hereafter is characterized by significant diamagnetic plasma response (smaller $B_{0z}$ inside the plasma) and, at the center, $\beta$ reaches up to nearly 0.4 (higher than in the previous case) to keep KAW electron Landau damping small. The antenna frequency is $\omega_0=1.53$ MHz, and, for the sake of simplicity, we assume $n_0 = m_0 = 2$. These choices move the resonant layer to $x_0=0.746$. In the following, we also assume
\begin{center}
\begin{multicols}{3}
\noindent
$n_0(0)=5\cdot10^{14}$ cm$^{-3}$\\
$J_{0z}(0)=0.7\cdot10^{4}$ statA\\
$B_{0z}(0)=1.6\cdot10^4$ G
\end{multicols}
\end{center}
while other parameters remain the same as in section \ref{equilibriumprofile}. As in chapter \ref{kawlinchap}, with the present launched antenna spectrum, diamagnetic effects can be neglected for the mode converted KAW pump. However, they are generally important for both CC and KAW sidebands. For the present application, we arbitrarily choose $\delta\hat B_{rh}/B_0=0.01$, corresponding, for the KAW pump (on the slow radial scale), to $\delta B_{\theta 0}/B_0\approx0.003$. The radial behavior of CC frequency and growth rate and the polarization are obtained with these parameters.

\begin{figure}
	\centering
	\includegraphics[scale=0.46]{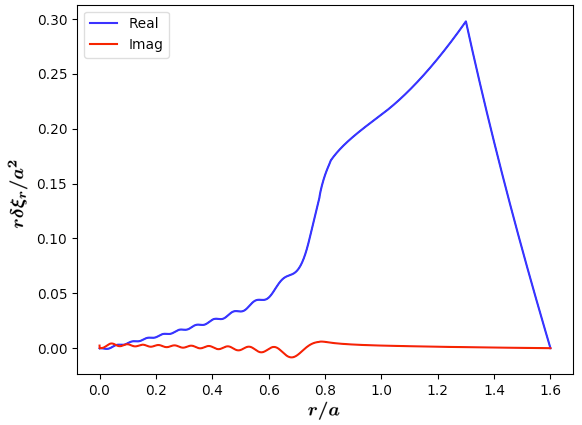}
	\includegraphics[scale=0.46]{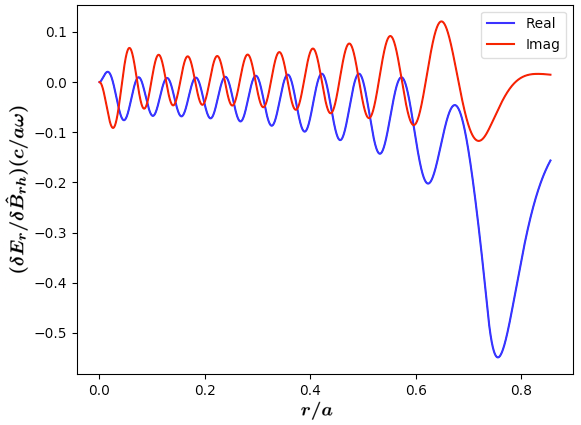}
	\caption{\emph{Fig.\ref{thetaplots1}} Radial displacement, normalized to $(-\delta \hat B_{rh}/B_0)$, and normalized radial electric field, for the $\theta$-pinch case (nominal damping).}
	\label{thetaplots1}
\end{figure}
\begin{figure}
	\centering
	\includegraphics[scale=0.457]{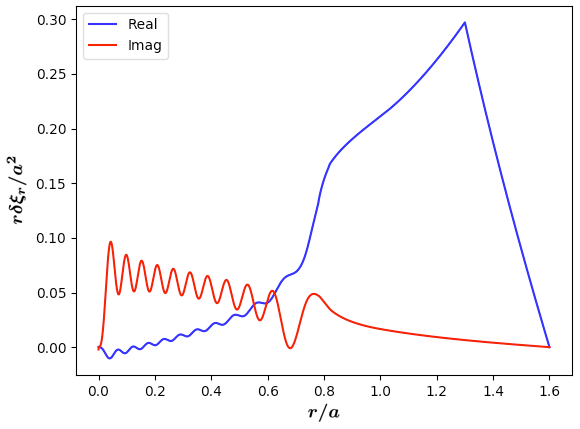}
	\includegraphics[scale=0.46]{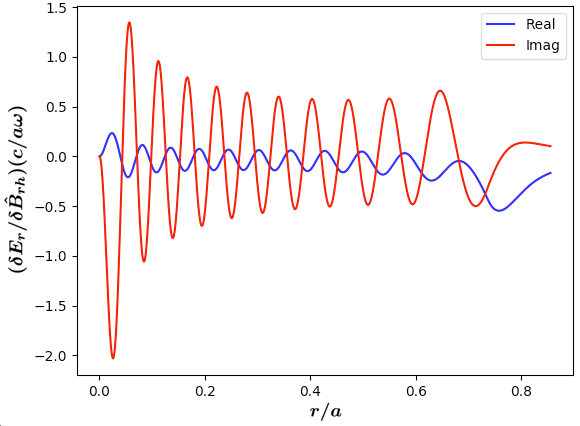}
	\caption{\emph{Fig.\ref{thetaplots2}} Radial displacement, normalized to $(-\delta \hat B_{rh}/B_0)$, and normalized radial electric field, for the $\theta$-pinch case ($10\%$ of nominal damping).}
	\label{thetaplots2}
\end{figure}
\begin{figure}
	\centering
	\includegraphics[scale=0.455]{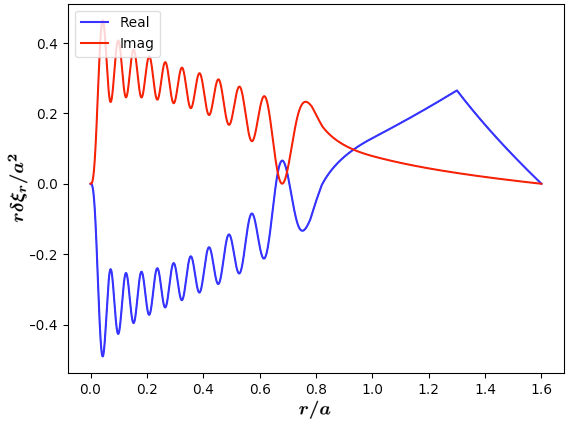}
	\includegraphics[scale=0.43]{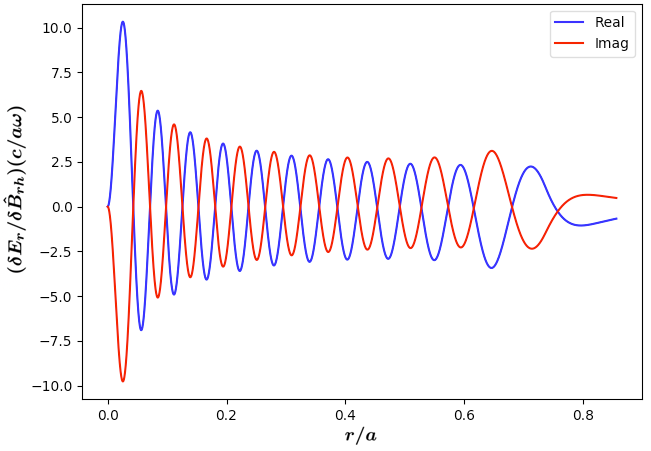}
	\caption{\emph{Fig.\ref{thetaplots3}}  Radial displacement, normalized to $(-\delta \hat B_{rh}/B_0)$, and normalized radial electric field, for the $\theta$-pinch case ($1\%$ of nominal damping).}
	\label{thetaplots3}
\end{figure}

The normalized radial displacement and radial electric field are computed for a nominal value of damping (fig.\ref{thetaplots1}), for $10\%$ of nominal value (fig.\ref{thetaplots2}), and for $1\%$ of nominal value (fig.\ref{thetaplots3}), all at the same frequency $\omega_0=1.53$ MHz: the second and third couple of figures are qualitatively identical, showing that the mode structure becomes essentially independent from damping (except for amplitude and phase effects) at about $10\%$ of the nominal value. In the following, we will adopt the aforementioned plasma parameters, corresponding to a $\theta$-pinch equilibrium, assuming the reduced damping case with $10^{-2}$ of the nominal damping to better illustrate the effects of nonlinear excitations of CCs.

\begin{figure}
	\centering
	\includegraphics[scale=0.8]{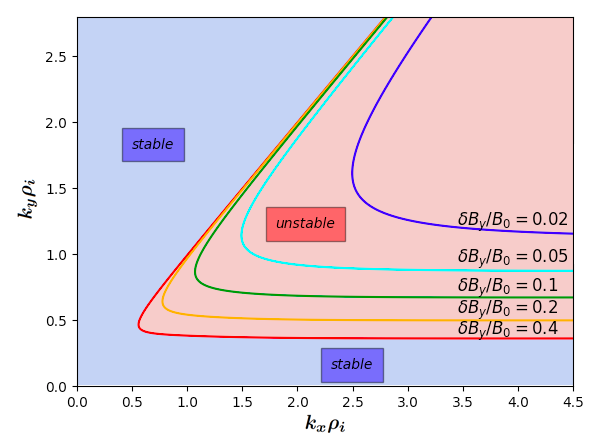}
	\caption{\emph{Fig.\ref{nodiam}} Marginal stability curves in the $(k_x\rho_i,k_y\rho_i)$-plane obtained from the present analysis in the uniform plasma limit and for $k_\parallel \rho_i = 0.02$, consistently reproducing the original results for different values of $\delta B_y/B_0$ studied in \cite{Zonca2015,Zoncarev}.}
	\label{nodiam}
	\vspace{.5cm}
	\includegraphics[scale=0.8]{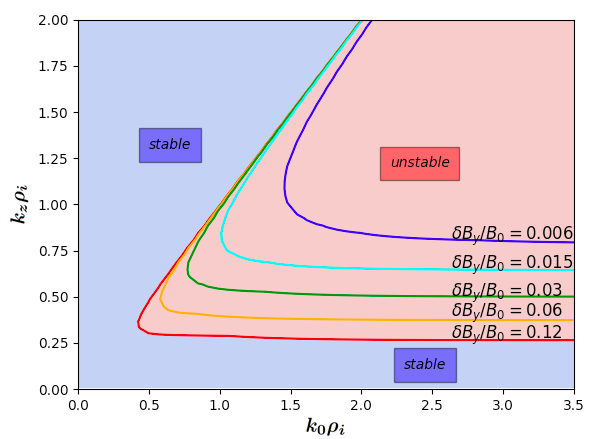}
	\caption{\emph{Fig.\ref{nodiam2}} Plot of marginal stability for the uniform case. Here $k_z$ is the wave number (predominantly in the $\theta$-direction) of the CC, while $k_0$ is that of the pump KAW (predominantly in the radial direction). Results are consistent with \cite{Zonca2015}, while quantitative differences are due to the value of $k_\parallel \rho_i = 0.003$. Note that the azimuthal KAW pump magnetic field fluctuation, $\delta B_{\theta 0}/B_0$, replaces $\delta B_y/B_0$ in \emph{fig.\ref{nodiam}}.}
	\label{nodiam2}
\end{figure}

Hereafter, we study the problem of the generation of a convective cell by modulational instability, due to nonlinear excitation by a mode converted KAW pump mode. As regards the dispersion relation, the equation that is solved is \eqref{finalcouple}, with \eqref{final1} and \eqref{final2} that only provide the definition of the coefficients entering in \eqref{finalcouple}. Correspondingly, the polarization vector is computed as solution of \eqref{finalcouple} at the obtained eigenvalue $\omega_z$\footnote{Bessel functions are computed by means of the scipy.special python library, while the function $F_1$ from \eqref{postsimpli2} is computed by quadrature using the integrate package from scipy. Finally, the roots of the dispersion function are obtained by the fmin routine from scipy.optimize, minimizing the absolute value of the dispersion function nearby a guess in the complex plane.}. We can restrict our analysis to a conventionally positive $\omega_0$ and a positive $\omega_{*i}$ (the ion diamagnetic frequency \eqref{diamfreq}) due to the fact that the CC dispersion relation is symmetric under the transformation $\omega_{*i}\mapsto-\omega_{*i}$ and $\omega_z\mapsto-\omega_z$. As a first test for our theoretical approach, we take the uniform plasma limit, that is $\omega_{*i}=0$, and recover the marginal stability curves obtained in \cite{Zonca2015} (here in fig.\ref{nodiam}). Direct connection with the uniform plasma slab of \cite{Zonca2015} is obtained, identifying the radial direction with $x$ and the azimuthal direction with $y$. Remember that $\omega_{*i}$, according to its definition \eqref{diamfreq}, depends on the CC azimuthal wave vector but also on the density gradient: so, for finite $k_y\rho_i$, $\omega_{*i}=0$ if we take the uniform plasma limit. Fig.\ref{nodiam} assumes $k_{\parallel 0} \rho_i = k_\parallel \rho_i = 0.02$, consistent with \cite{Zonca2015}, and previous results are exactly reproduced as expected. Fig.\ref{nodiam2}, meanwhile, assumes $k_{\parallel 0} \rho_i = k_\parallel \rho_i = 0.003$, consistent with typical values of the $\theta$-pinch equilibrium considered in the present analysis and with the aforementioned wave spectrum launched by the external antenna. It is evident that the behavior of the modulational instability is qualitatively the same as in \cite{Zonca2015} in the uniform plasma limit, upon replacing  $k_x \rho_i \mapsto k_0 \rho_i$ and $k_y \rho_i \mapsto k_z \rho_i$, the main quantitative difference being due to the different values of $k_\parallel \rho_i$. The plots in figs.\ref{nodiam}-\ref{nodiam2} show the upper and low limits of instability domain: the upper limit is given by $|k_0|\approx |k_z|$ (or, in the first figure, $|k_x|\approx |k_y|$), namely, in the uniform limit, modulational instability can't generate a CC whose wavelength is shorter than the pump wavelength (see \cite{Zonca2015} for further discussion about this point). This result is the main motivation for the conditions maximizing the modulational instability given by \eqref{simplifications}, \eqref{postsimpli1} and \eqref{postsimpli2}. In fact, $\mathbf{k}_z$ must simultaneously satisfy the condition $|k_z| <  |k_0|$ as well as of being perpendicular to $\mathbf{k}_0$ (predominantly radial, consistent with the mode converted KAW pump) and to the ambient magnetic field. In what follows, we demonstrate that this condition is not qualitatively modified by the inclusion of diamagnetic effects. 

\begin{figure}
	\centering
	\includegraphics[scale=0.74]{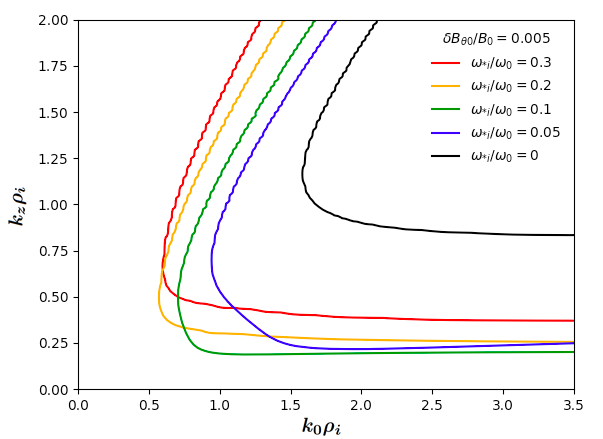}
	\caption{\emph{Fig.\ref{deltabcurves}} Marginal stability curves for the CC modulational instability as a function of plasma nonuniformity, accounted for by the local value of $\omega_{*i}/\omega_0$; in particular, $\omega_{*i}/\omega_0=0$ (black line) represents the uniform plasma limit. $\delta B_{\theta 0}/B_0 = 0.005$ is fixed and the black curve is consistent with the iso-lines displayed in \emph{fig.\ref{nodiam}}.}
	\label{deltabcurves}
\end{figure}

Let us now focus on the more general non-uniform plasma case, looking at the diamagnetic effect on marginal stability curves, CC real frequency and modulational instability growth rate, and at the related symmetry breaking effects. In fig.\ref{deltabcurves} we see the result for different values of $\omega_{*i}$, showing the crucial quantitative effect of equilibrium plasma non-uniformity. Note that, similar to previous considerations made for taking the proper physically meaningful uniform plasma limit, we can vary independently $k_{ z} \rho_i$ and $\omega_{*i}$ by acting on the density gradient. For the present $\theta$-pinch cylindrical plasma equilibrium, the range of $\omega_{*i}$ values corresponds to $|m_z| = {\cal O}(10) \gg |m_0|$. Thus, as anticipated above, diamagnetic effects are important for CC and KAW sidebands but not for the KAW pump, consistent with equations \eqref{simplifications}, \eqref{postsimpli1} and \eqref{postsimpli2}. Looking at the upper part of the diagram, we see that, differently from the uniform case, CC modulational instability can occur at wavelengths shorter than that of the pump KAW: in fact, part of the region $|k_z|>|k_0|$ gives access to instability. Actually, non-uniformity also affects the lower bound; thus, in general, it extends the region of modulational instability both for lower and for higher CC wavelengths.

\begin{figure}
	\centering
	\includegraphics[scale=0.8]{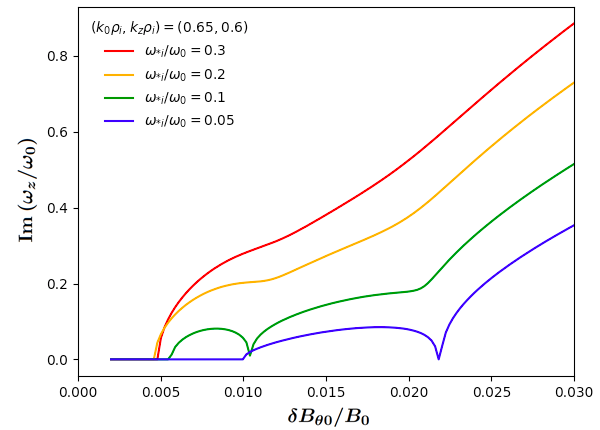}
	\caption{\emph{Fig.\ref{imfreqgrow}} Imaginary CC frequency vs. pump KAW amplitude.}
	\label{imfreqgrow}
	\vspace{1.5cm}
	\includegraphics[scale=0.8]{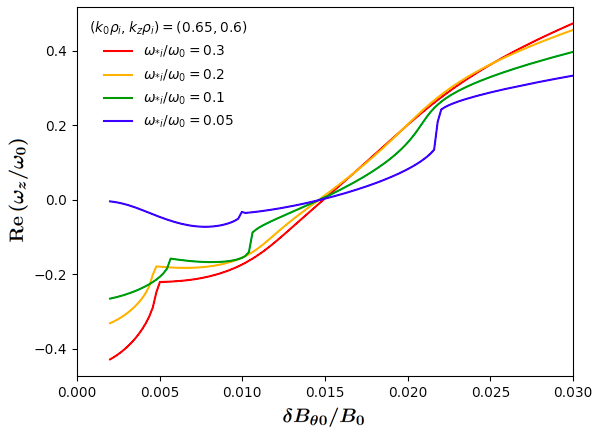}
	\caption{\emph{Fig.\ref{realfreqgrow}} Real CC frequency vs. pump KAW amplitude.}
	\label{realfreqgrow}
\end{figure}

Next, we discuss the dependence of the CC real frequency, growth rate and polarization on the amplitude of the KAW pump, $\delta B_{\theta 0}/B_0$, which is connected with the radial electric field as $\delta B_{\theta 0}/B_0 =  (k_{\parallel 0}/\omega_0) \sigma_0 (c/B_0) \delta E_{r 0}$. The input quantities for the nonlinear local dispersion relation are $k_0\rho_i$ and $k_z\rho_i$, the amplitude of the pump wave $\delta B_{\theta 0}$, the CC diamagnetic frequency $\omega_{*i}/\omega_0$, and $k_{\parallel 0} \rho_i = k_\parallel$. The solution of the dispersion relation is the CC frequency $\omega_z/\omega_0$ and polarization $\delta \hat \psi_z/\delta \hat \phi_z$. In the very low dissipation limit that we are studying the dispersion relation is polynomial with real coefficients, so with real or complex conjugate roots: a conjugate couple is made by a stable solution and an unstable solution which is the one of our interest. Notably, from fig.\ref{imfreqgrow}, representing the growth rate $\text{Im}(\omega_z/\omega_0)$, we see that for low diamagnetic frequency we actually have two thresholds, represented by two "elbows" of the curves, due to different branches of the dispersion relation and to the complex interplay of nonlinearity and nonuniformity; these elbows fade into flexes for higher diamagnetic frequencies (orange and red lines). For $\omega_{*i}/\omega_0=0.3$ the change in slope of the growth rate vs. pump amplitude is almost inappreciable. Fig.\ref{realfreqgrow}, meanwhile, shows that the CC near marginal stability rotates in the electron diamagnetic direction (whose sign is opposite with respect to the ion diamagnetic frequency, taken here as positive), consistently with the results in \cite{Kaw1983}. However, for increasing instability the CC tends to rotate in the ion diamagnetic direction. This confirms the complex interplay of nonlinearity and nonuniformity, and that the KAW and drift-wave branches mix and couple in the nonlinear regime.

\begin{figure}
	\centering
	\includegraphics[scale=0.55]{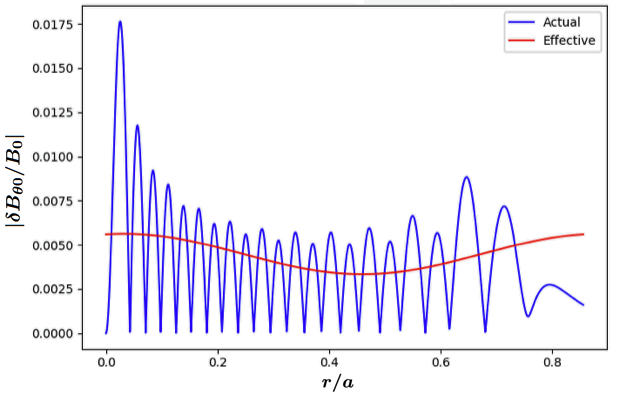}
	\caption{\emph{Fig.\ref{kaweffective}} Normalized effective KAW azimuthal magnetic field vs. the radial coordinate $r/a$.}
	\label{kaweffective}
\end{figure}

\begin{figure}[h!]
	\centering
	\includegraphics[scale=0.8]{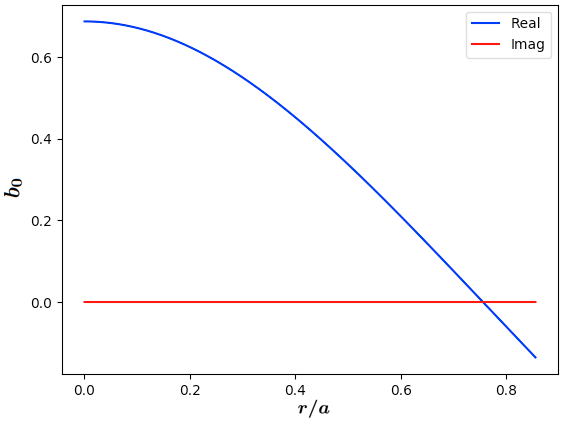}
	\caption{\emph{Fig.\ref{modinstsolplot3}} Normalized pump KAW radial wavenumber as a function of radial position.}
	\label{modinstsolplot3}
\end{figure}

\begin{figure}
	\centering
	\includegraphics[scale=0.8]{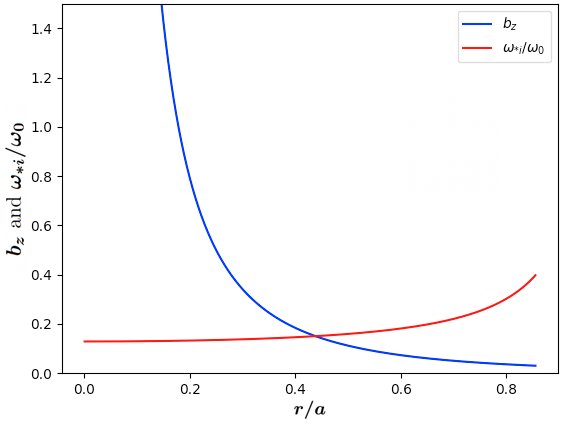}
	\caption{\emph{Fig.\ref{modinstsolplot4}} Normalized CC azimuthal wavenumber and diamagnetic frequency as a function of radial position.}
	\label{modinstsolplot4}
\end{figure}

\begin{figure}
	\centering
	\includegraphics[scale=0.6]{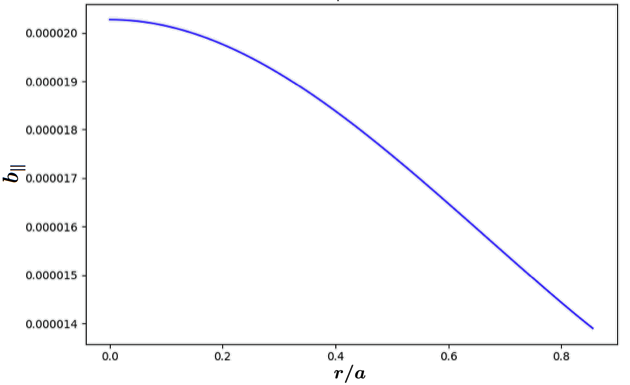}
	\caption{\emph{Fig.\ref{modinstsolplot5}} Normalized KAW parallel wavenumber as a function of radial position.}
	\label{modinstsolplot5}
\end{figure}

\begin{figure}
	\centering
	\includegraphics[scale=0.8]{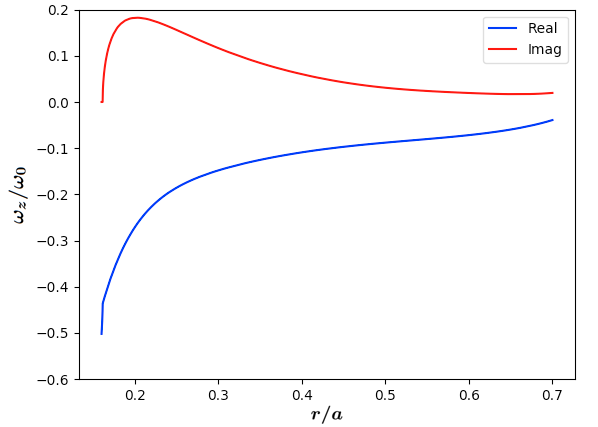}
	\caption{\emph{Fig.\ref{modinstsolplot7}} Normalized CC frequency and growth rate as a function or radial position.}
	\label{modinstsolplot7}
\end{figure}

\begin{figure}
	\centering
	\includegraphics[scale=0.8]{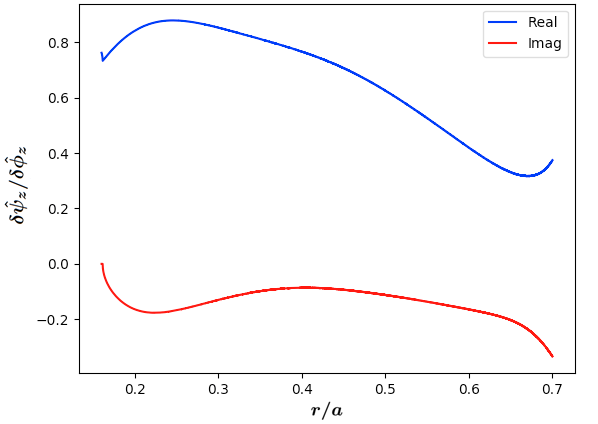}
	\caption{\emph{Fig.\ref{modinstsolplot8}} Normalized CC polarization as a function or radial position.}
	\label{modinstsolplot8}
\end{figure}

Finally, we compute the radial dependence of modulational instability, that is, the growth rate and frequency vs. the radial position, along with the polarization of the CC. From fig.\ref{thetaplots3} it is clear that the pump KAW is dominated by its short radial scale, typical of the mode converted fluctuation structure. However, it is also clear that the most "effective" CC excitation is characterized by relatively short azimuthal wavelength and relatively long radial scale, as argued above. Thus, the "effective" KAW amplitude at long radial scale is that shown in fig.\ref{kaweffective}, obtained from fig.\ref{thetaplots3} for reduced dissipation ($1\%$ of the nominal value). For the KAW pump, $b_k$ is almost purely real, as we expected, and in the range of interest for the CC modulational excitation, in agreement with previous results. In fig.\ref{modinstsolplot3} we can recognize the location of the resonant surface, where the value of $b_0$ becomes negative. The behavior of $b_0$ is essentially linear until close to the magnetic axis, where the effect of cylindrical geometry becomes important. In fig.\ref{modinstsolplot4}, $b_z$ goes as $1/r^2$, as it reflects the dominant behavior of the azimuthal wave number. Meanwhile, the normalized parallel wave number, $b_\parallel = k_\parallel^2 \rho_i^2$, is shown in fig.\ref{modinstsolplot5}. As a result, going from the resonant layer to the magnetic axis, we have a first zone where $b_{0}>b_{z}$, namely the pump KAW radial wavelength is shorter than the CC azimuthal wavelength. Then, starting at $r/a\approx 0.2$, the situation is already reversed and $b_{z}$ dominates. In the light of the stability diagram of fig.\ref{deltabcurves}, we expect this fact to have important consequences on the radial dependence of the CC growth rate. It is essentially in the region $b_0\gtrsim b_z$ that we expect the strongest growth rate, as we see from fig.\ref{modinstsolplot7} (between $r/a=0.2$ and $r/a=0.4$, approximately). Ultimately, fig.\ref{deltabcurves} suggests that CC growth rate should be suppressed near axis, because of increasing $b_z$ and decreasing $\omega_{*i}$. Growth rate should also be reduced while approaching the resonant layer since $b_k$ goes to zero (even though $\omega_{*i}$ is increasing) and we need FLR to excite CCs. The polarization of the CC is shown in fig.\ref{modinstsolplot8} and confirms the finding of \cite{Zonca2015,Zoncarev} that, in general, electrostatic and/or magnetostatic CCs are not self-consistent nonlinear solutions produced by modulational instability and the polarization state must be properly accounted for and determined case by case.

\begin{figure}
	\centering
	\includegraphics[scale=0.8]{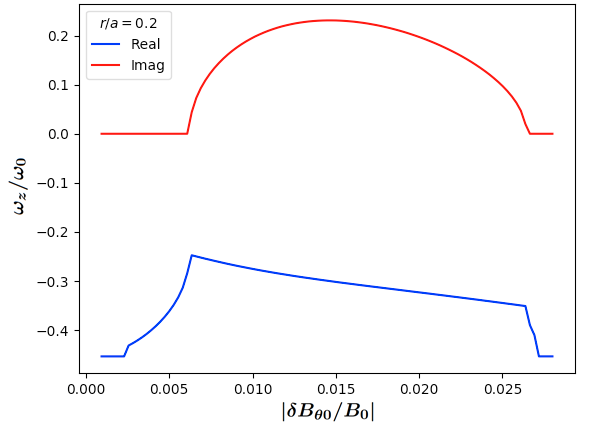}
	\caption{\emph{Fig.\ref{modinstsolplot9}} Normalized CC frequency and growth rate vs. KAW pump strength at $r/a = 0.2$.}
	\label{modinstsolplot9}
\end{figure}

\begin{figure}
	\centering
	\includegraphics[scale=0.8]{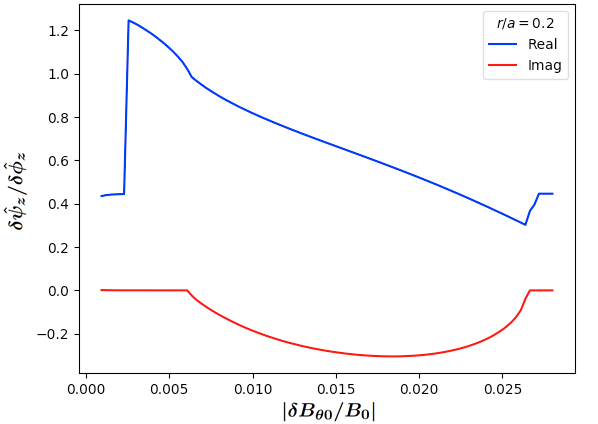}
	\caption{\emph{Fig.\ref{modinstsolplot10}} Convective cell polarization vs. KAW pump strength at $r/a = 0.2$.}
	\label{modinstsolplot10}
\end{figure}

\begin{figure}
	\centering
	\includegraphics[scale=0.8]{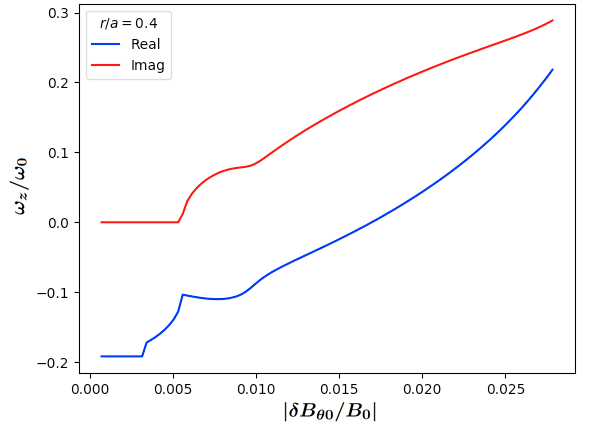}
	\caption{\emph{Fig.\ref{modinstsolplot11}} Normalized CC frequency and growth rate vs. KAW pump strength at $r/a = 0.4$.}
	\label{modinstsolplot11}
\end{figure}

\begin{figure}
	\centering
	\includegraphics[scale=0.8]{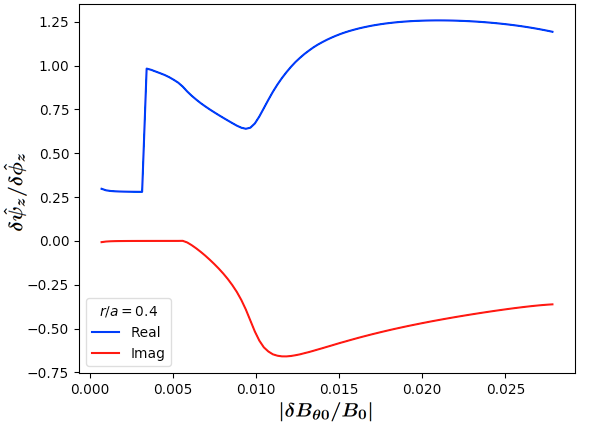}
	\caption{\emph{Fig.\ref{modinstsolplot12}} Convective cell polarization vs. KAW pump strength at $r/a = 0.4$.}
	\label{modinstsolplot12}
\end{figure}

In order to understand how the CC radial structure can be controlled externally, let us analyze the behavior of the local CC frequency, growth rate and polarization vs. the amplitude of the KAW pump at different radial locations. Figs.\ref{modinstsolplot9} and \ref{modinstsolplot10} illustrate these behaviors at $r/a = 0.2$, thus, in a region where $b_0 \lesssim b_z$. Consistent with previous findings in figs.\ref{imfreqgrow}-\ref{realfreqgrow}, the CC is effectively excited only in a limited range of KAW pump strength. Meanwhile, figs.\ref{modinstsolplot11}-\ref{modinstsolplot12} show the same behaviors at $r/a = 0.4$, where the CC growth vs. KAW pump strength monotonically increases as expected for $b_0 > b_z$. Again, these results demonstrate the complex interplay of nonlinearity and plasma nonuniformity and suggest that controlling the external antenna perturbation amplitude is a very efficient way to control the CC radial structure.

\begin{figure}
	\centering
	\includegraphics[scale=0.8]{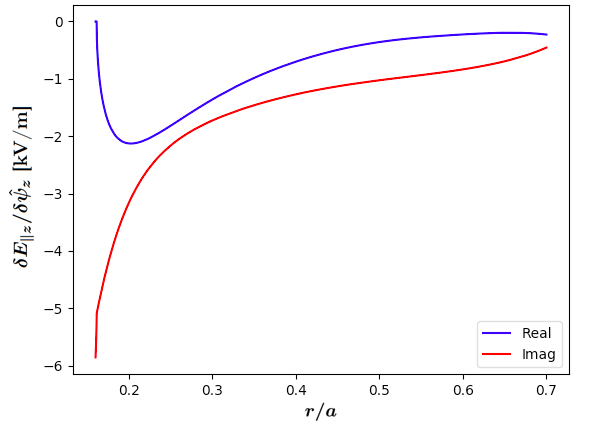}
	\caption{\emph{Fig.\ref{Parefield}} Normalized CC parallel electric field as a function of $r/a$ expressed in units of \emph{kV/m}.}
	\label{Parefield}
\end{figure}

The excitation of CCs is accompanied by the formation of inductive parallel electric field structures, which are readily derived in the form \al{
\delta E_{\parallel z}=\text{i}\frac{T_ik_{\parallel0}\omega_z}{e\omega_0}\delta\hat{\psi}_z.
} For the present case, the radial structure of $\delta E_{\parallel z}$ in units of kV/m and normalized to $\delta \hat \psi_z$ is given in fig.\ref{Parefield}. The role of parallel electric field is very important in space physics, viz. in Earth's magnetosphere, where it can significantly impact particle acceleration \cite{Zonca2015,Zoncarev}. The present study suggests that the parallel electric field values that are typically produced in laboratory plasmas for present parameters are rather small, in the order of a few tens of V/m, consistent with the typically low values observed in high-temperature magnetized plasmas of fusion interest; nonetheless, our study also suggests typical scalings with plasma parameters, which could be used for generating stronger electric fields; e.g., higher electron temperature and shorter parallel wavelength of the launched antenna wave spectrum.

\section{Summary}
In this chapter we have performed a theoretical nonlinear analysis of KAWs in a cylindrical plasma equilibrium, focusing on the spontaneous generation of convective cells by a large KAW and examining a configuration which maximizes nonlinear coupling and minimizes Landau damping; then, we have computed the numerical solutions. Earlier results of convective cell excitation by KAWs in uniform plasmas \cite{Zonca2015} have been recovered in the proper limiting case. The effect of plasma nonuniformity has been shown to importantly affect the convective cell stability both qualitatively and quantitatively via the diamagnetic response \cite{Chen2022,Kaw1983}. As new novel results, we have shown that the convective cell rotates in the electron diamagnetic direction near the critical excitation threshold, that the convective cell growth is significantly stronger (typically up to an order of magnitude) than in uniform plasmas, and that the unstable parameter region is significantly broader. Furthermore, we have shown the subtle interplay between nonlinearity and plasma nonuniformity. As a result, the plasma self-organization can be controlled by fine-tuning the amplitude of the antenna driven mode converted KAW. Finally, we have computed self-consistently the convective cell (generally mixed) polarization and radial structure of the generated inductive parallel electric field.

\chapter{\textsc{conclusions and outlook}}

In this Ph.D. thesis work we have studied the physics of shear and kinetic Alfv\'en waves (SAWs and KAWs) in a cylindrical non-uniform magnetized plasma. This is a well-known classical problem of fusion plasma physics, which we have revisited and proposed under new light and with new results.

In the first part, we have presented an in-depth review of the linear ideal magneto-hydro-dynamic (MHD) and gyrokinetic (GK) analyses of propagation and absorption of waves launched by an external antenna. The process of resonant wave absorption, at the radial position of the nonuniform plasma where the launched wave frequency matches that of the local SAW continuous spectrum, has been illustrated as the time asymptotic generation of increasingly shorter wavelengths (consistent with the classical approach of existing literature \cite{Grad1969,Hasegawa1974}), eventually breaking the MHD long wavelength assumption and requiring a GK analysis. This natural tendency of a SAW to generate localized radial singular structures in the ideal MHD model corresponds, in the GK description, to a complete transformation of the wave nature from a macroscopic (SAW) to a microscopic (KAW) fluctuation, which exists on the typical spatial scale of the thermal ion Larmor radius \cite{Hasegawa1975,Hasegawa1976,Zoncarev}. In particular, we have shown that the SAW Poynting flux is progressively and wholly converted to the KAW Poynting flux on the Airy scale, intermediate between the system macro-scale and the ion Larmor radius \cite{Hasegawa1976}. This explains why, when the mode converted KAW is strongly absorbed within a few radial wavelengths from the mode conversion layer, the power absorbed by the plasma in the GK description is essentially the same as that calculated using ideal MHD. In other words, the antenna impedance remains the same,  as noted in \cite{Hasegawa1976,Itoh1982,Itoh1983}, regardless of the model adopted for computing it.

The second part of this work focused on the linear and nonlinear behaviors of weakly damped mode converted KAW, that is on the case where the micro-scale fluctuation may propagate up to the magnetic axis without significant absorption. This case has little relevance, if any, for the classical problem of plasma heating, but it has great interest for analyzing the nonlinear behaviors of KAWs in non-uniform cylindrical plasmas. As such, this problem is the core of the novel contribution of the present Ph.D. thesis research. In the linear limit, we have demonstrated that the plasma region inside the mode conversion layer behaves as a "resonant cavity"; that is, when the antenna frequency matches a given discrete spectrum, the plasma response is stronger, exhibiting the clear behavior of a resonantly driven and weakly damped oscillator. Not surprisingly, this behavior, demonstrated here for the first time for the KAW spectrum, is qualitatively similar to that of mode converted electron Bernstein waves, described in \cite{Buchsbaum1966}. This behavior also is of particular interest for our scope of investigating nonlinear behaviors induced by large amplitude KAWs. In fact, by proper tuning of the antenna frequency, mode converted KAW can be pumped at significantly high amplitude, time asymptotically, even with a relatively small antenna power. Thus, after a brief review of the parametric decay process underlying nonlinear wave excitation, we have focused on the spontaneous generation of convective cells (CCs) by modulational instability of a weakly damped large amplitude mode converted KAW in a non-uniform cylindrical plasma equilibrium; in particular, we have focused on a $\theta$-pinch equilibrium, with purely axial ambient magnetic field, in order to maximize nonlinear wave-wave couplings. Earlier results have been recovered in the limiting case of uniform plasmas \cite{Zonca2015}. Meanwhile, as new novel results, we have discussed and demonstrated qualitative and quantitative effects of equilibrium non-uniformity, accounted for by the diamagnetic plasma response. First of all, the convective cell preferentially rotates in the electron diamagnetic direction near the critical excitation threshold, consistent with the prediction for magnetostatic CCs \cite{Kaw1983}: this means that the parity is broken for rotations about the symmetry axis of the cylindrical equilibrium, and that plasma rotation of the self-organized nonlinear plasma equilibrium can be controlled by the choice of the antenna launched wave spectrum. Furthermore, the modulational instability growth rate is significantly stronger (typically up to an order of magnitude) than in uniform plasmas, and the unstable parameter region is significantly broader: this is analogous to the similar effect discussed recently in \cite{Chen2022} for KAW parametric decay instability and, thus, confirms the crucial role of symmetry breaking wave-wave couplings in reactor-relevant fusion plasmas. As a novel result of this work, we have also shown that, due to the subtle interplay between nonlinearity and plasma non-uniformity, the self-organized nonlinear plasma equilibrium can be controlled by fine-tuning the amplitude of the mode converted KAW; in fact, local CC rotation frequency, growth rate and polarization crucially depend on diamagnetic plasma response as well as on the spectrum of both pump KAW and CC. This is a consequence of cylindrical geometry and causes, e.g., the CC growth rate not to generally have a monotonic behavior vs. the mode converted KAW amplitude. Finally, we have self-consistently computed the convective cell polarization, which is generally mixed, and the radial structure of the generated inductive parallel electric field.

The implications of the novel findings of this Ph.D. thesis work are diverse \cite{DeFabrizio2022a,DeFabrizio2022b} and all connected with the fact that Alfvén waves, as fundamental electromagnetic oscillations in magnetized plasmas, can importantly contribute to heating and transport processes of charged particles. Having adopted a nonuniform magnetized plasma 
equilibrium in cylindrical geometry is of fundamental importance since it addresses the essential physics elements that must be included into realistic models. In particular, our analytical and numerical results on the generation of convective cells can serve as test-bed for gyrokinetic and fully kinetic or hybrid codes. In fact, the present work shows the importance of accurately accounting for multiple spatiotemporal scales on the same footing for the correct description of the 
underlying physics, which can serve as precious verification test. At present, comparisons are in progress within the CNPS\footnote{Center for Nonlinear Plasma Science, https://www.afs.enea.it/zonca/CNPS/} collaboration network of our findings against TRIMEG  \cite{Lu2019,Lu2021} and STRUPHY \cite{Holderied2021} numerical simulation results. As a further and future step, comparisons will be extended to the gyrokinetic electron and fully kinetic ion description of GeFi code \cite{Lin2005}. This effort will imply the first step in the implementation of GeFi in realistic toroidal fusion plasmas equilibria as well as the verification of the validity limits of the GK reduced description, which is the foundation of TRIMEG and of other GK codes.

Despite the simplified cylindrical geometry, adopted in this work, the novel results discussed here bear important implications for toroidal fusion plasmas applications. We introduced the key assumption of weak KAW absorption to investigate the properties of wave structures propagating farther away from the mode conversion layer and to shed light into the corresponding nonlinear physics in nonuniform plasmas. In toroidal plasmas, similar conditions could be realized and externally controlled, viz. in the weak magnetic shear core regions of tokamaks and stellarators \cite{Chu2006}. Controlling the finite parallel electric field by magnetic field configuration would allow KAWs to reach magnetic axis and efficiently drive plasma current in the near-axis region by electron Landau damping of non-symmetric fluctuation spectrum.

Furthermore, it is worth noting that KAW spectrum in the toroidal fusion plasma core is expected to be nearly isotropic due to parametric decay processes enhanced by plasma nonuniformity \cite{Chen2022} and strongly excited by energetic particles \cite{Chen2016,Zhao2011,Zonca2014}. Thus, zonal field structures predominantly depending on the radial flux coordinate could be readily excited, while poloidally varying CCs would be suppressed due to the incompatibility of $k_\parallel = 0$ with finite magnetic shear. In other words, the situation would be opposite than here, where CCs dominate in the simplified $\theta$-pinch configuration and nonlinear excitation of zonal field structures is weaker. However, other than these notable but not so crucial differences, similarities of the case discussed in this work with the near-axis region in the core of toroidal plasmas is evident, as evident is the relevance of symmetry-breaking effects, connected with plasma nonuniformity, on convective transport and current drive \cite{Chen2022}.

Looking further ahead, it is worthwhile making a further comment on possible applications of the present findings due to the remarkable applications of KAWs in laboratory, space, and astrophysical plasmas. Our research results suggest an ideal "table-top" plasma experiment where, by means of an external antenna, one can control the formation of self-organized plasma states resulting from the nonlinear interaction between KAW and CC. Given the relatively simple cylindrical geometry, a proper set of diagnostics could be conceived with the proper spatiotemporal resolutions to actually measure the underlying physical processes and the corresponding characteristic spatial and temporal scales. In particular, this approach may apply to the generation of parallel electric fields with applications to particle acceleration and transport. In other words, the study proposed in this Ph.D. thesis work could result into a simple and useful paradigm for verifying and validating nonlinear physics, supported by theory, by use of advanced kinetic (GeFi), gyrokinetic (TRIMEG), and hybrid (STRUPHY) codes as well as of experimental measurements.

\cleardoublepage

\printindex
\end{document}